\documentclass[preprint]{aastex}

\usepackage{natbib}
\shorttitle{3-D Radiation Transfer in Young Stellar Objects}
\shortauthors{Whitney et al.}

\begin{document}


\title{Three-Dimensional Radiation Transfer in Young Stellar Objects}


\author{B. A. Whitney\altaffilmark{1,2}, T. P. Robitaille\altaffilmark{3}, J. E. Bjorkman\altaffilmark{4}, R. Dong\altaffilmark{5},  M. J. Wolff\altaffilmark{2}, K. Wood\altaffilmark{6}, \& J. Honor\altaffilmark{1}}


\altaffiltext{1}{University of Wisconsin, 475 N. Charter St., Madison, WI 53706, bwhitney@astro.wisc.edu}
\altaffiltext{2}{Space Science Institute, 4750 Walnut Street, Suite 205, Boulder, CO 80301}
\altaffiltext{3}{Max-Planck-Institute for Astronomy, K\"onigstuhl 17, 69117 Heidelberg, Germany}
\altaffiltext{4}{Ritter Observatory, MS 113, Department of Physics and Astronomy, University of Toledo, Toledo, OH
    43606-3390}
\altaffiltext{5}{Department of Astrophysical Sciences, Princeton University, Princeton, NJ 08544}
\altaffiltext{6}{School of Physics \& Astronomy, University of St Andrews, 
North Haugh, St Andrews, Fife, KY16 9AD}


\begin{abstract}
We have updated our publicly available dust radiative transfer code (\textsc{Hochunk3d}) to include new emission processes
and various 3-D geometries appropriate for forming
stars.   The 3-D geometries include warps and spirals
in disks, accretion hotspots on the central star, fractal clumping density enhancements, and misaligned inner disks.  Additional
axisymmetric (2-D) features include gaps in disks and envelopes, ``puffed-up inner rims" in disks, multiple bipolar
cavity walls, and
 iteration of disk vertical structure assuming hydrostatic equilibrium.
We include the option for simple power-law envelope geometry, which combined with fractal clumping,
and bipolar cavities, can be used to model evolved stars as well as protostars.
We include non-thermal emission from PAHs and very small grains, and external illumination from
the interstellar radiation field.
The grid structure was modified to allow multiple dust species in each cell; based on this, a simple
prescription is implemented to model dust stratification.  

We describe these features in detail, and show example calculations of each.  
Some of the more interesting results include the following:
1)  Outflow cavities may be more clumpy than infalling envelopes. 2) PAH emission in high-mass stars may be a better indicator of evolutionary stage than the broadband SED slope; and related to this, 3) externally illuminated clumps and high-mass
stars in optically thin clouds can masquerade as YSOs.  4) Our hydrostatic equilibrium models suggest that
dust settling is likely ubiquitous in T Tauri disks, in agreement with previous observations.

\end{abstract}


\keywords{circumstellar matter---dust, extinction---polarization---radiative transfer---stars: formation---stars: pre-main sequence}



\section{Introduction}

Protostars generally consist of young stars surrounded by disks, infalling envelopes and outflows.  In low-mass protostars, the central star or stars do not reach the main sequence until well after the envelope and most of the disk has dispersed, whereas high-mass protostars may already be undergoing core hydrogen burning during the early envelope accretion stage.
Recent high-spatial resolution, multi-wavelength and time-resolved observations have revealed three-dimensional
structures in disks such as warps, spiral density patterns, gaps, and misaligned inner disks that can confirm and challenge theories of planet formation
\citep{heap00, golimowski06, ahmic09,grady09,morales11,dong12a,dong12b,grady13}.  
Observations of protostellar envelopes hope to determine, among other things, the dynamics of early collapse, including how quickly disks form and fragment, and how  outflows interact with the infalling envelopes and eventually disperse them \citep{reipurth01,bally07,tobin08,chiang10,tobin10,tobin11,tobin12,loinard12}.  Because low-mass protostars are more numerous and have longer timescales than high-mass, their evolutionary timescales are reasonably well-understood, though it is of great interest to know how these timescales (e.g., disk survival lifetimes) 
vary in different environments.  For high-mass protostars, it is of interest to learn more about their evolutionary timescales:  how long-lived are their disks, and do they form planets?   

Thus even though we have had an explosion of sensitive infrared imaging, photometry, and spectroscopy of protostars from e.g., the {\it Spitzer} Space Telescope, Hubble Space Telescope, Herschel Space Observatory, and several  ground-based observatories, we still have many interesting science questions to explore with these datasets.   We developed a 2-D dust radiation transfer code for protostars in anticipation of the {\it Spitzer} Space Telescope data \citep[Papers I, II, \& III, respectively]{whitney03p2,whitney03p1,whitney04b}.  These were used to classify evolutionary stages on many large datasets \citep[for example,][]{whitney08,sewilo10,povich10,povich11}.  Other groups have also developed publicly available radiation transfer codes suited for protostars \citep{wolf03_mc3d,dullemond04,pinte06, pinte09,robitaille11}.   

Data provided by several observatories have revealed geometric structures and physical processes that were not included in our earlier codes:  Polycyclic Aromatic Hydrocarbon (PAH) molecule and very small grain (VSG) emission, external illumination by the interstellar radiation field, and 3-D structures.  This paper describes our new modifications and initial results using the  code, called \textsc{Hochunk3d}\footnote{available at \url{http://gemelli.spacescience.org/$\sim$bwhitney/codes} and \url{http://www.astro.wisc.edu/protostars}}.  \S2 describes the numerical improvements and \S3 describes the physics enhancements with initial results using the code.  
\S4 concludes with a  summary of the results using the new code.




\section{Description of the Code}

The code described in Paper I solves the radiation transfer equation for any material in radiative equilibrium, in axisymmetric geometries illuminated by a central source \citep[for a review of Monte Carlo radiation transfer see][] {whitney11}.  The geometry incorporated into the code to model protostars
and T Tauri stars includes an accretion disk \citep{shakura73}, infalling envelope \citep{ulrich76, terebey84}, and bipolar outflow cavity.  The analytic formulae used to describe these geometries are given in Paper I.  Papers II, III, and \citet{robitaille06} illustrate how the code can be used to model a range of evolutionary stages.  Here we give a brief description of some numerical improvements to the code, and then in \S\ref{physics} describe
the physics enhancements and scientific results.

The \textsc{Hochunk3d} code has been converted to Fortran 95, which allows a specification of 1-D, 2-D, or 3-D grids at runtime.  The code is parallelized so it can be run on multiple processors on one machine, or on multiple machines in a network.
As in previous versions, the code outputs Spectral Energy Distributions (SEDs) at a number of viewing angles ($\theta$ and $\phi$) specified in the input file.  An optional raytracing algorithm (called ``peeling-off'') produces a high signal-to-noise SED and broadband images in a specified direction; this has been expanded to allow multiple viewing angles.   A set of broadband filter functions is included in the code.   The instructions manual describes how to add new filters.  Movies and light curves can be made for a sequence of viewing angles.  The images have intensity units of MJy/sr.  

The output can be formatted in ASCII and unformatted fortran files as before, or in FITS format.  FITS output uses much less disk
space and is not machine dependent.    If the FITS output is specified, there is an option to create an image cube where the third dimension is wavelength.  Thus a movie can be made as a wavelength sequence.  Plotting programs for examining input and output are available with the program distribution, including ones that convolve images with instrumental resolution.   The images presented in this paper are 299x299 pixels with no
convolution of instrumental point-spread-functions.
Our code distribution includes all of the models presented in this
paper as examples.

Certain features are not included in our latest distribution because they may are easily implemented in the
new radiation transfer code, \textsc{Hyperion}, developed by \citet{robitaille11}.  These include multiple stellar sources (both finite-size and point-source).  The \textsc{Hyperion} code includes the option for various grids: cartesian, spherical-polar, cylindrical, and adaptive cartesian (octree and adaptive mesh refinement).  The cartesian and adaptive cartesian are probably better suited for incorporating density grids from dynamical simulations.   Another advance in the \textsc{Hyperion} code is the use of raytracing to produce higher signal-to-noise at the longest wavelengths.  Finally, the \textsc{Hyperion} code has 3-D diffusion in the dense regions of the axisymmetric disk, whereas \textsc{Hochunk3d} is using 1-D diffusion in either the radial or polar direction depending on the grid cell location.  \textsc{Hyperion} uses a python interface for generating input and output.   \textsc{Hochunk3d} uses text input, and the supplied plotting programs use IDL.

We tested the code as each new feature was implemented.   We have developed three independent versions of our codes, including  the \textsc{Hyperion} code \citep{robitaille11}, and tested between them.  The \textsc{Hyperion} code has been benchmarked with other codes as well \citep{robitaille11,pinte09}.


\section{Physics Enhancements and Initial Science Results}
\label{physics}

\subsection{Temperature Correction}

Our code now includes two options for temperature correction: The \citet{bjorkman01} method corrects the temperature and emission spectrum in each grid cell as photons are absorbed, and does not require iteration.  The \citet{lucy99} method very efficiently
calculates temperature by summing photon path lengths instead of absorptions in a grid cell.  Photons are re-emitted immediately, as in \citet{bjorkman01} but the temperature of the grid cell is not updated until an iteration of the simulation is complete.  The temperature usually converges in a short number of iterations (3-5 is typical); this efficiency is due to the fact that
flux is conserved exactly across all surfaces \citep{lucy99}.
The two temperature algorithms give identical results (within noise). 
\citet{whitney11} summarizes in more detail the differences between these
methods and the relative efficiencies in different situations.  In short, their relative efficiencies depend on the
number of grid cells and the optical depth in the grid.  In 3-D simulations with many grid cells, fewer photons are absorbed per grid cell so it takes more photons to calculate an accurate temperature than to compute a clean output spectrum.   The Lucy method is more efficient here because
the number of photons that simply pass through a grid cell is orders of magnitude larger than the number that are
absorbed.   In the 2-D simulations we have set up for protostars, the two methods have similar efficiencies.   
Our default is to use the Lucy method, since it has similar run-times, and since iteration is required for both the PAH/VSG emission
 (\S\ref{s_pah}) and the vertical hydrostatical equilibrium solution (\S\ref{s_hseq}).

\subsection{Gridding Structure and Dust Settling} \label{grid_dust}

The grids are spherical-polar ($r,\theta,\phi$) as in previous versions. The temperature and density grids now have an added dimension (in addition to the 3 spatial dimensions), which allows for multiple dust species.  The code is set up currently for 8 different dust types to be distributed in various or all locations; this number can be expanded arbitrarily to incorporate, for example,
different grain sizes or species.    
This makes it easy to implement a simplified version of dust stratification, which leads to larger grains in the midplane than the upper layers 
\citep{dalessio01,wood02_hh30,dalessio06,furlan05}.
Previously we allowed for
different grain models in different regions of the disk, with a larger-grain model in the high-density
regions.   
Now we allow for two disks (and the envelope) to co-exist, each with different grain properties.  The user can specify the geometry of
each disk; for example, making the one with the larger grains have a flatter profile to mimic dust stratification.  
Table \ref{t_dust} shows the 8 different dust types, their locations
in the protostar, and the dust files we use.
New dust models with format compatible with our code can be computed using a publicly available code.\footnote{available at \url{http://github.com/hyperion-rt/bhmie}}

\begin{figure}
\epsscale{1.0}
\includegraphics[angle=0,width=3.2in]{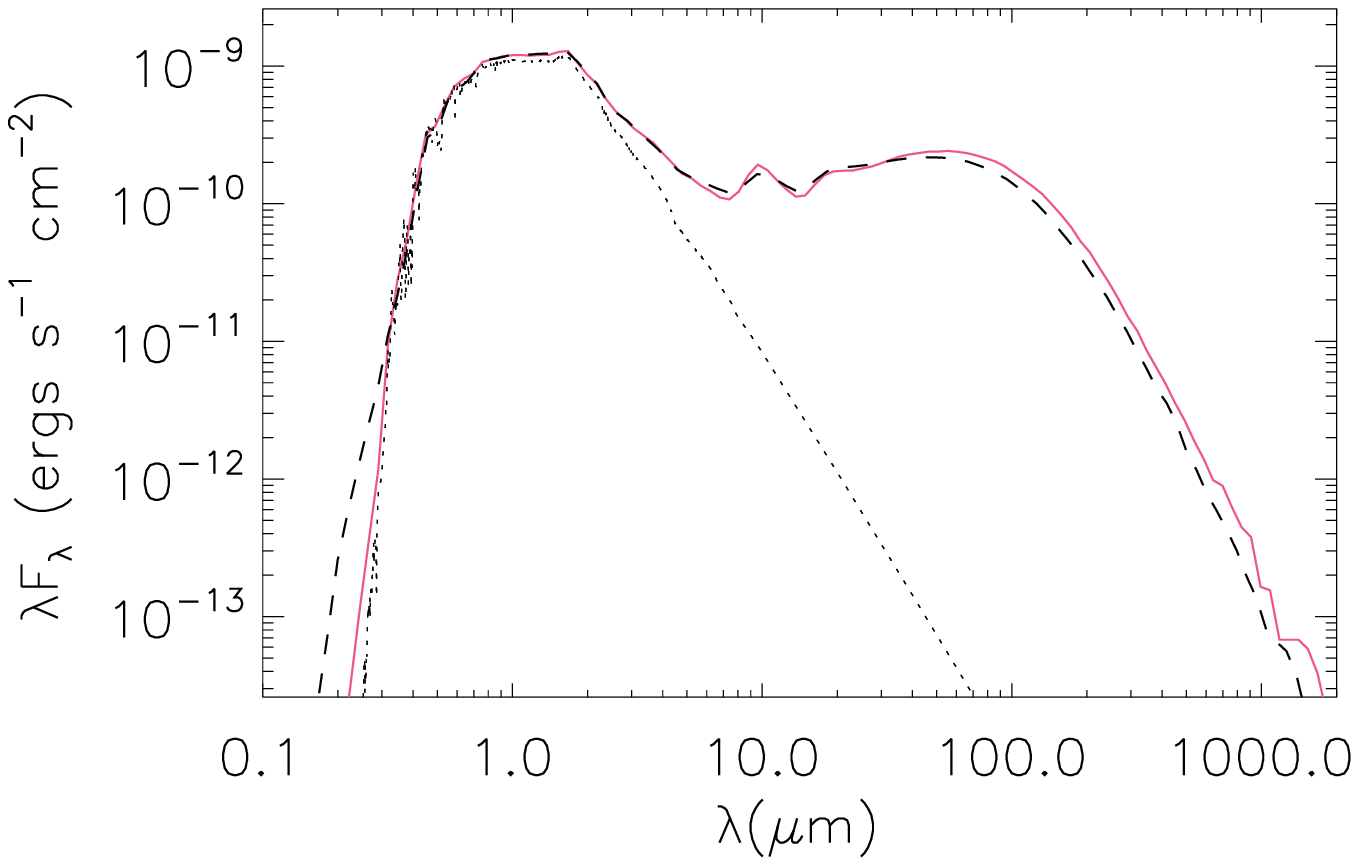}
\includegraphics[angle=0,width=3.2in]{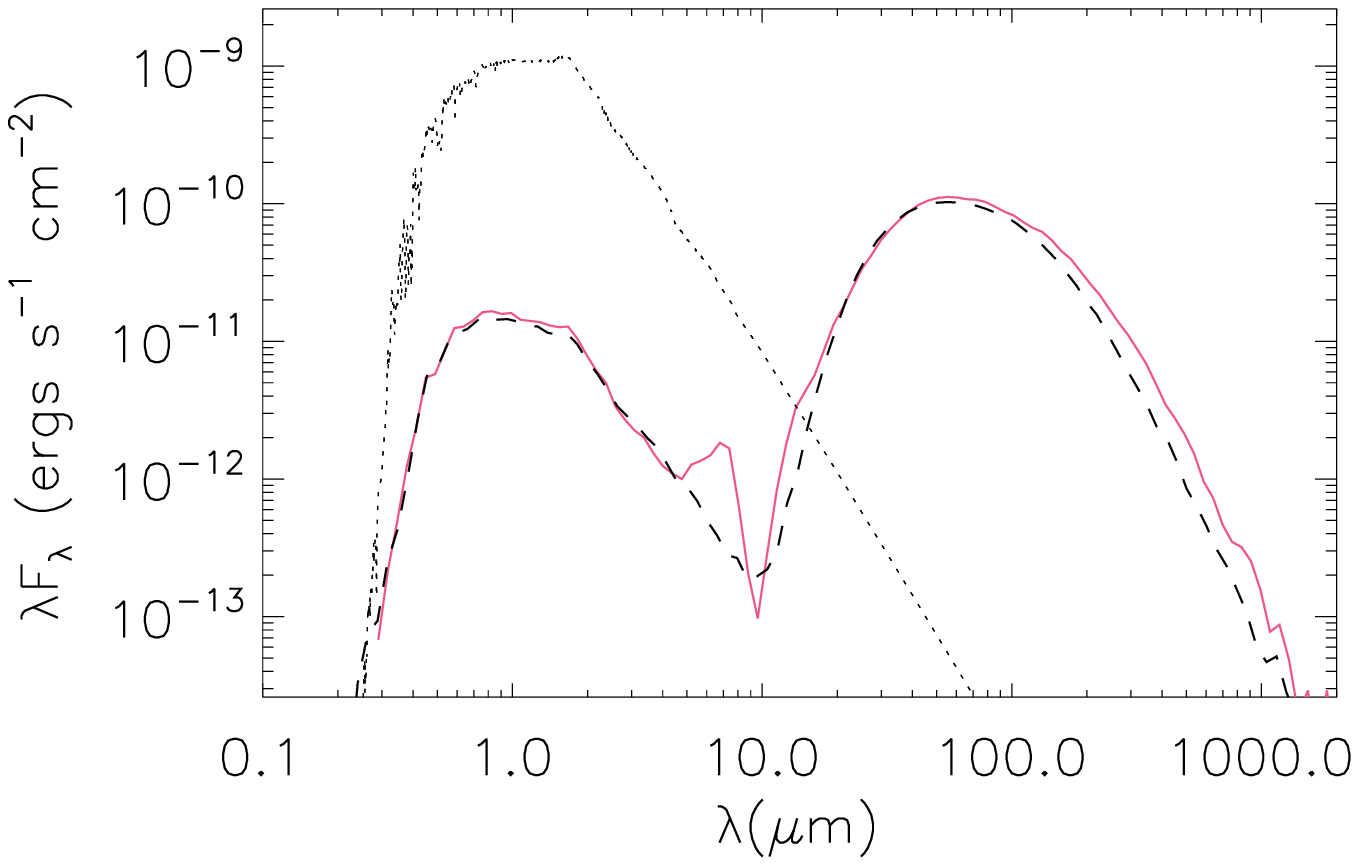}
\caption{Top:  T Tauri + disk (Class II) model, new code compared to Paper II version at viewing angles of 18\arcdeg (left) and 81\arcdeg(right).  The pink solid line is computed with the new code, and the black dashed line is plotted with the old code.  The black dotted line is the input stellar spectrum.   The newer model (pink) has a lower disk accretion rate so the ultraviolet flux is lower.  At longer wavelengths, the differences are due to the different dust distributions in the disk in the newer model  (\S\ref{grid_dust}).
\label{f_sed_modcII}}
\end{figure}

Figure \ref{f_sed_modcII} shows a comparison of a T Tauri star + disk model with the one in Paper II (the Class II model in Figure 3b).  
The model parameters are chosen to be typical for a low mass T Tauri star:
$T_\star=4000$K, $L_\star=1 L_\sun$, $M_{\rm disk} = 0.01 M_\sun$, $\dot M_{\rm disk} = 4.6\times10^{-9} M_\sun/$yr. 
Disk accretion contributes an additional 0.03 $L_\sun$ bringing the total luminosity to 1.03 $L_\sun$.
This disk accretion rate is lower than the one used in Paper II ($7.5\times10^{-9} M_\sun/$yr).
This shows in the shortest wavelengths, where the Paper II model has higher ultraviolet flux.
We do not expect an exact match to the SED in Paper II because, as stated above, we now use two dust disks with different scale heights overlaid on each other rather than having the dust separated into 2 regions of a single disk.
To account for the effect of dust stratification, the disk with larger grains has half the scale height as the disk with smaller
grains.
The plotted fluxes assume a distance to the source of 140 pc.

\begin{figure}
\epsscale{1.0}
\plotone{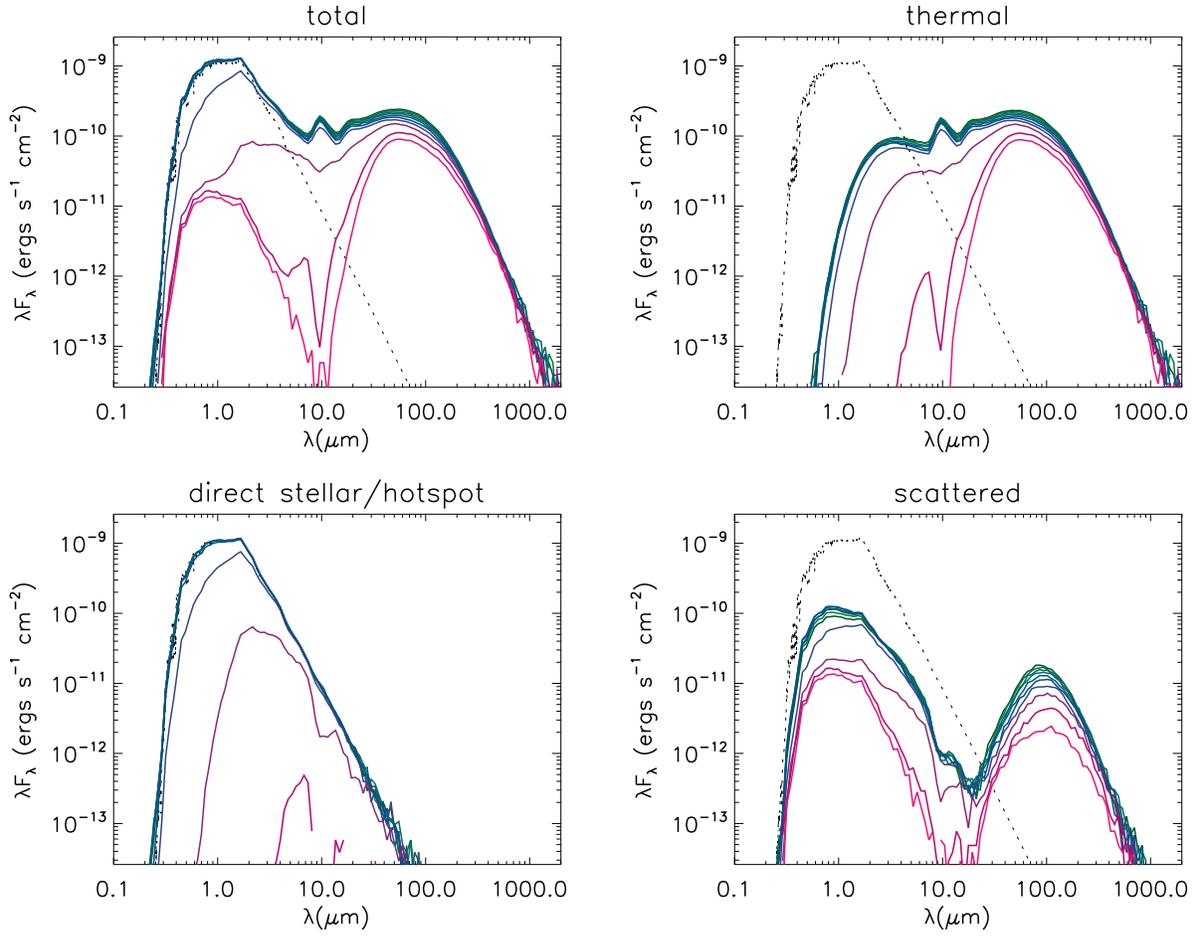}
\caption{SEDs at 10 viewing angles for the T Tauri + disk model (pole-on in green, edge-on in pink).  The different panels show different components of the SEDs as shown in the labels.
\label{f_sed4t_modcII}}
\end{figure}

\begin{figure}
\epsscale{1.0}
\plotone{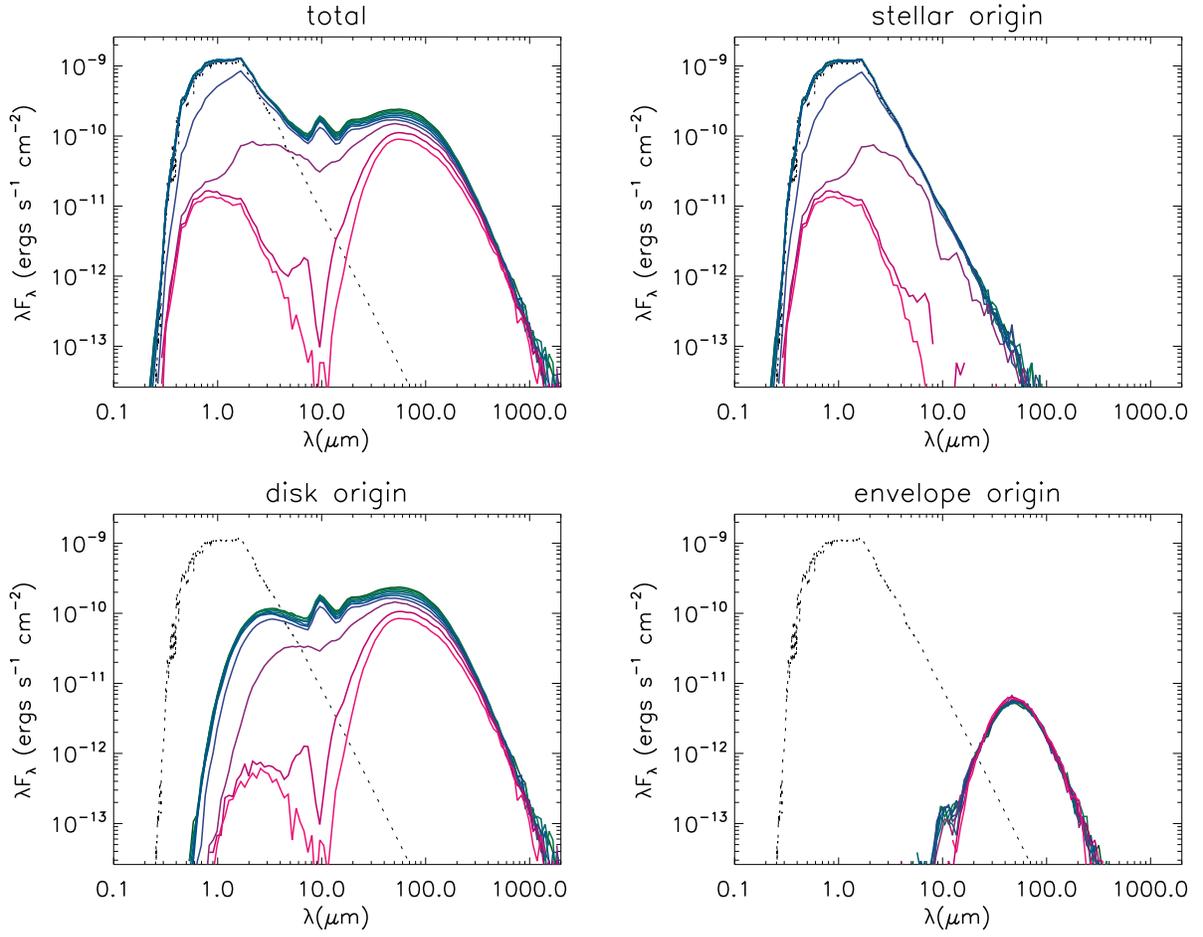}
\caption{SEDs at 10 viewing angles for the T Tauri + disk model (pole-on in green, edge-on in pink).  The different panels show different origins of the emerging photons, as shown in the labels.
\label{f_sed4o_modcII}}
\end{figure}

Figure \ref{f_sed4t_modcII} shows different emission components of the SED from this disk model:  total, thermal, stellar/hotspot, and scattered.  This provides  insight into, for example, at what wavelengths scattering, direct stellar, or thermal can dominate the emission.  Note that the scattering bump at long wavelengths is due to the large grains in the disk which have a significant scattering albedo at long wavelengths.
Figure \ref{f_sed4o_modcII} shows different origins of the outgoing photons:  total, stellar, disk and envelope.  While all of the photons originally came from the star or disk accretion (or external illumination if that option is chosen), if they were absorbed and re-emitted, the origin refers to the place in which the final re-emission took place. 

We have re-run the 6 models from Paper II, with the parameters tuned to get good agreement with the SEDs in Paper II.   Note that these are not meant to be best-estimate parameters of various evolutionary stages, but are just examples.  

\subsection{PAH/VSG emission and Phantom YSOs} \label{s_pah}

Polycyclic Aromatic Hydrocarbons (PAHs) are large molecules.  Both these and very small grains (VSGs) are not in thermal equilibrium with the radiation field; however, their emissivities can be computed based on the specific energy, $\dot{A}$, absorbed in each grid cell.  Following \citet{wood08}, \citet{robitaille11}, and \citet{robitaille12}, we use pre-computed emissivity tables for several different values of specific energy.  In the Monte Carlo method, a photon has an interaction when a randomly sampled optical depth is reached (computed based on the opacity of the medium); the particular interaction is then sampled based on the relative probabilities of each, which is directly related to their opacities.  The PAH/VSG emission is simply another interaction like scattering or thermal emission.  The PAH/VSG photon is emitted from the emissivity spectrum, based on the specific energy of the radiation field.  This method requires iteration to determine $\dot{A}$ in each grid cell; however, we are already iterating  for the temperature calculation, so the PAH calculation adds no noticeable additional cpu time.   The complicated physics of PAH/VSG emission is incorporated into the calculation of the emissivity tables.   

We use energy density tables calculated by \citet{draine_li07} for a fixed illuminating spectral shape. The actual illuminating spectral shape will be different, but this approximation is mitigated by quantifying the radiation field using specific energy, which is the  integral of the mean intensity multiplied by the opacity of the grains.  This measures the radiation absorbed by the grains.  In effect, the spectral shape of the radiation field is taken into account to first order.  This is discussed in more detail in \citet{robitaille12}, where they note that the ionization state assumed for the PAHs is the bigger source of error.
Because of this approximation, our implementation is appropriate for such applications as interpreting evolutionary stages of protostars that emit PAHs/VSG, but is not appropriate for investigating the detailed nature of the PAHs (e.g., ionization stage).
A more appropriate code for analyzing the physics of PAH emission in protostellar disks/envelopes may be the \textsc{Radmc-3d} code\footnote{available at \url{http://www.ita.uni-heidelberg.de/$\sim$dullemond/software/radmc-3d/}} \citep{dullemond04}. 
 
\citet{siebenmorgen10} proposed that low-mass T Tauri stars do not emit PAH emission because high-energy (FUV, EUV, X-ray) emission from magnetospheric accretion onto the shocked stellar surface destroys the PAHs and VSGs in the surrounding disk/envelope.  In addition, the weaker radiation field in T Tauri stars gives fewer positively ionized PAHs \citep{visser07}.
The low-mass models in our code distribution therefore include no PAH/VSG emission (however, this can easily be added in by the user).  

\begin{figure}
\epsscale{1.0}
\includegraphics[angle=0,width=3.2in]{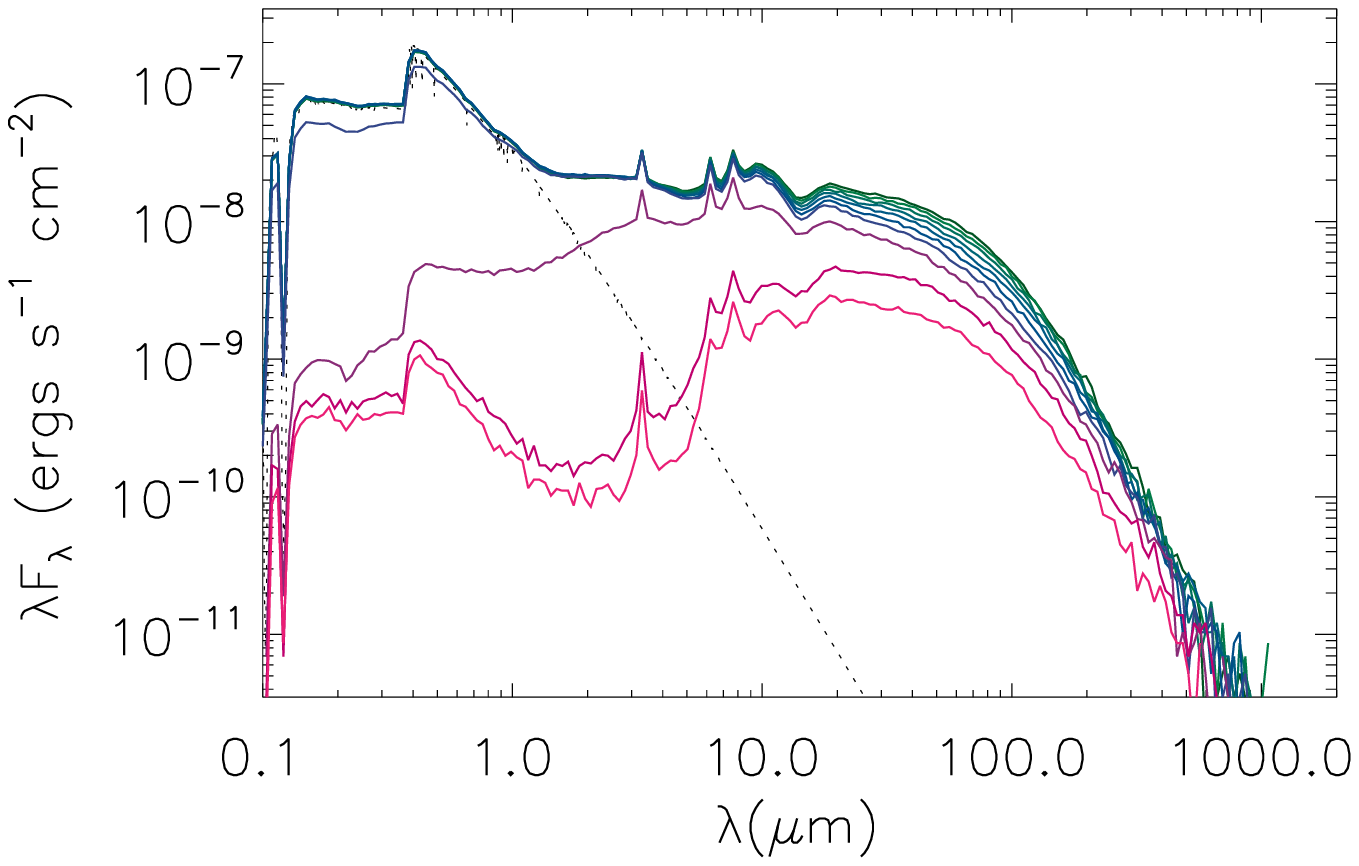}
\includegraphics[angle=0,width=3.2in]{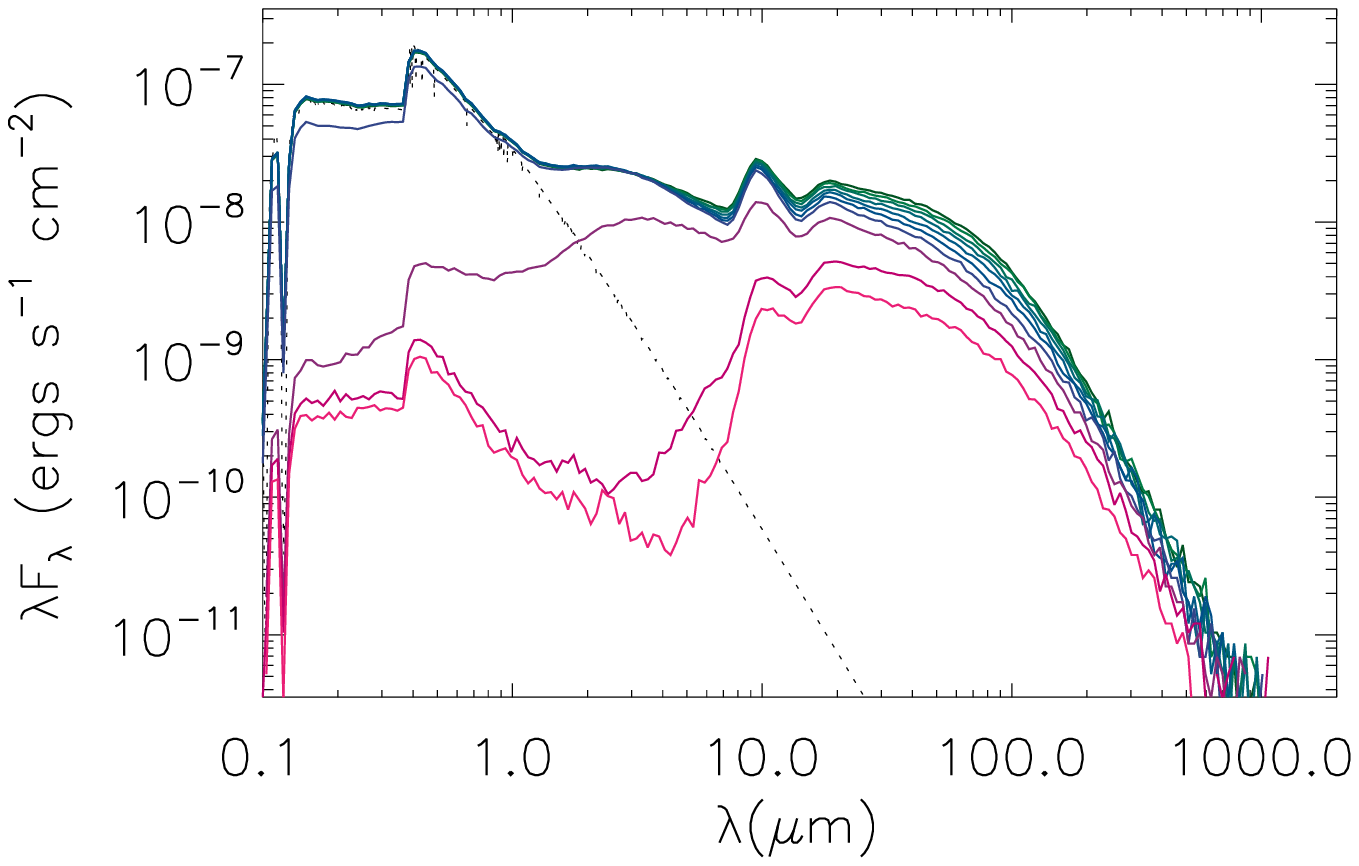}
\includegraphics[angle=0,width=3.2in]{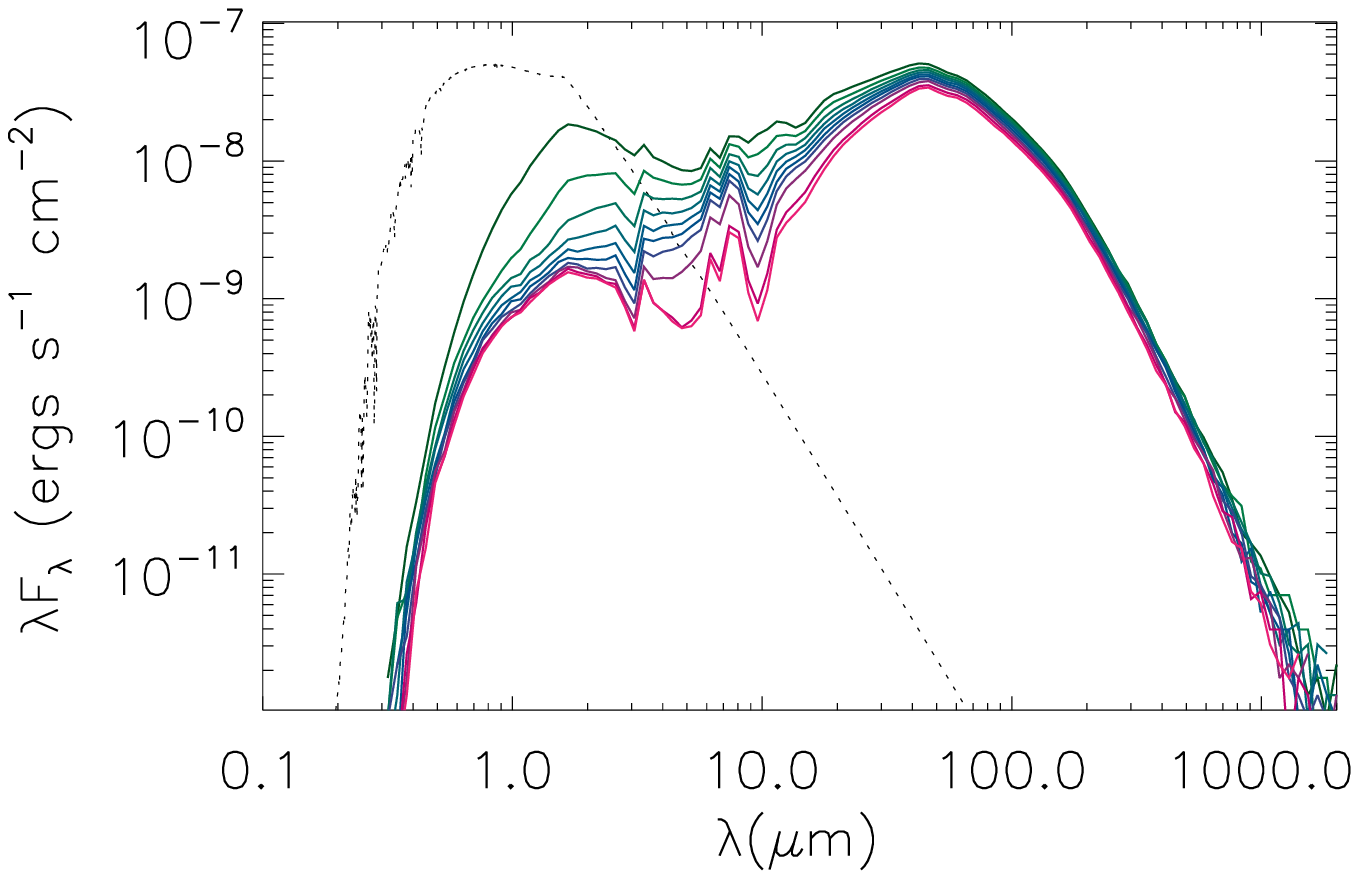}
\includegraphics[angle=0,width=3.2in]{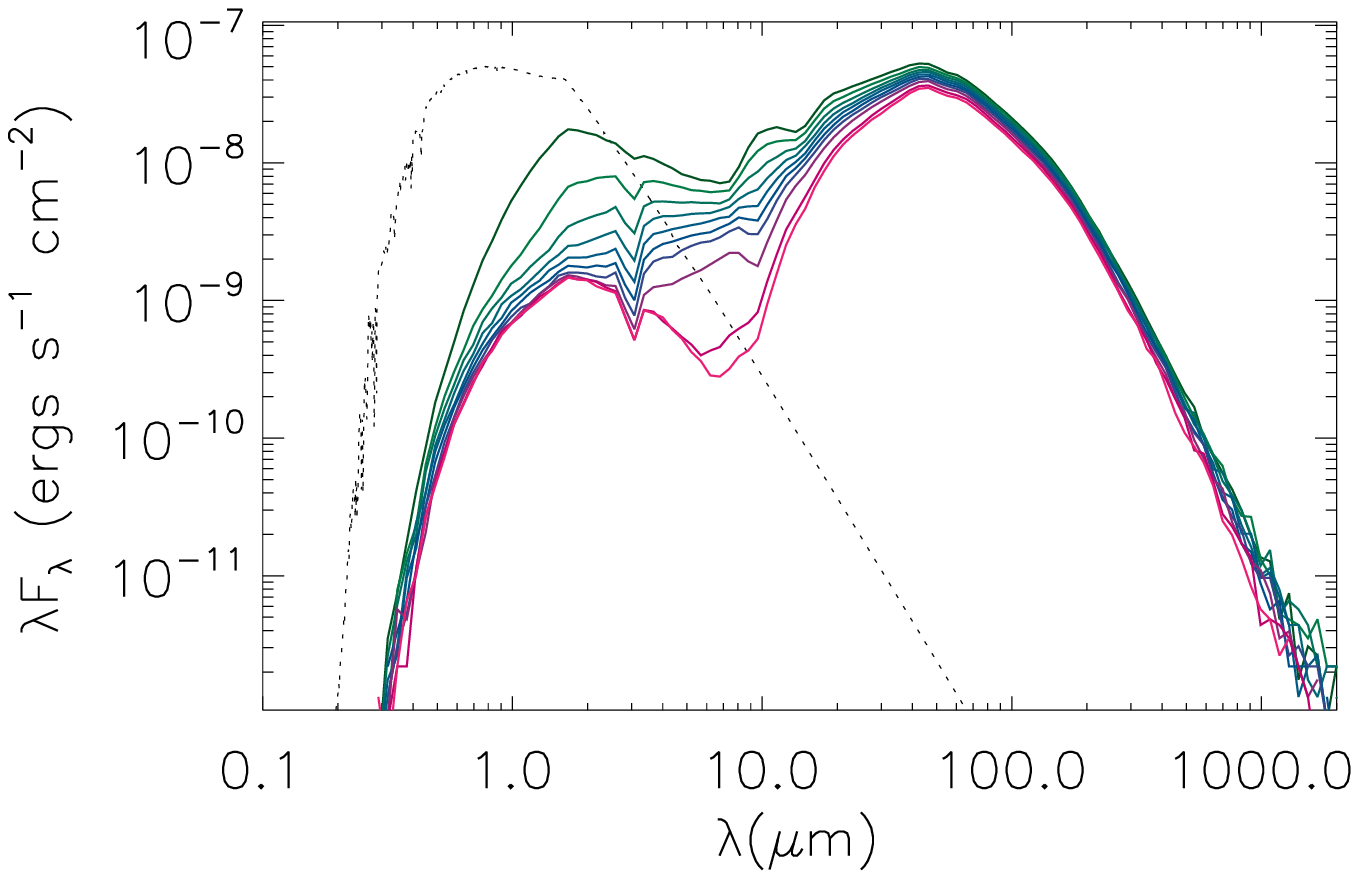}
\caption{Top:  AeBe star+disk with PAH emission (left) and without (right). 
Bottom:  embedded AeBe protostar with PAH emission (left) and without (right) for 10 viewing angles (pole-on in green, edge-on in pink).  The black dotted line is the input stellar spectrum. 
\label{f_AeBe_sed}}
\end{figure}

\citet{dullemond07}
investigated the effects of sedimentation and PAH abundance in AeBe disks.  Since our
dust models and stratification parameters are different, we can only do a qualitative comparison to their models (Figure 2 in 
their paper).  
The model parameters are:
$T_\star=10000$K, $R_\star= 2.5 R_\sun$, $M_{\rm disk} = 0.01 M_\sun$, $\dot M_{\rm disk} = 1.2\times10^{-8} M_\sun/$yr. 
The system luminosity is 56 $L_\sun$.
We assume a distance to the source of 100 pc to give the same flux as \citet{dullemond07}.
Our models, shown in the top panels of Figure \ref{f_AeBe_sed}, produce similar SEDs.  Our spectral resolution is lower so the PAH features are not as sharp.  

The bottom panels in Figure \ref{f_AeBe_sed} show an embedded AeBe protostar.  Since this is a younger source,
the stellar and envelope parameters are different:  $T_\star=4,500$K, $R_\star= 8 R_\sun$, $M_{\rm disk} = 0.05 M_\sun$, $\dot M_{\rm env} = 1\times10^{-5} M_\sun/$yr. 
Both the embedded and the AeBe disk sources show relatively stronger PAH emission in the more edge-on viewing angles.   
This is because the warm {\it thermal} dust from the inner regions of the disk and envelope are more obscured from view at
edge-on viewing angles, whereas the PAH emission can be excited as long as the higher energy (near-UV)
photons have an unobscured path in certain directions (usually the polar regions) and upper layers of the disk.

\begin{figure}
\epsscale{1.0}
\includegraphics[angle=0,width=3.2in]{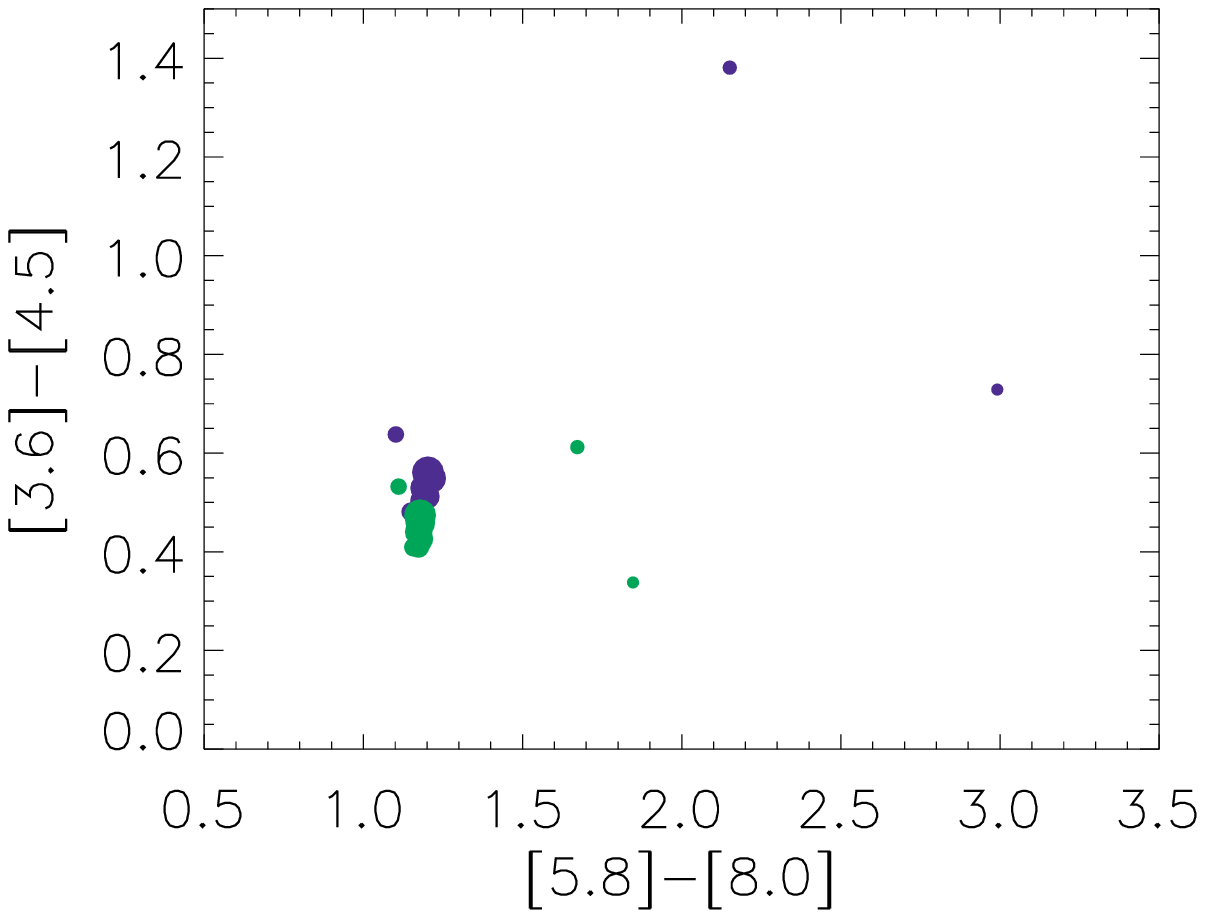}
\includegraphics[angle=0,width=3.2in]{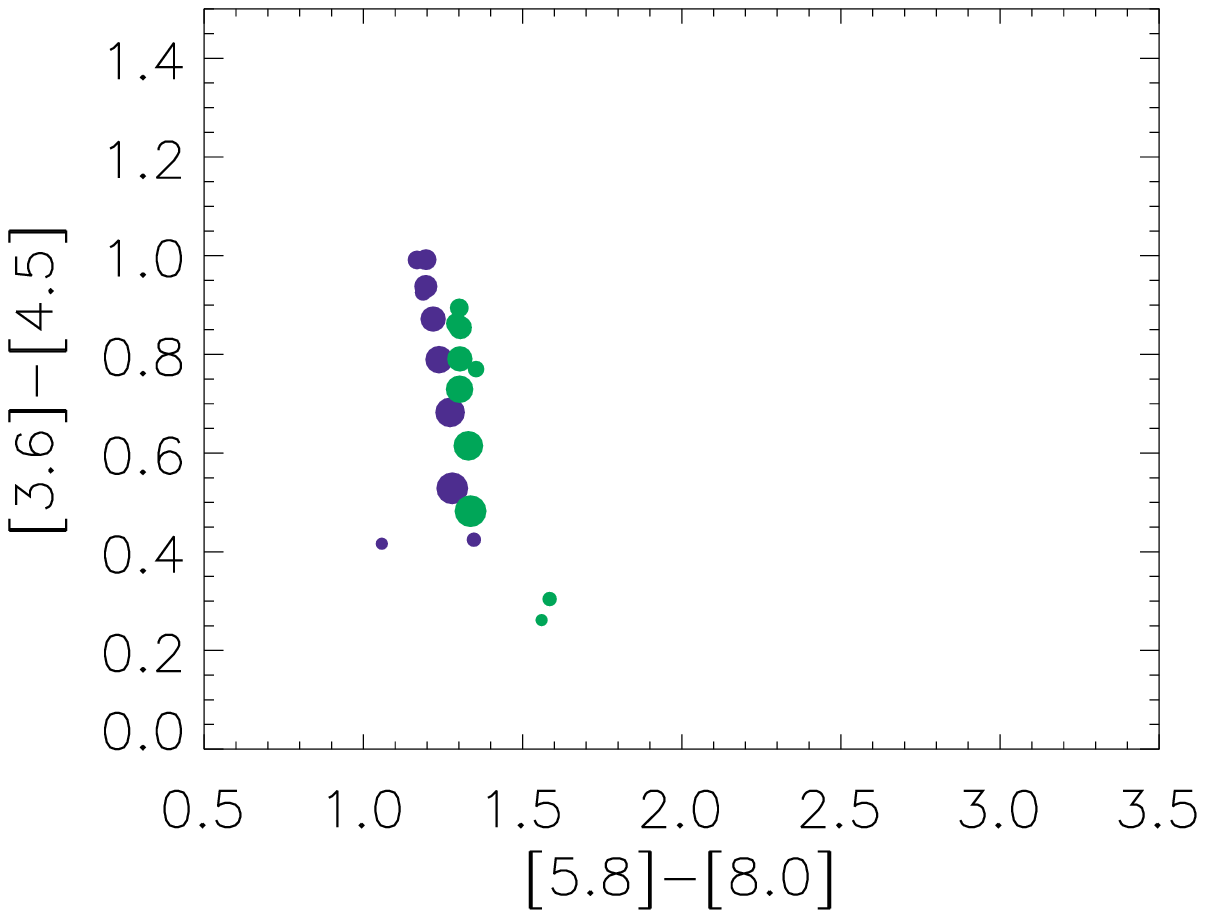}
\caption{Color-color plots of the AeBe models.  At left are the disk models, and at right are the embedded protostar models.  Green symbols show the models with PAH/vsg emission and blue without.  The symbol size is related to viewing angle, with the smallest edge-on and the largest pole-on.  The models with PAH emission are slightly redder at [5.8]-[8.0] and bluer at [3.6-[4.5] for a given viewing angle.  For the edge-on models (small symbols), the effect is more pronounced.
\label{f_AeBe_cc}}
\end{figure}

Of interest to us is whether the PAH emission affects the mid-IR colors of these objects.  The {\it Spitzer} Space
Telescope has observed thousands of YSOs, and color-color-diagram (CCD) and color-magnitude-diagram (CMDs) are a common analysis tool.  Figure \ref{f_AeBe_cc}
shows CCDs of the AeBe disk and embedded source models.  For the most part, the colors are similar with and without PAH emission.
The only exceptions are the edge-on sources, as explained above; however, most edge-on sources are faint and may escape detection.
This suggests that analysis that doesn't include PAH emission may be valid in some cases.  However, more models with different
parameters should be run to confirm this.

\begin{figure}
\epsscale{1.0}
\includegraphics[angle=0,width=3.2in]{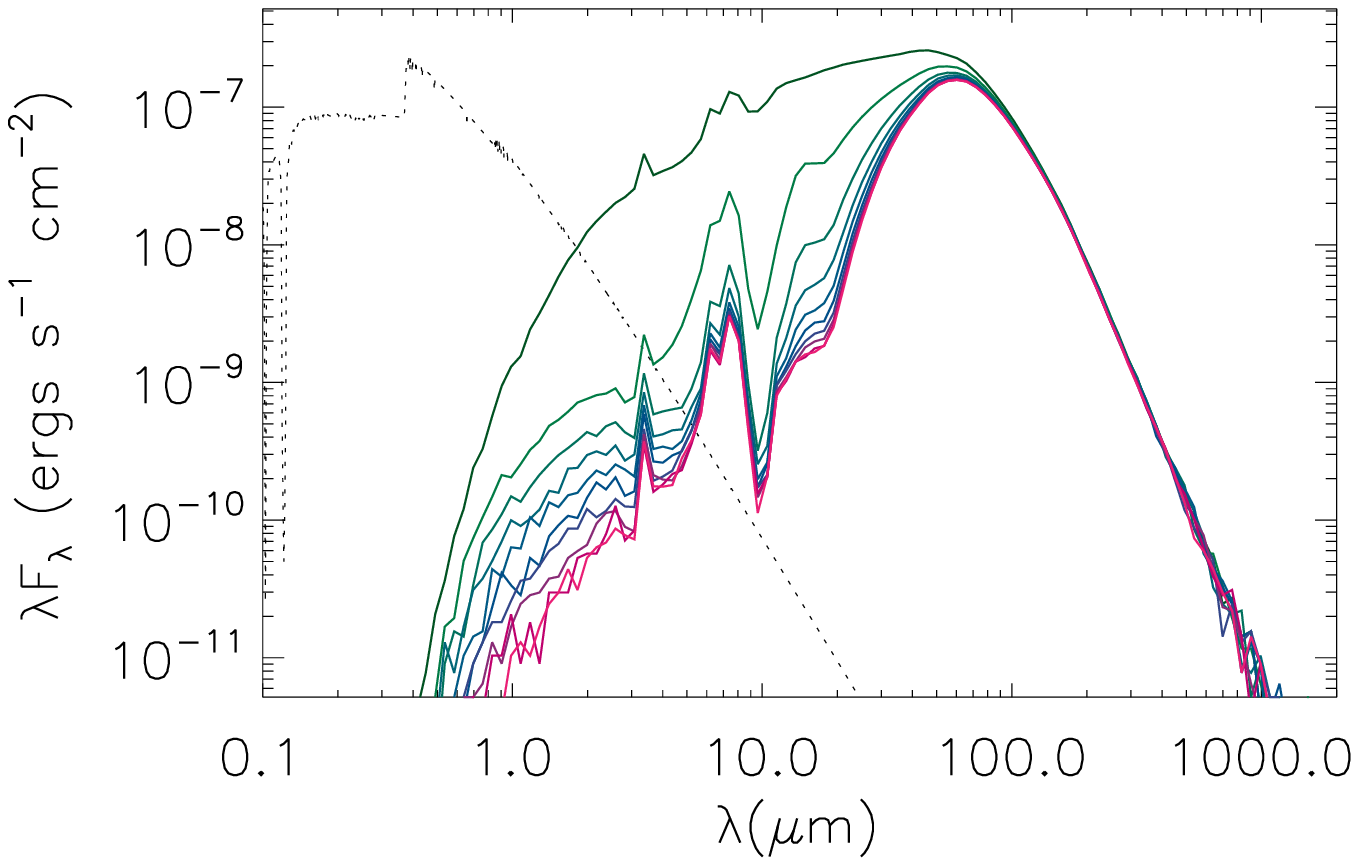}
\includegraphics[angle=0,width=3.2in]{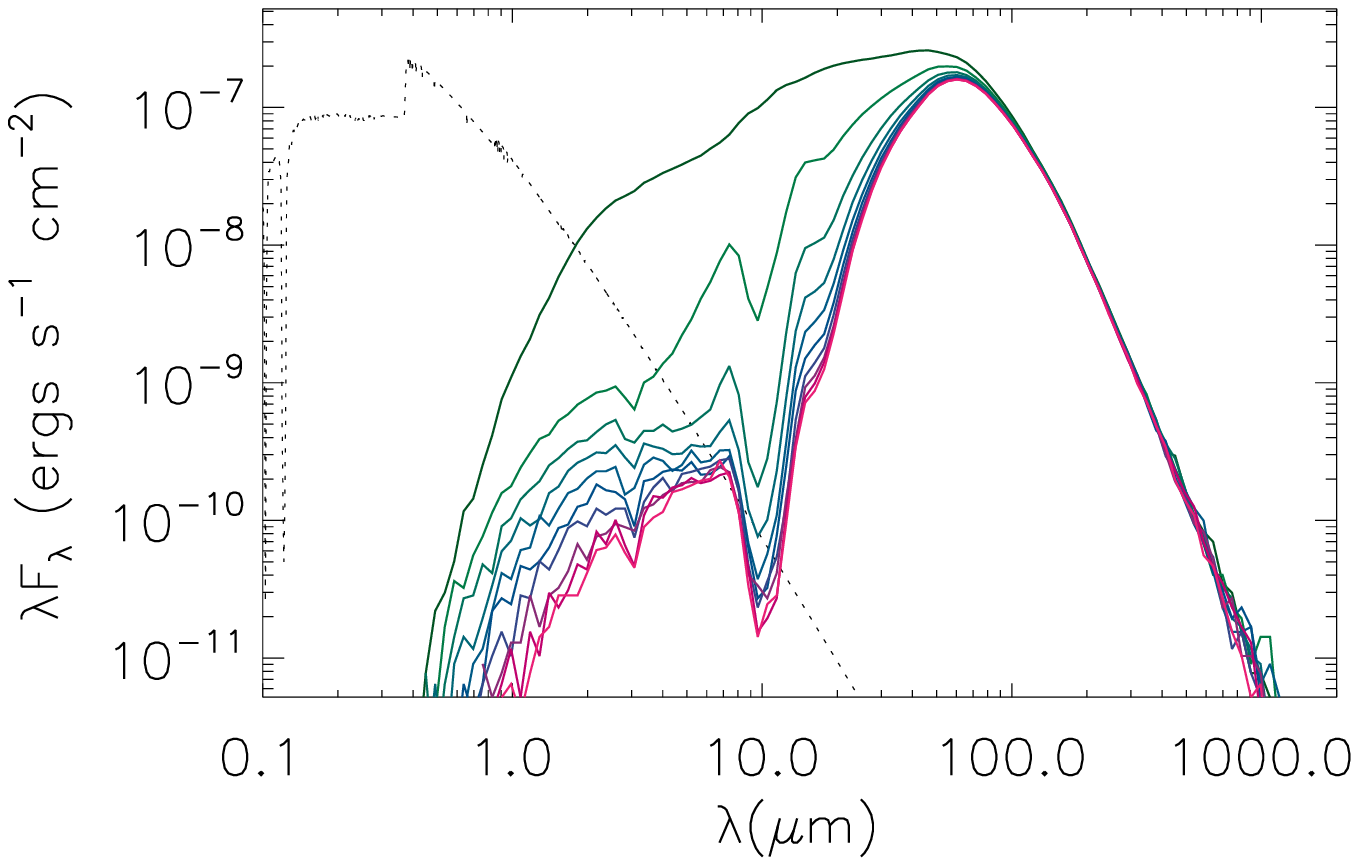}
\caption{Massive embedded YSO (MYSO) with PAH emission (left) and without (right).  The different colored lines are as described in Figure \ref{f_AeBe_sed}.
\label{f_hmpo_sed}}
\end{figure}

\begin{figure}
\epsscale{1.0}
\includegraphics[angle=0,width=3.2in]{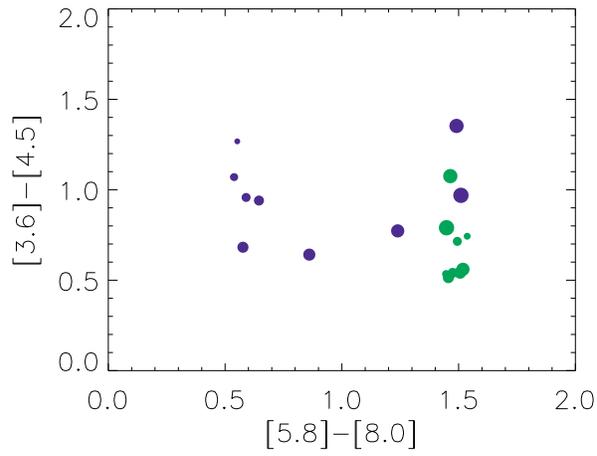}
\caption{Color-color plots of the high-mass YSO.   The symbols are as in Figure \ref{f_AeBe_cc}.  The model with PAH emission is generally redder at [5.8]-[8.0] and bluer at [3.6-[4.5] for a given viewing angle.
 \label{f_hmpo_cc}}
\end{figure}

Our more massive embedded source model, in contrast to the embedded AeBe model, shows relatively strong PAH features at all viewing angles except pole-on, as shown in Figure \ref{f_hmpo_sed}, likely due to the higher stellar temperature producing
more UV photons.
The relevant parameters for this model are:
$T_\star=10,000$K, $R_\star= 28 R_\sun$, $M_{\rm disk} = 0.1 M_\sun$, $\dot M_{\rm env} = 1.\times10^{-3} M_\sun/$yr. 
The color-color plots also show this effect (Figure \ref{f_hmpo_cc}) with the PAH sources appearing redder by up to 1 magnitude.   
Massive protostars have been observed with and without PAH emission \citep{woods11,gibb00}.  This is probably due to both the temperature
of the central source (a younger source will be cooler), and the amount of obscuration in the polar regions that can 
extinguish UV photons and thereby prevent stimulation of the PAHs.

\begin{figure}
\epsscale{1.0}
\includegraphics[angle=0,width=3.2in]{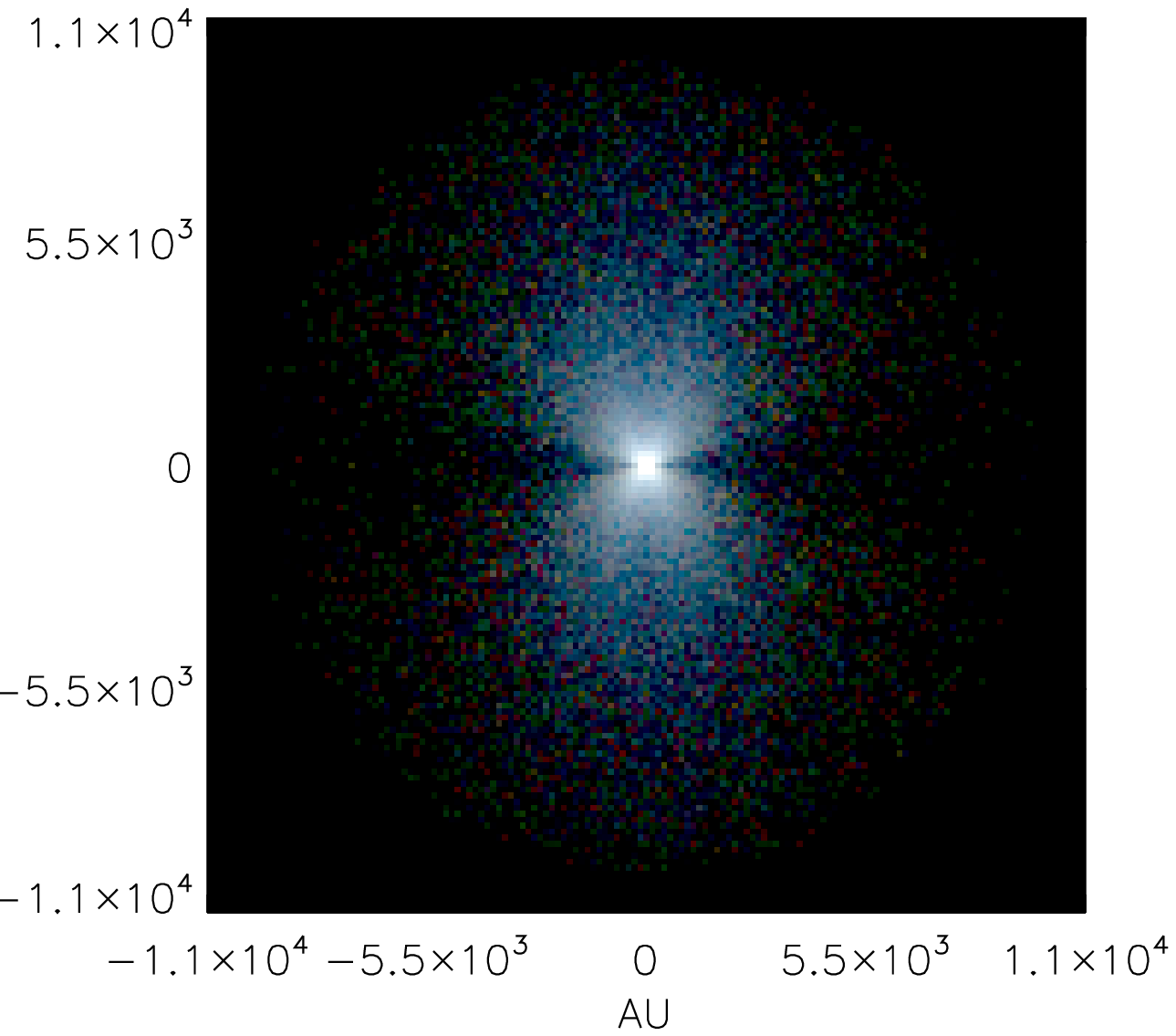}
\includegraphics[angle=0,width=3.2in]{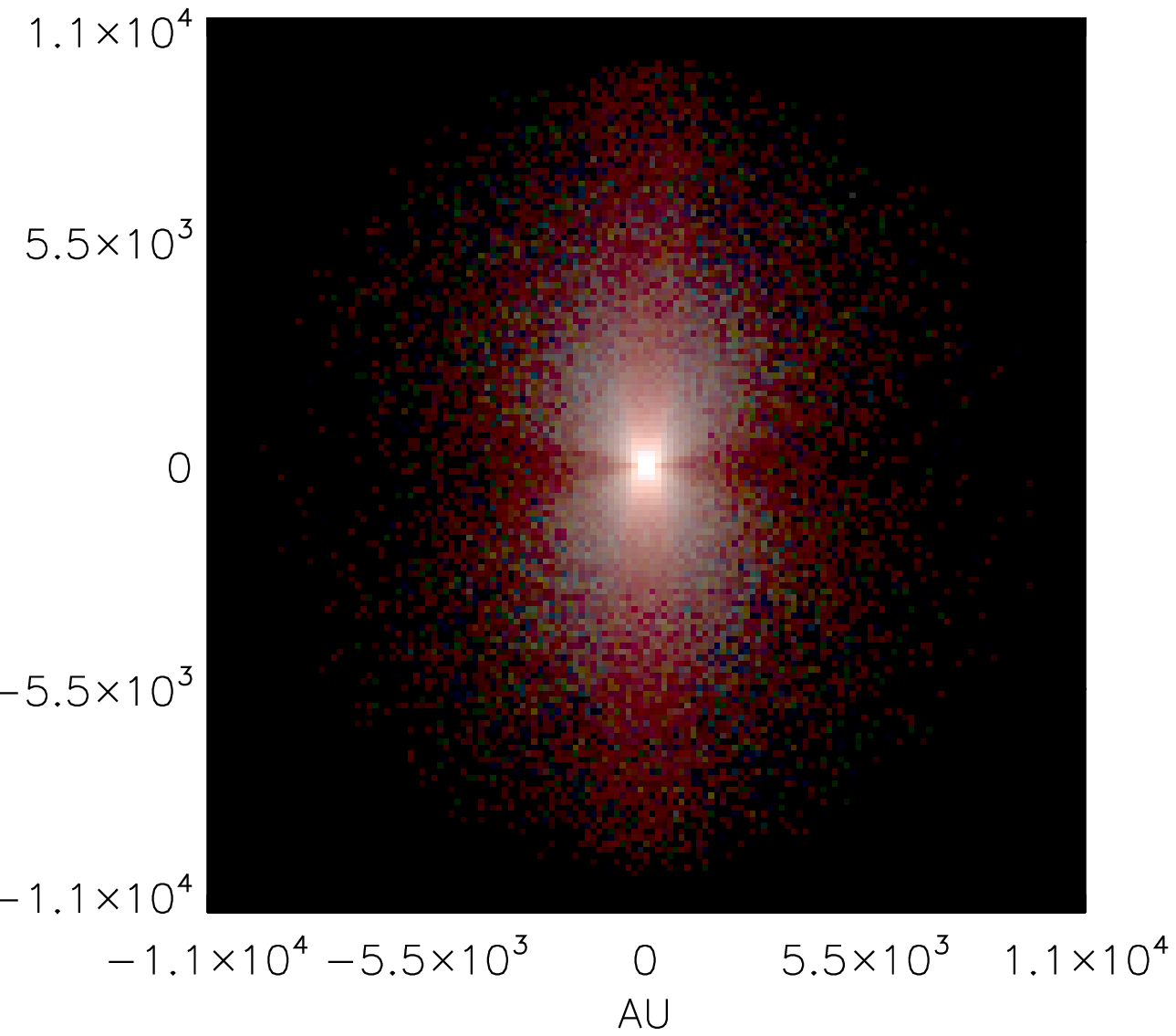}
\includegraphics[angle=0,width=3.2in]{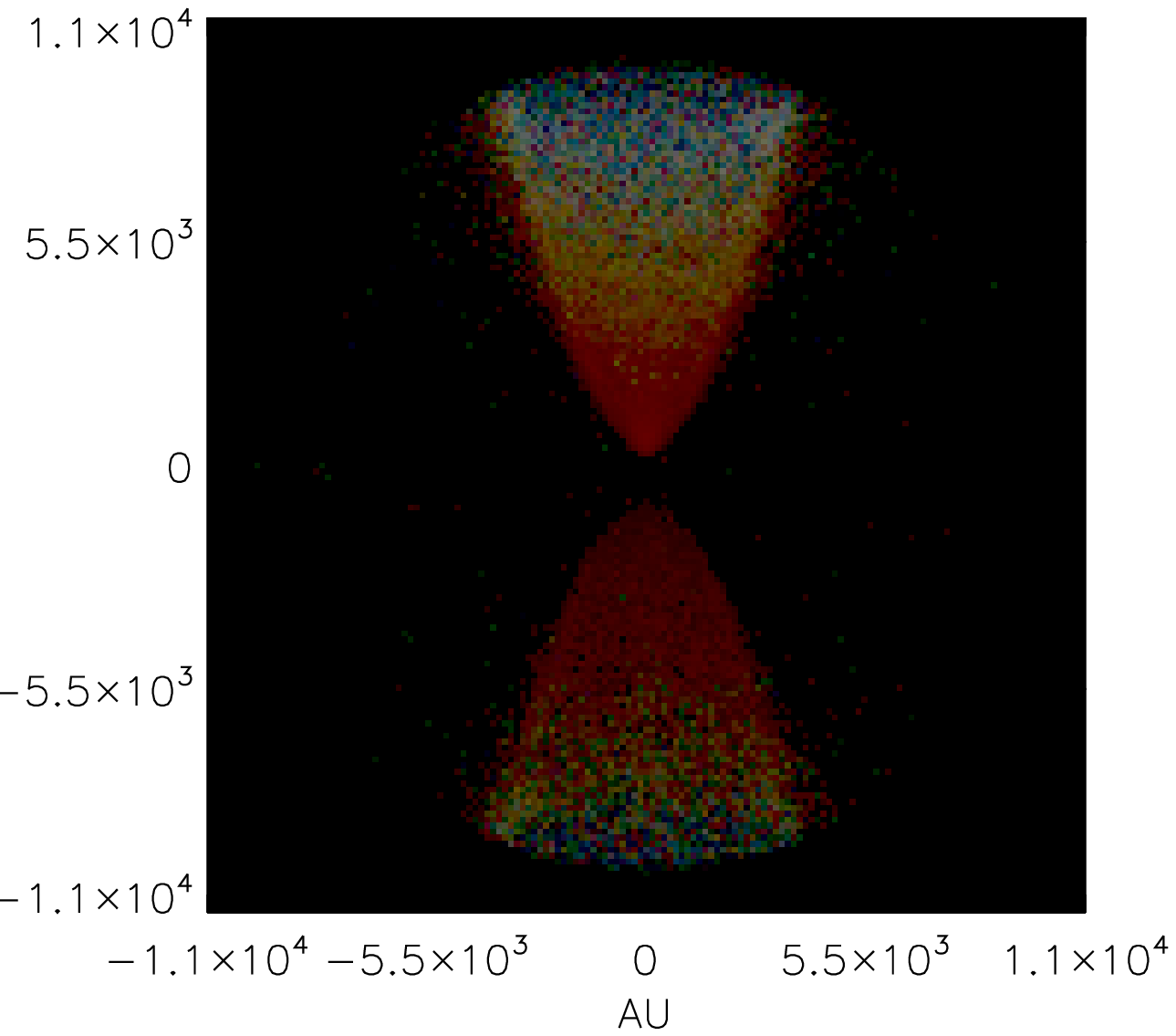}
\includegraphics[angle=0,width=3.2in]{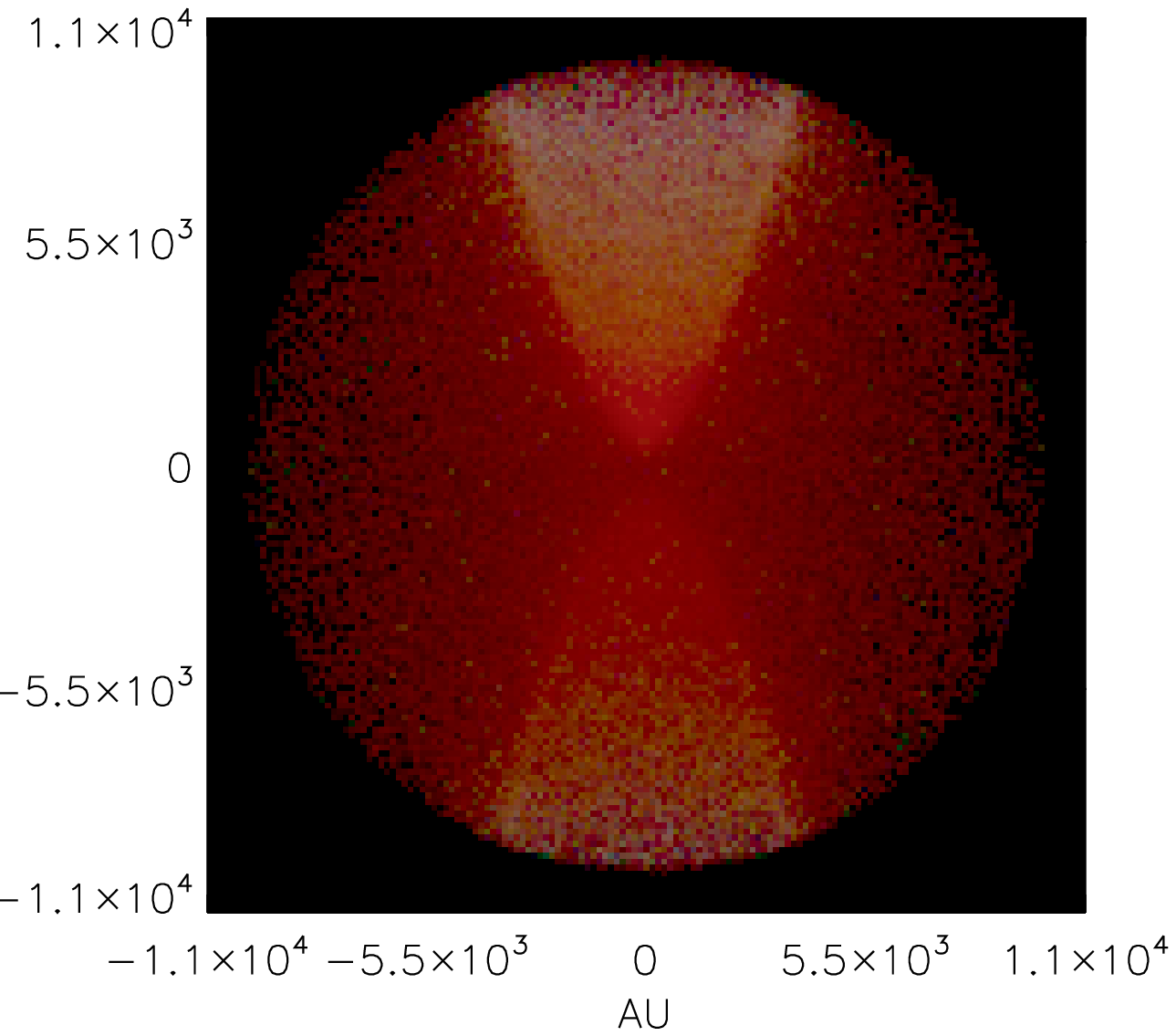}
\caption{3-color images (3.6 $\mu$m in blue, 4.5 in green, 8.0 in red) of the embedded AeBe protostar models are shown in the top two panels, without PAH emission on the left, with on the right.  The MYSO models are shown at  bottom, without PAH emission on the left, with on the right.  The MYSO model is more embedded, showing redder thermal colors than the AeBe protostar (left panels).  The MYSO model with PAH emission is dominated by PAH emission (at right) because most of the warm thermal emission from the inner disk and envelope is extincted by the envelope.  Enough ultraviolet emission escapes the cavity regions to excite the PAH grains in the cavity and outer envelope.  This is also true for the AeBe model.
\label{f_AeBe_hmpo_img}}
\end{figure}

The images of the embedded AeBe and Massive YSO models are shown in Figure \ref{f_AeBe_hmpo_img}.  These show that the 
Massive YSO is  more embedded than the AeBe one, with much redder thermal emission (the left two panels, without PAH emission).
In both models, the PAH emission lights up the cavities and outer envelope.

\begin{figure}
\epsscale{1.0}
\includegraphics[angle=0,width=3.2in]{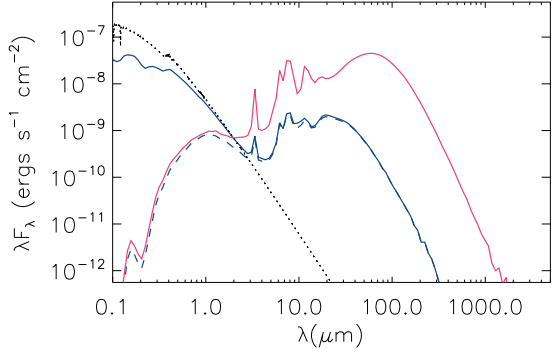}
\caption{
The SED of a B2 star illuminating a large optically thin dust cloud (described in the text).  The blue solid line shows the source viewed through a 4000 AU aperture; the blue dashed line has a foreground extinction of 4 magnitudes.    The pink line is for the same foreground extinction and viewed through an entire aperture of 200,000 AU.  The SEDs with foreground extinction appear to have an IR excess and resemble the shape of the embedded protostars in previous figures.
}.
\label{f_hmms_sed}
\end{figure}

High-mass stars of {\it any} evolutionary stage will heat up  dust in the vicinity of the star whether associated with the star or not.
\citet{lamerscassinelli99} show how to calculate the optically thin equilibrium temperature for dust heated by a central
source (their \S7.4.2).  For typical ISM dust properties, the radius $r_d$ at which the dust is heated to a given temperature $T_d$ is given by
\begin{equation}
\frac{r_d}{R_\star} = 0.5 \left ( \frac{T_d}{T_\star} \right )^{-2.5}.
\end{equation}
From this we can calculate the ``sphere of influence'' of a massive star.  For example, a B2 V star will heat dust above 30 K
within a radius of $\sim0.7$ pc.   Thus, a main sequence star in the vicinity of a molecular cloud, such as one recently formed, will heat up its surroundings.
Figure \ref{f_hmms_sed} shows the SED of a B2 V star illuminating an optically thin dust cloud with an outer radius of 200,000 AU.    The density of the dust cloud is  $7 \times 10^{-22}$gm/cm$^3$ (which corresponds to a number density of molecular hydrogen of 200 cm$^{-3}$), and the A$_V$ of the cloud is  0.5; the total mass of the envelope is 40 $M_\sun$ due to the large radius.  The blue solid line shows the source viewed through a 4000 AU aperture, or about 2$\arcsec$ at a distance of 2 kpc, a typical distance to nearby spiral arms in our Galaxy.  The blue dashed line has a foreground extinction of 4 magnitudes.  This SED shape begins to resemble that of an embedded source.  The pink line is for the same foreground extinction and viewed through an entire aperture of 200,000 AU.  This object at the distance of the Magellanic clouds (50-60 kpc) could masquerade as a protostar.  In fact, \citet{sewilo12} have identified several YSO candidates that are likely high-mass main sequence stars in the vicinity of a molecular cloud.

How can we distinguish a main-sequence star illuminating optically thin dust from an embedded source based only on the SED?  In comparing the SEDs in Figures \ref{f_hmpo_sed} and \ref{f_hmms_sed}, we can see differences in the PAH/silicate features.  
In the embedded source, the PAH emission is weak relative to the thermal when viewed pole-on, and is combined with silicate absorption near edge-on.  If not heavily extincted by interstellar dust, the main sequence star should have a bright optical source associated with it.   IR imaging will easily distinguish the two types of objects in our Galaxy, but in the Magellanic clouds they may not be sufficiently resolved. 

\subsection{External Illumination by the Interstellar Radiation Field}\label{isrf}

The external interstellar radiation field (ISRF) can heat up a disk or envelope and produce thermal as well as non-thermal emission.  We use the Galactic value computed near the solar neighborhood, given in Table A3 of \citet{mathis83}, in units of $4 \pi J_{\nu,ISRF}$, where $J_{\nu,ISRF}$ is the average intensity \citep{chandra60}.  The input ISRF can be scaled by an arbitrary factor and extincted by intervening dust.  The ISRF illuminates the YSO from the outer radius of the envelope with an isotropic angular dependence.   The luminosity of the ISRF is 
\begin{equation}
L_{ISRF} = {\rm ISRF\_SCL} * (4 \pi R_{max}^2)  \int_0^\infty ( \pi J_{\nu,ISRF}) e^{-\tau_\nu} d\nu,
\end{equation}
where $e^{-\tau_\nu}$ is the extinction function calculated from a standard interstellar extinction law and the input $A_V$, $ISRF\_SCL$ is a user-specified scale factor, and $R_{max}$ is the outer radius of the envelope.
We tested this with a model calculating the mean intensity of the solar neighborhood.
For computation of the model images, we count only the interacting photons in the photon summations.  This is equivalent to doing background subtraction of images.  

External illumination can dominate in two ways:  if the central source luminosity is very low, and/or if the outer radius of the envelope is very large.   This occurs for very low-luminosity protostars and for large clumps or starless cores,
as well as molecular clouds illuminated by ``cloud shine'' \citep{foster06}.

\begin{figure}
\epsscale{1.0}
\includegraphics[angle=0,width=3.2in]{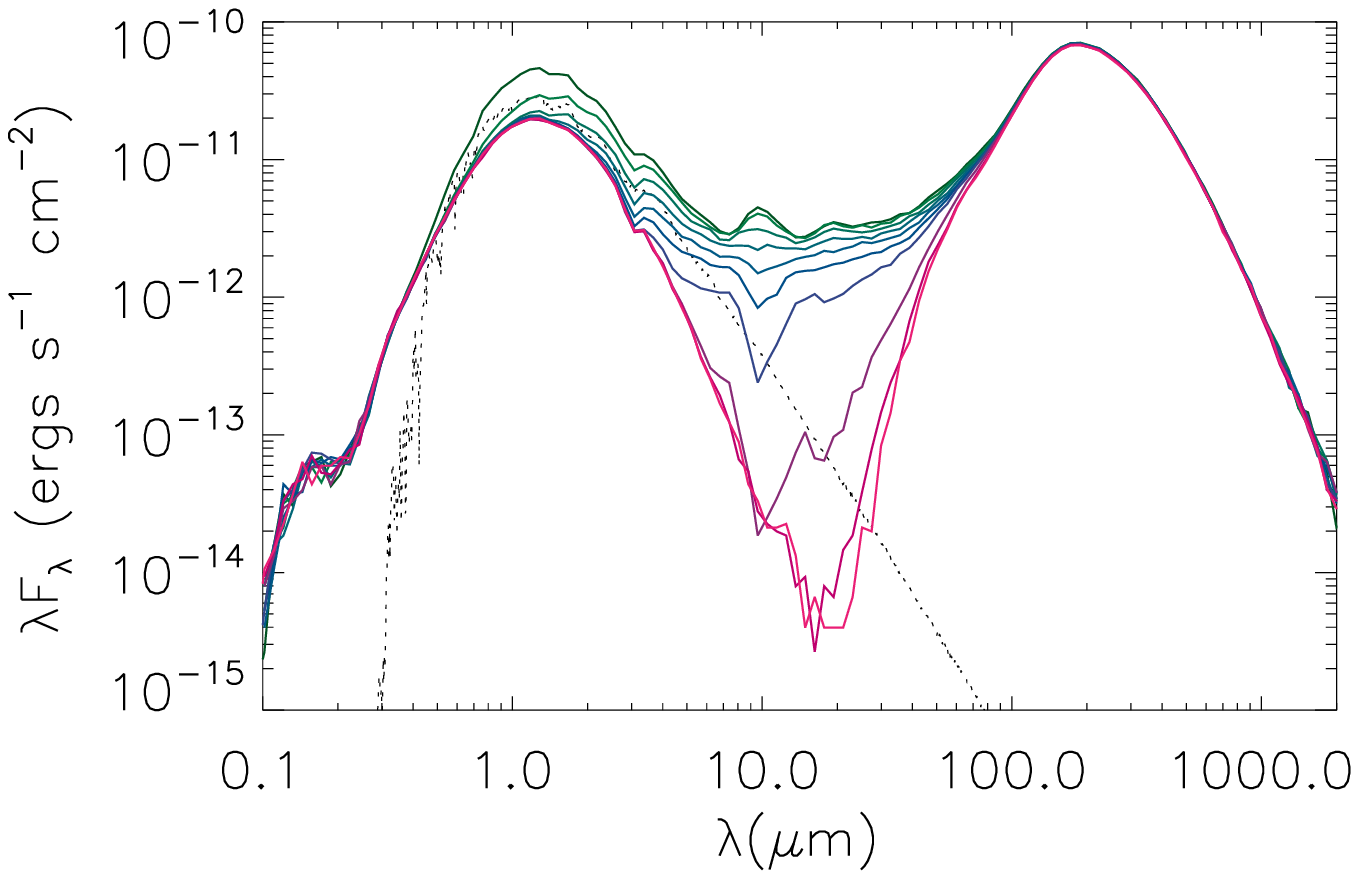}
\includegraphics[angle=0,width=3.2in]{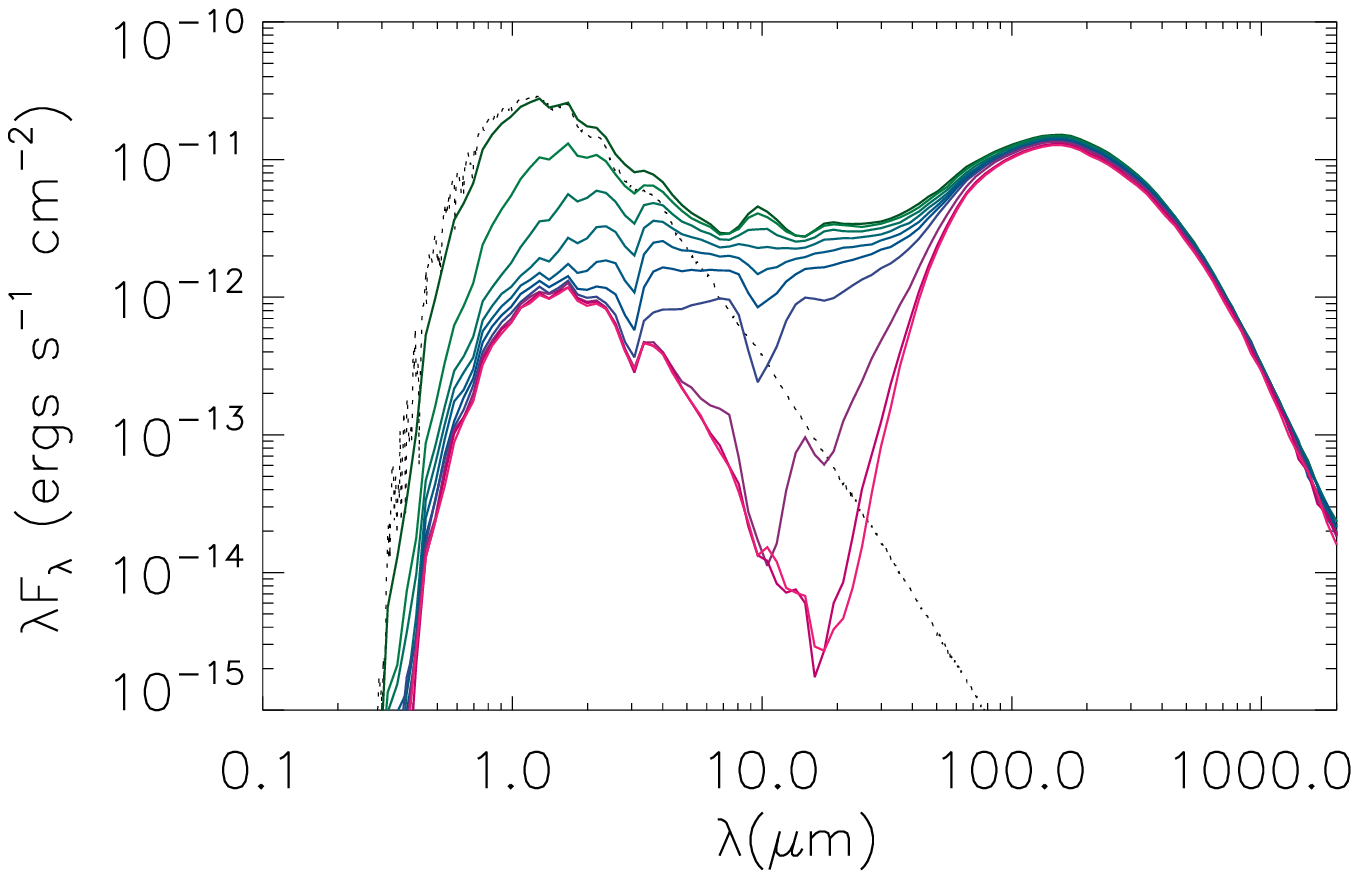}
\caption{Very Low Luminosity Object (VeLLO) with external illumination by the interstellar radiation field (left), and without (right), viewed at a distance of 100 pc.    The different colored lines are as described in Figure \ref{f_AeBe_sed}.
\label{f_vello_sed}}
\end{figure}

\begin{figure}
\epsscale{1.0}
\includegraphics[angle=0,width=3.2in]{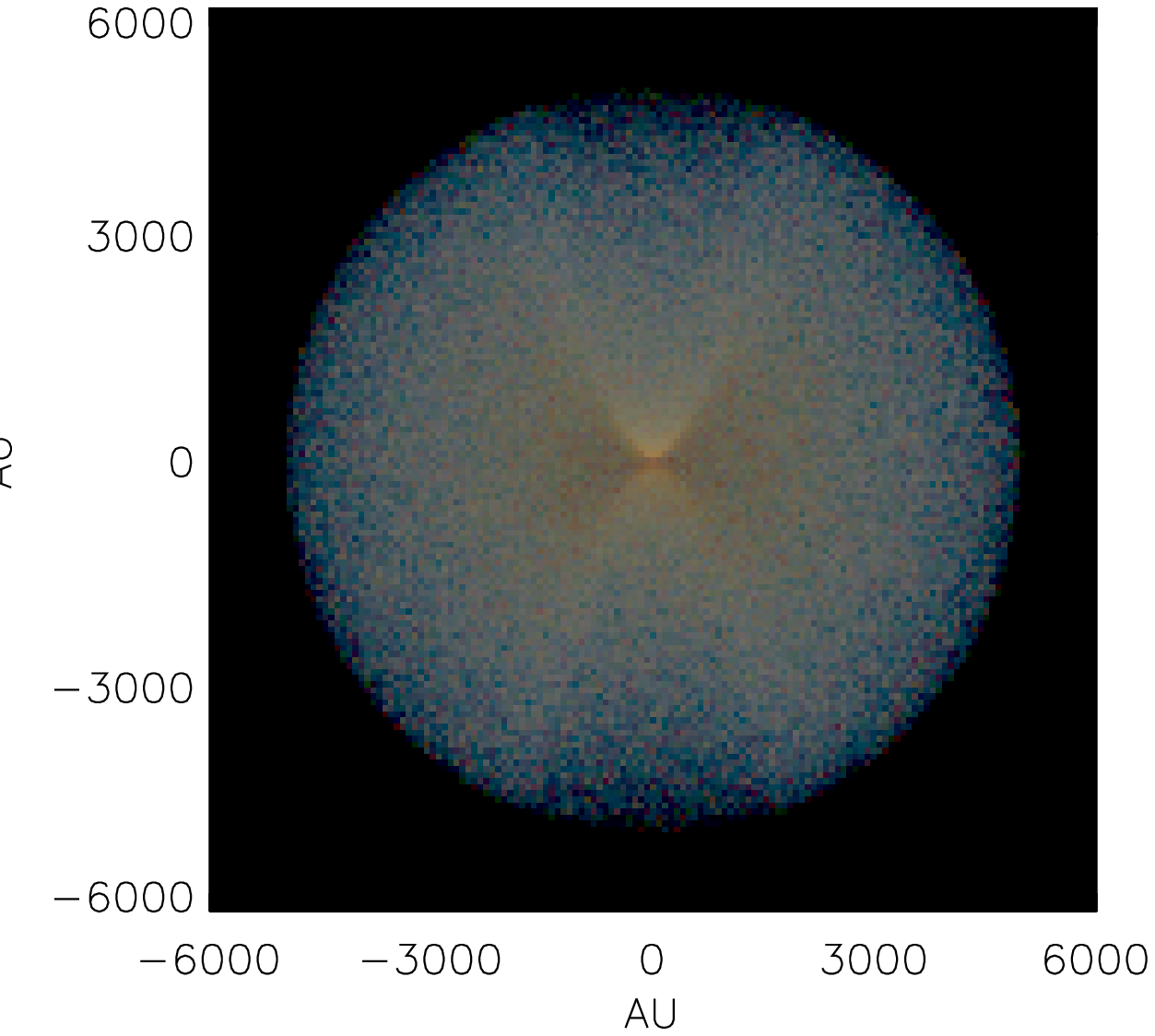}
\includegraphics[angle=0,width=3.2in]{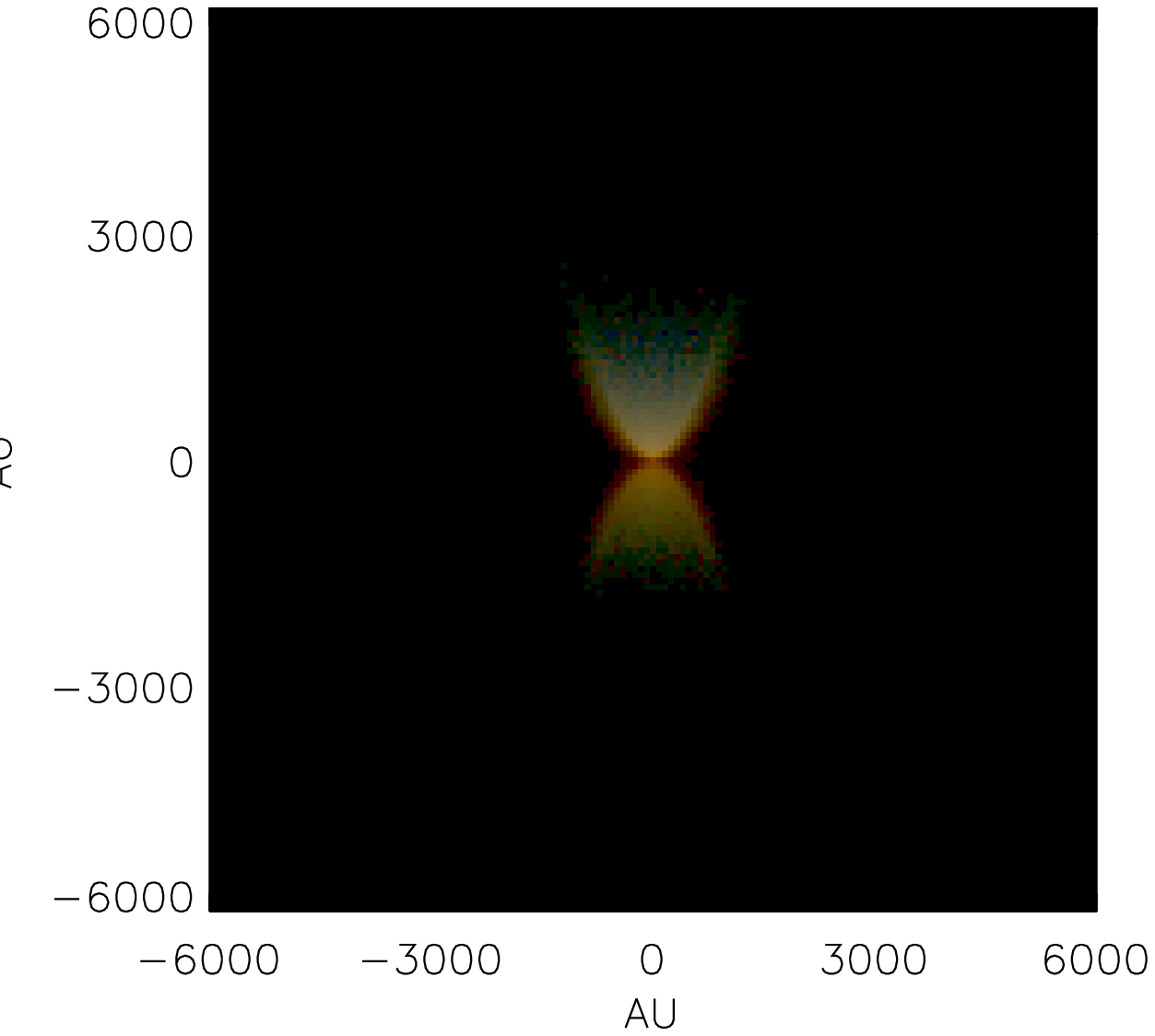}
\caption{Near-infrared 3-color (JHK) images of the VeLLO models: with external illumination by the interstellar radiation field (left), and without (right). 
\label{f_vello_img}}
\end{figure}

Several Very Low Luminosity Objects (VeLLOs) were uncovered by the {\it Spitzer} Space Telescope in dense cores previously thought to be starless \citep{young04, dunham06, huard06, bourke06, dunham08, lee09, terebey09, dunham10}.   Due to their low intrinsic luminosity, heating of the outer envelope from an external radiation field can dominate the luminosity of these sources.  We show a model of such a source, both with and without external illumination, in Figures \ref{f_vello_sed} and \ref{f_vello_img}.  The intrinsic luminosity of the source is 0.01 $L_\sun$ and its outer radius is 5000 AU.  The illumination from the average Galactic Interstellar Radiation Field (ISRF) on the outer radius of the envelope increases the luminosity to 0.07 $L_\sun$.   Figure \ref{f_vello_sed} shows the SED of the two models (with and without ISRF), and Figure \ref{f_vello_img} shows near-IR 3-color images.  The ISRF increases the scattered spectrum (UV through near-IR) and the far-infrared flux, due to the additional scattering and absorption (heating) in the outer envelope.  The scattered spectrum is especially enhanced in the edge-on viewing angles.  

\begin{figure}
\epsscale{1.0}
\includegraphics[angle=0,width=3.2in]{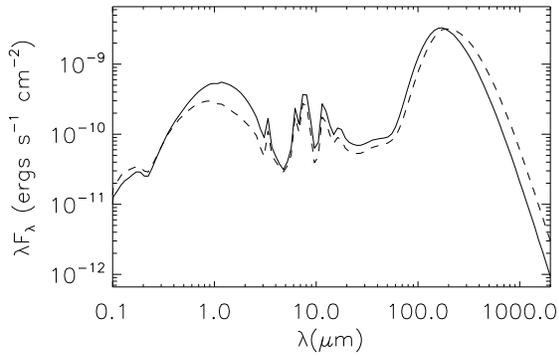}
\caption{SEDs for externally illuminated cloud cores, viewed at a distance of 500 pc.  The solid line is for a smooth-density core (with $\rho \sim r^{-1.5}$), and the dashed line is the spherically averaged SED of a model that has 3-D fractal clumping in the envelope, with the same average radial density power law.  \label{f_clump_sed}}
\end{figure}
\begin{figure}
\epsscale{1.0}
\includegraphics[angle=0,width=3.2in]{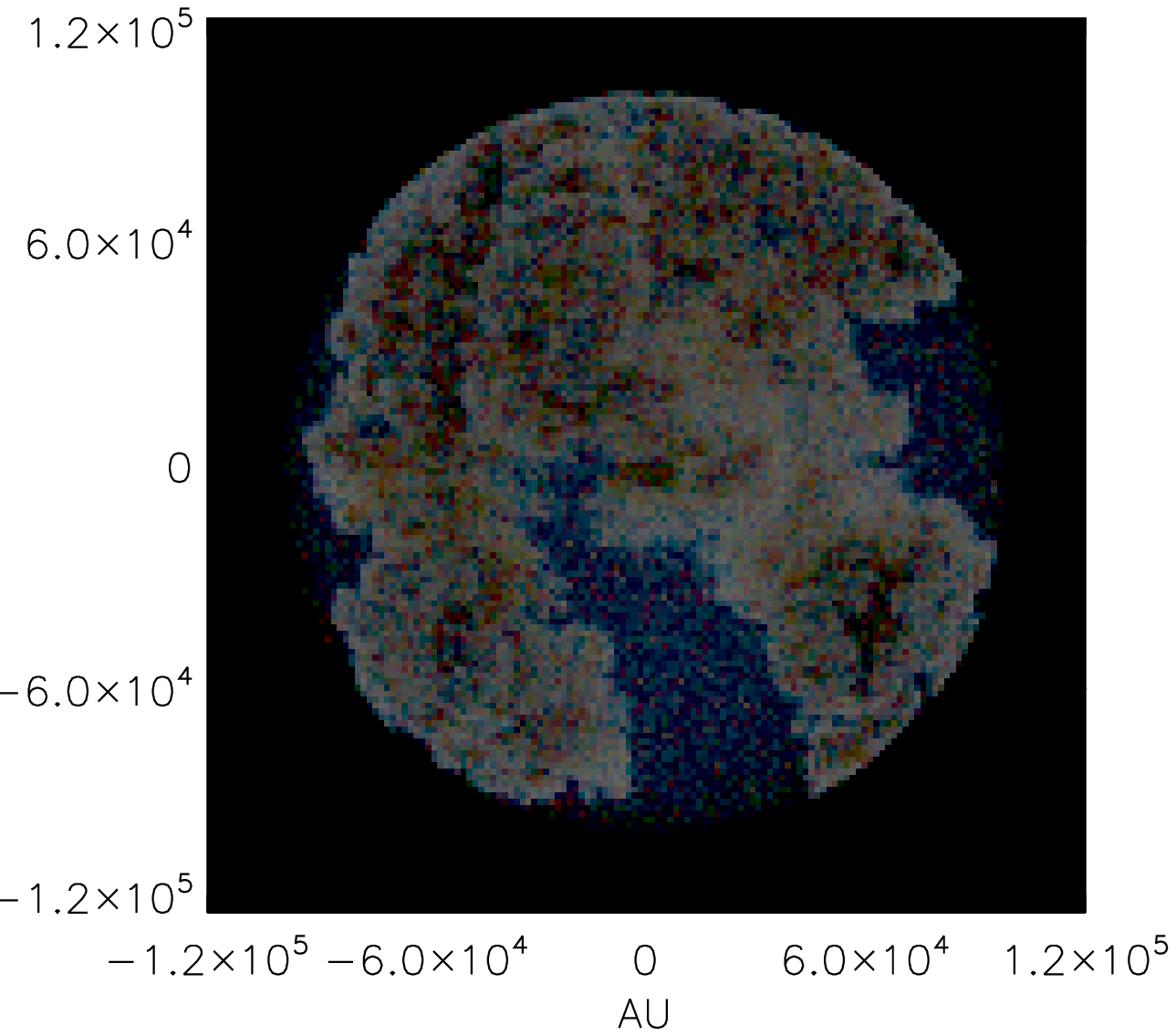}
\includegraphics[angle=0,width=3.2in]{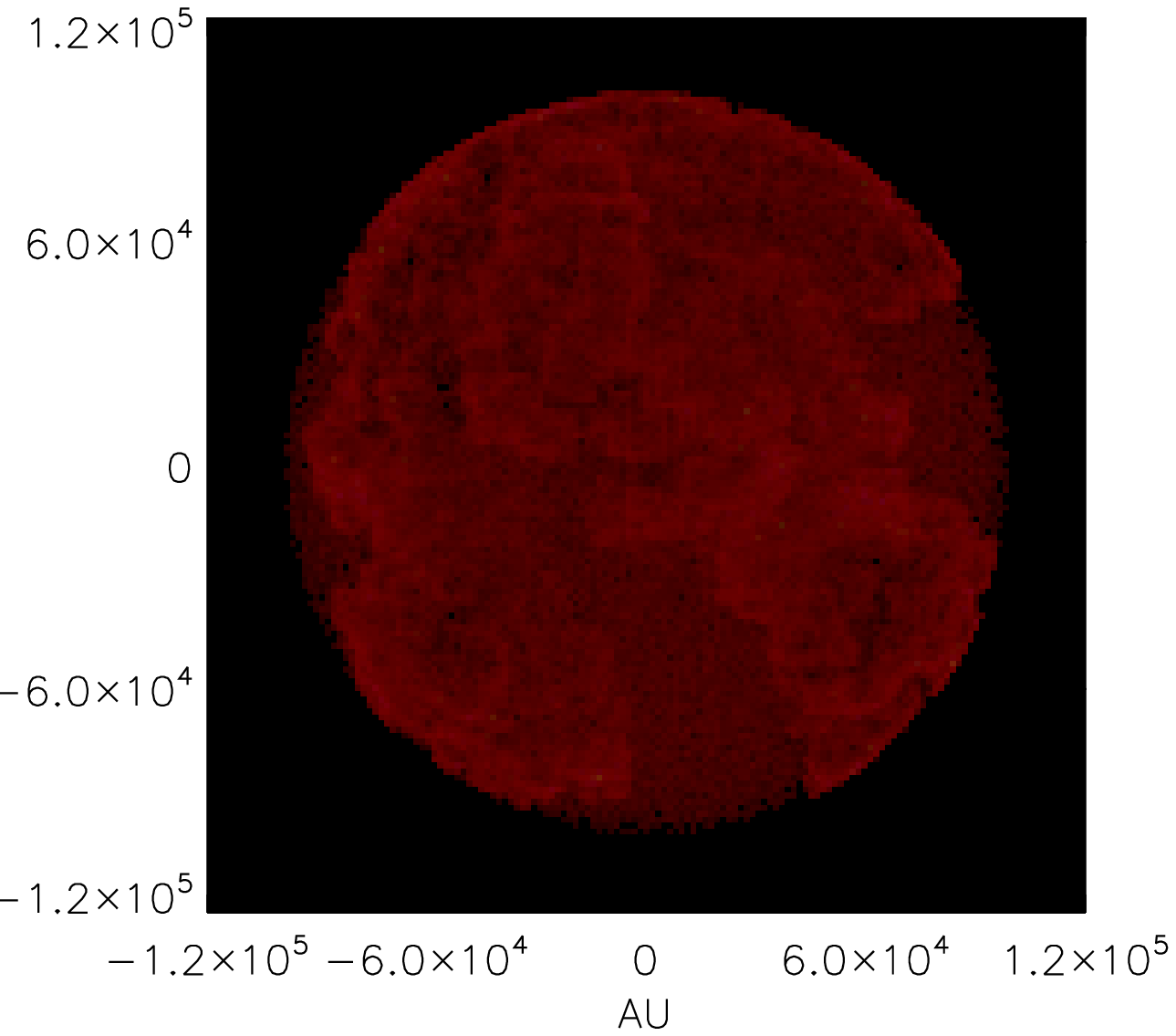}
\caption{Left: Near-infrared 3-color (JHK) image of the clumpy externally illuminated clump.  Right:  IRAC 3-color (3.6-4.5-8.0 $\mu$m) image of the same clump.
\label{f_clump_img}}
\end{figure}

Figures \ref{f_clump_sed} and \ref{f_clump_img} show SEDs and images of externally illuminated clumps with no internal luminosity source.  These are centrally condensed cores ($\rho \sim r^{-1.5}$) with an outer radius of 100,000 AU, and an envelope mass of 230 $M_\sun$.  We computed 2 models, one with a smooth density distribution, and the other with fractal clumping (\S\ref{s_fractal}).  The SEDs are similar.  The luminosities of the models are 36 $L_\sun$.  Images of the clumpy model are shown in the near-IR and at {\it Spitzer} IRAC wavelengths \citep[3.6-8 $\mu$m;][]{fazio04} in Figure \ref{f_clump_img}.  The IRAC images are red, dominated by the PAH emission at 8 $\mu$m.  Several objects in the LMC and SMC have been identified with similar features and these may be externally illuminated clumps \citep[Seale et al., in prep]{sewilo12}.

\subsection{Option for Power-Law Envelope}

The previous version of the code included the envelope geometry of a rotating sphere in free-fall gravitational collapse \citep{ulrich76}.   This version includes that as well as a simple power-law formation, $\rho = \rho_1 r^{-a}$, where $\rho_{1}$ is the fiducial density at 1 AU (units of g/cm$^3$) and $a$ is the density exponent.  This can be combined with bipolar cavities (Paper I), fractal clumping (\S\ref{s_fractal}), and gaps (\S\ref{s_gaps_etc}).  This modification makes the code more general.  For example, it can be used to model evolved stars (AGBs, carbon stars, proto-planetary nebulae).

\begin{figure}
\epsscale{1.0}
\includegraphics[angle=0,width=3.2in]{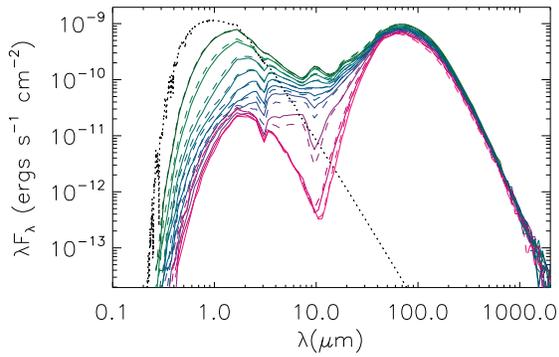}
\caption{The SEDs of our standard Class I model with the Ulrich solution (solid lines) are compared to a power-law envelope (dashed lines; $\rho \sim r^{-1.5}$).   The flux is scaled to a distance of 140 pc.
\label{f_sed_powlaw}}
\end{figure}

\begin{figure}
\epsscale{1.0}
\includegraphics[angle=0,width=3.2in]{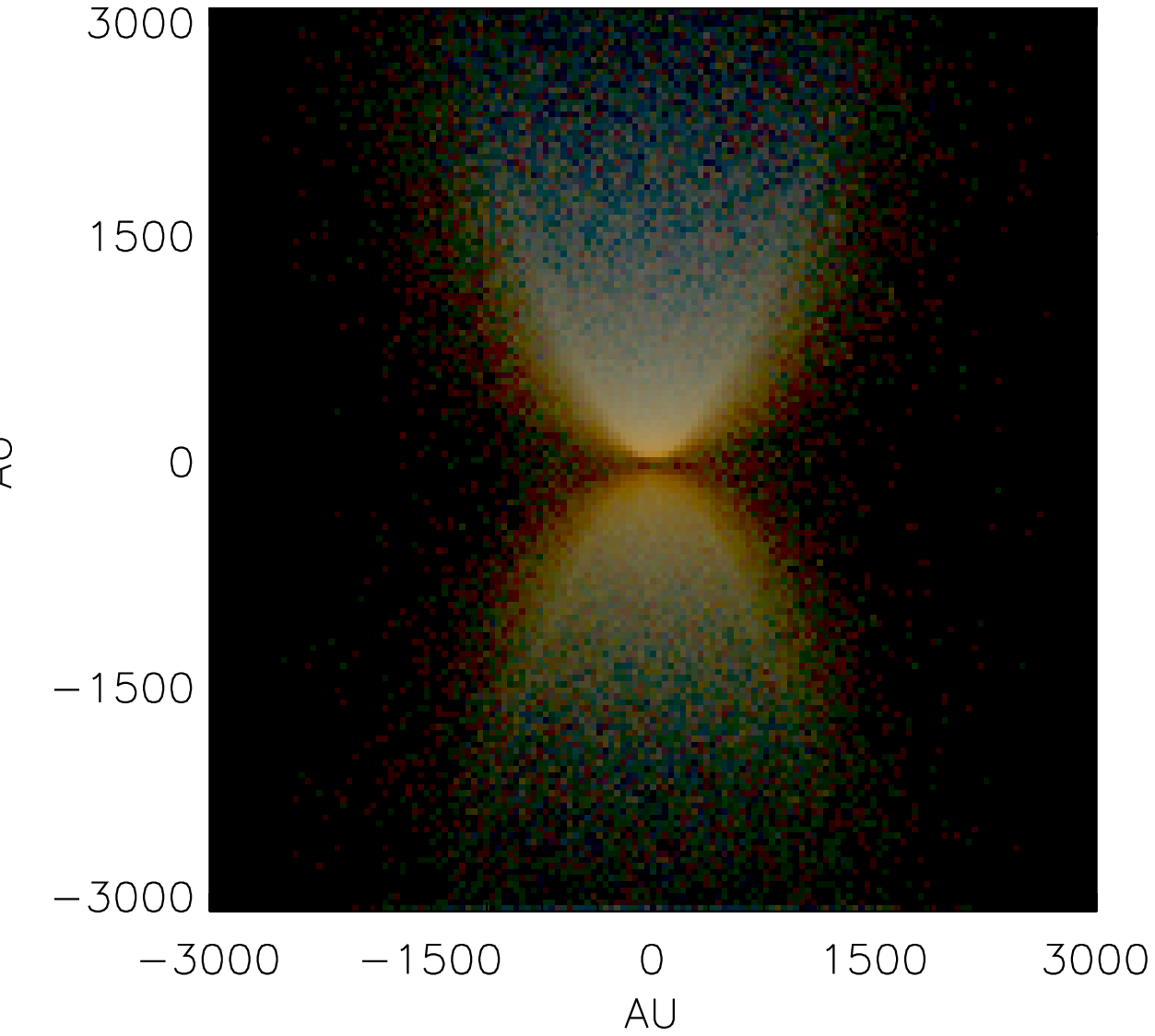}
\includegraphics[angle=0,width=3.2in]{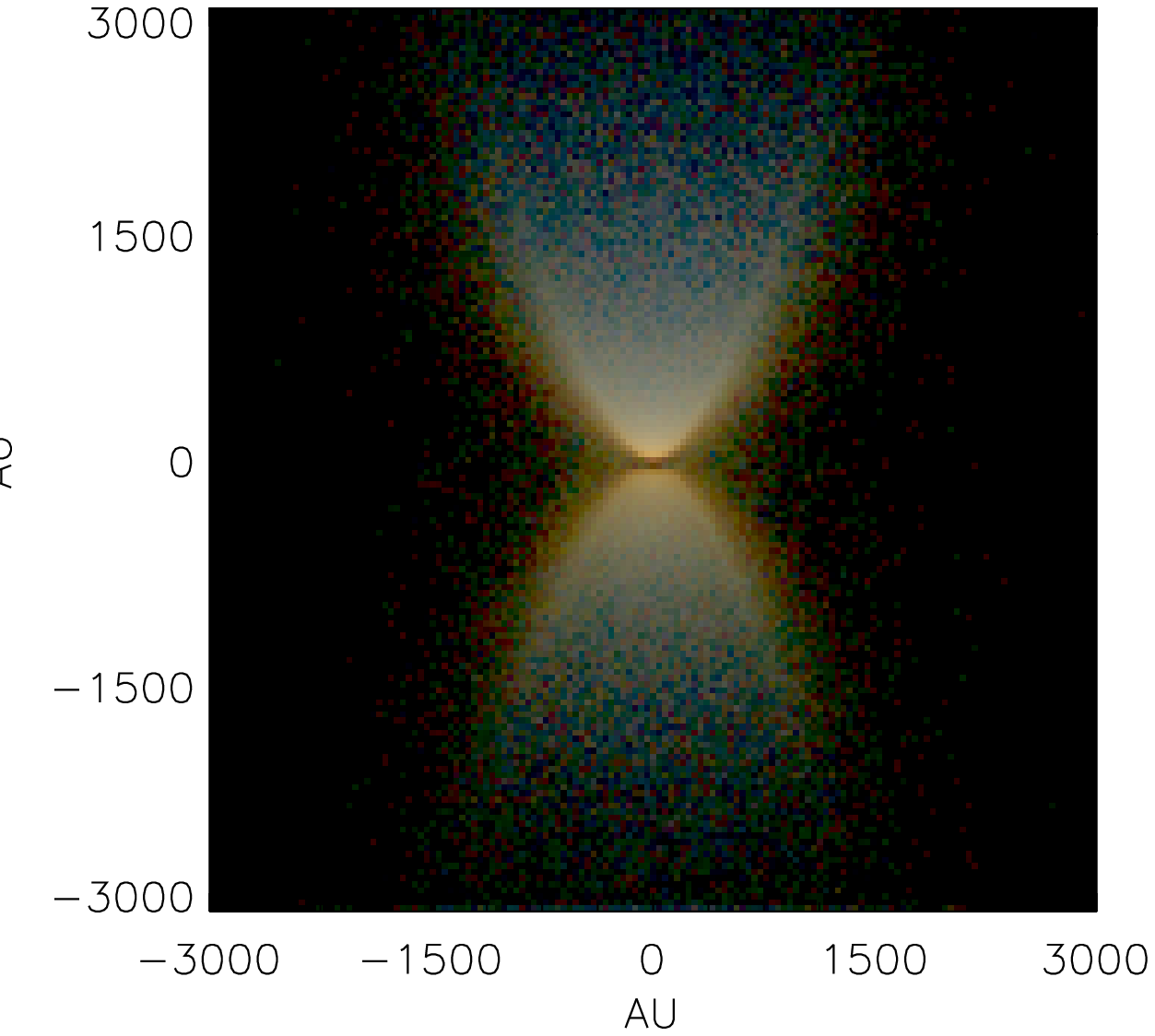}
\caption{Left: Near-infrared 3-color (JHK) image of our standard Class I model with the Ulrich envelope density solution.  Right:  Near-IR image of a similar model with a power-law envelope density  ($\rho \sim r^{-1.5}$).   The disk size in both models is 200 AU.
\label{f_3col_powlaw}}
\end{figure}

Figure \ref{f_sed_powlaw} shows example SEDs for a power-law envelope compared to the Ulrich solution (rotationally-flattened
gravitational free-fall envelope) for the low-mass Class I model similar to that presented in Paper II.  The SEDs are very similar.  The main difference occurs in the 1-10 $\mu$m region where the Ulrich solution model is more extincted.  This is also evident in the near-IR images shown
in Figure \ref{f_3col_powlaw}.  The Ulrich solution has equatorial flattening in the envelope and this produces a redder image in the midplane and more extinction to the central source, as shown in the SED and images.  This effect becomes more noticeable in the images as the centrifugal radius ($R_c$) is increased.  

\subsection{Multiple Cavity Walls}

The pressure in bipolar outflows can cause the cavities to expand horizontally as well as vertically, causing pile-up of the material at the interface of the cavity and envelope, or thickened cavity walls.  To mimic this effect, we define two cavity surfaces
with different density profiles specified within each.  If only one cavity surface is desired, simply specify the same density profile in both.   The outer cavity surface can be described as a streamline shape or a polynomial.  The streamline shape is described
in \citet{whitney93}: at large distances it appears conical; closer to the source it is curved, and intersects the disk at a radius that depends on the opening angle chosen for the hole and the centrifugal radius.   The inner cavity surface is a polynomial shape, 
\begin{equation}
z_1 = z_{01} + a_1 \varpi^{ex_1},
\end{equation}
where $\varpi$ is the cylindrical radius;  $z_{01}$, $ex_1$, and the opening angle of the cavity defined at the outer radius, $\theta_1$, are input by the user;  and $a_1$ is calculated based on $\theta_1$.
If a polynomial shape is specified for the outer cavity wall, it is described by the same function with different input variables ($z_{02}$ and $ex_2$.
Inside each cavity surface the density follows a power-law:  $\rho = \rho_{cav1} r^{-ex_f}$, and $\rho = \rho_{cav2} r^{-ex_f}$.  For simplicity the power-law exponent is the same within each surface, with only the fiducial density changing.  Modifying the code for other cavity structures is straightforward.

\begin{figure}
\epsscale{1.0}
\includegraphics[angle=0,width=3.2in]{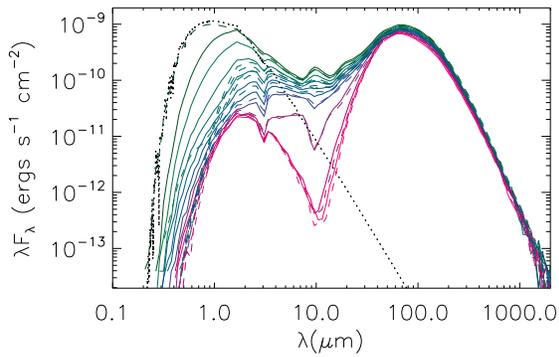}
\caption{The SEDs of our standard Class I model (solid lines) are compared to a model with less density in the cavity and enhanced density in the wall boundary between the cavity and envelope (dashed lines).  
\label{f_sed_2walls}}
\end{figure}

\begin{figure}
\epsscale{1.0}
\includegraphics[angle=0,width=3.2in]{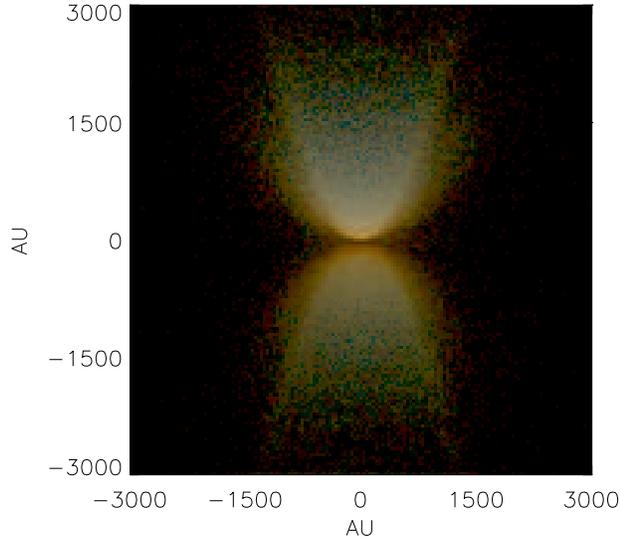}
\caption{Near-infrared 3-color (JHK) image of a Class I model with less density in the cavity and enhanced density in the boundary between the cavity and envelope than our standard Class I model (Figure \ref{f_3col_powlaw}, left). 
\label{f_3col_2walls}}
\end{figure}

Some observed objects  appear to show strong delineation of the cavity walls, for example, the HST NICMOS images of CoKu Tau/1 and DG Tau B \citep{padgett99}.  
The brightness of these features depends strongly on the optical depth through cavity, walls, and envelope.  If the optical depth is too high, the cavity will appear filled in due to forward scattering from the near wall.  Figures \ref{f_sed_2walls} and \ref{f_3col_2walls} show SEDs and near-IR images for a model with enhanced density in the wall between the outflow cavity and envelope, and decreased (to zero) density in the outflow cavity, compared to the standard Class I model (Figure \ref{f_3col_powlaw}, left).  The SEDs are slightly modified from the standard Class I model, especially pole-on where the optical extinction along the cavity direction is different between the two models.  

\subsection{Fractal Clumping}

\label{s_fractal}

We incorporate fractal clumping, following a recipe described by \citet{mathis02} and \citet{elmegreen97}.  We have used this algorithm previously to model scattered light nebulae \citep{mathis02}, UCHII regions \citep{indebetouw06} and extragalactic super star clusters \citep{whelan11}.  
The fractal clumping is incorporated as a fractional variation on the density structure of the disk/envelope/outflow,
specified by an input variable which is the ratio of clumped to smooth density.
Currently, to simplify input parameters, the code is set up to give a fractal dimension  of the clumping $D=2.6$, and
a power spectrum exponent of $\beta = 2.8$, in agreement with observed values in the ISM \citep[and references therein]{stanimirovic99, elmegreen96}.  

\begin{figure}
\epsscale{1.0}
\includegraphics[angle=0,width=3.2in]{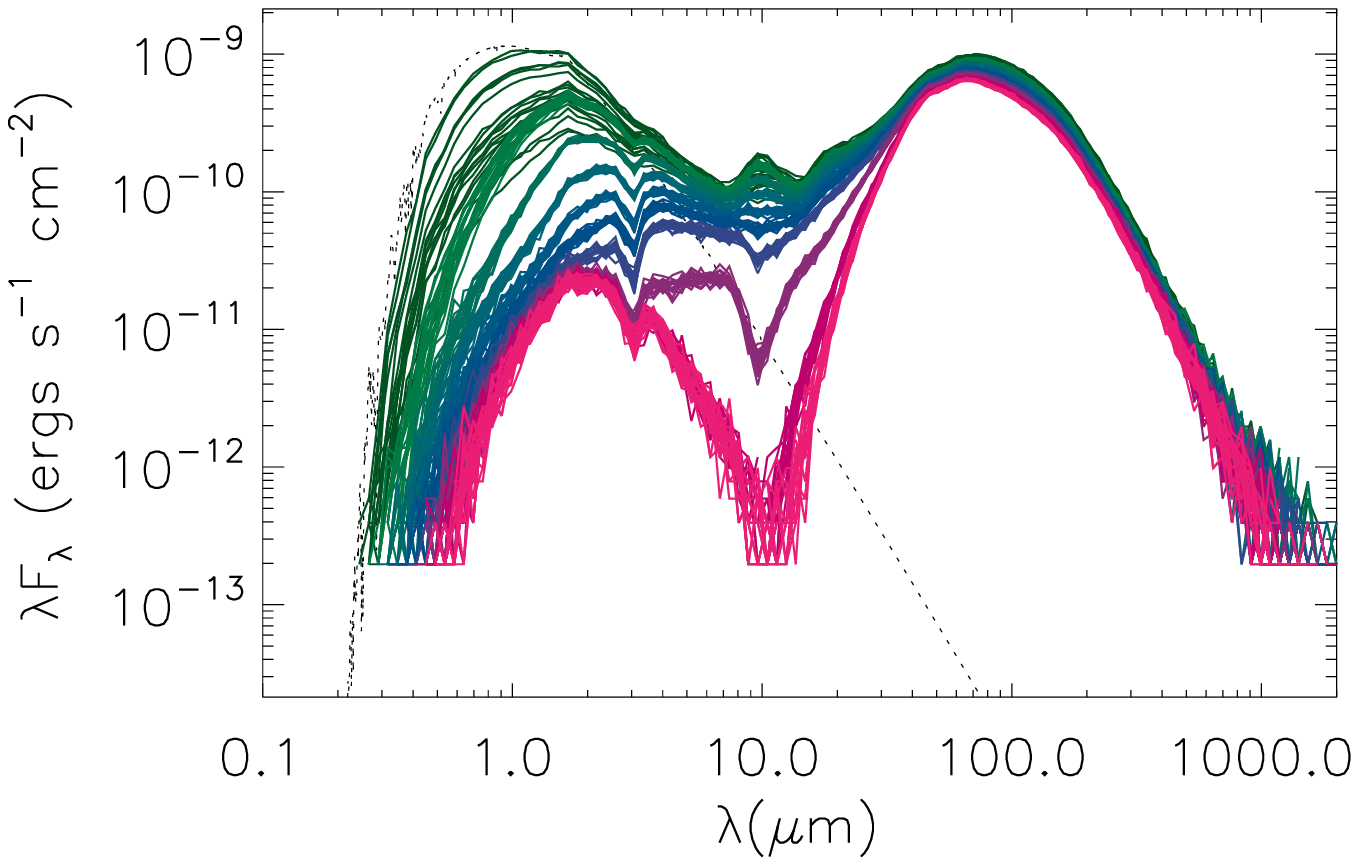}
\includegraphics[angle=0,width=3.2in]{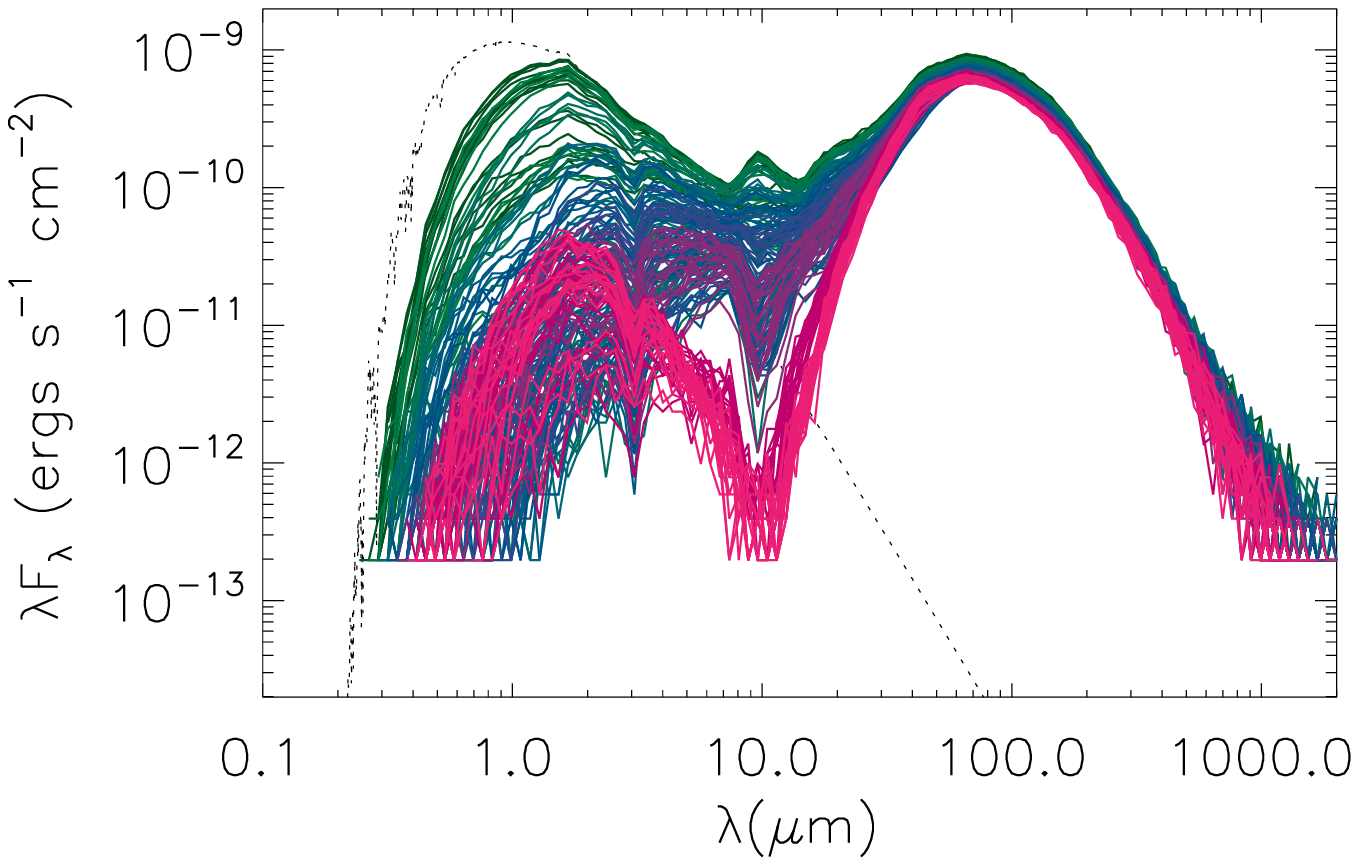}
\caption{The SEDs of a Class I model with fractal density variations in the outflow only (left), and throughout the circumstellar environment (right).   200 viewing angles are shown.  The polar viewing angles are the same as in previous figures, but each polar viewing angle has 20 azimuthal angles which are slightly different due to clumpiness.  The model at right exhibits more clumpiness throughout, so the variations in the SED are greater with viewing angle.    
\label{f_sed_fractal}}
\end{figure}

\begin{figure}
\epsscale{1.0}
\includegraphics[angle=0,width=3.2in]{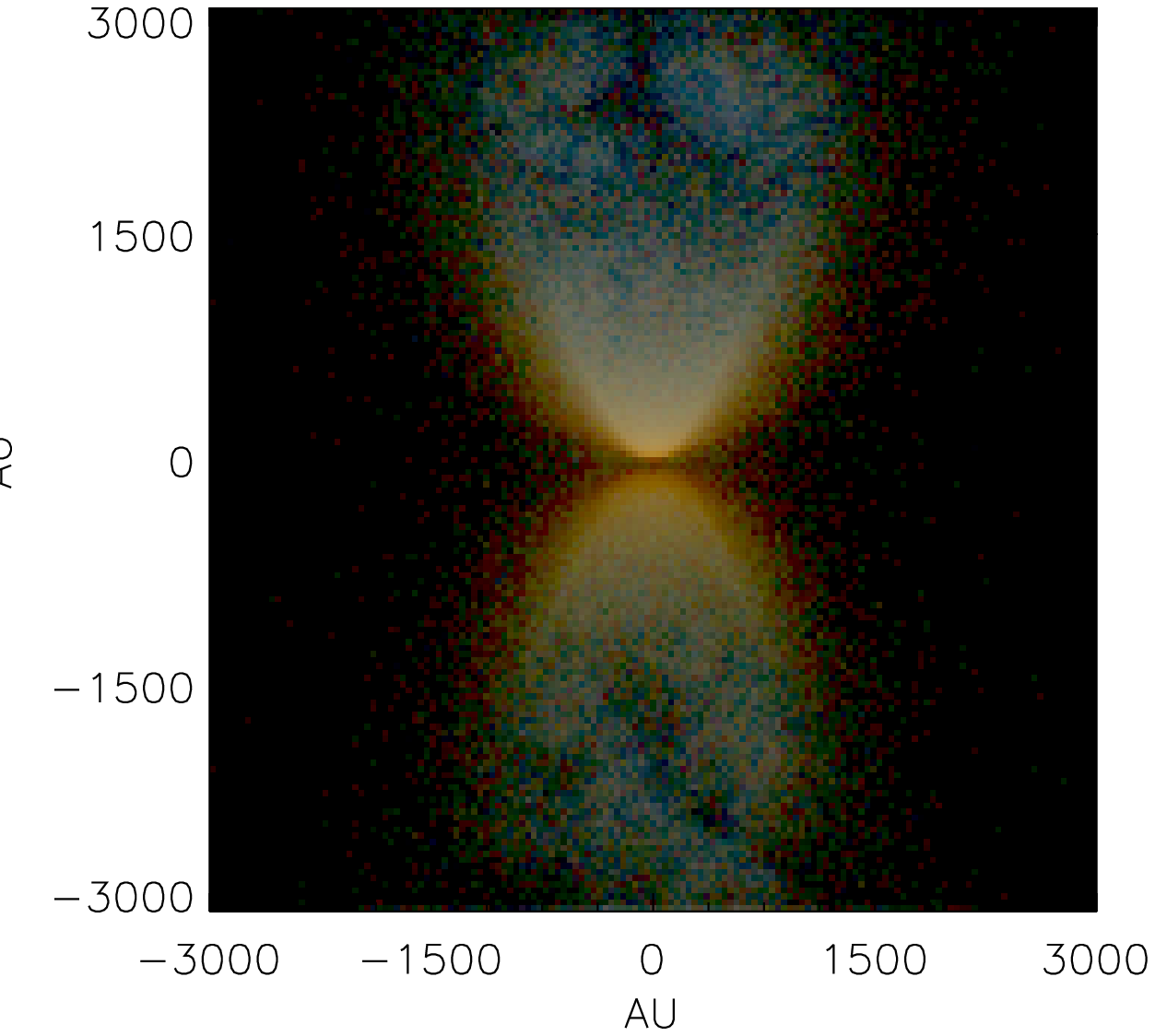}
\includegraphics[angle=0,width=3.2in]{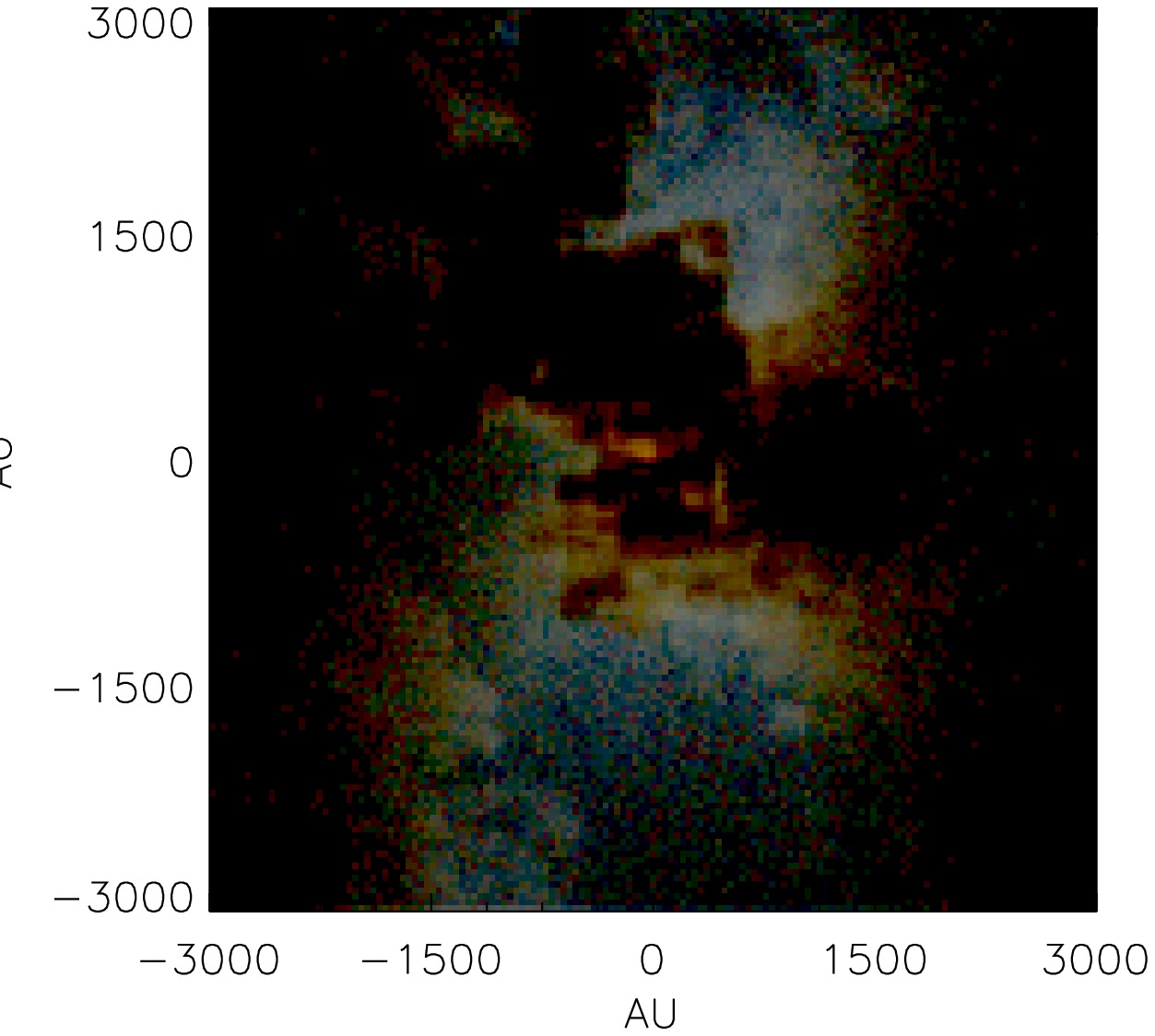}
\caption{
Near-infrared 3-color (JHK) image of a Class I model with fractal clumpiness in the outflow only (left), and clumpiness throughout the nebula (right).  At left, the clumps are blue because they are seen in scattered light in the outflow cavity.  At right, many of the clumps are dark because they are seen in the foreground, absorbing the light scattered in the bipolar cavity.  A smooth component is required to give somewhat of the standard axisymmetric geometry typical of protostars.  The ratio of clumped to smooth density is 0.5 at right and 0.9 at left (in just the outflow). 
\label{f_3col_fractal}}
\end{figure}

Observed images of protostars show evidence for clumpiness in their outflows and possibly envelopes as well.   The six Class I/II sources observed by \citet{padgett99} at high spatial resolution with HST NICMOS are examples.  We added fractal
clumpiness to our Class I model and show the SEDs and near-IR images in Figures \ref{f_sed_fractal} and \ref{f_3col_fractal}.  We first included clumpiness throughout the entire circumstellar disk/envelope/outflow, and found that the images lose most of the axisymmetrical features unless the ratio of clumpy to smooth density was 0.05 or less.  Most observed protostars, even those with apparent clumpiness, show typical axisymmetric structures of disks and bipolar outflows.  Many show more clumpiness in the outflow cavities.   Perhaps the motions in the outflows are more turbulent.  On the other hand, \citet{tobin10, tobin11, tobin12} show evidence for complex structure in Class 0 protostars in both imagery and velocity structure.   Further investigations of model images compared to data can place limits on the smooth to clumpy ratio in both the envelope and outflows.   Figure \ref{f_clump_img} shows an example model with clumps in the outflow only (left) vs clumps throughout the circumstellar environment (right).  The outflow clumpiness has a clumpy to smooth ratio of 0.9.  The model with clumps throughout has a ratio of 0.5.     

\subsection{Stellar Accretion Hot Spots and Inner Disk Warps}
\label{s_hotspot}

In the magnetospheric accretion model, thought to be appropriate in low-mass T Tauri stars (pre-main sequence stars surrounded by disks and possibly envelopes), material accretes through the disk until it reaches the disk truncation radius, where it then flows along magnetic field lines and crashes onto the stellar surface.  Idealized models assume a dipole magnetic field on the star.  In these models, the material  flows along these field lines and shocks onto two ``hot spots'' on the star where the dipole field is the strongest. The dust sublimates inside the dust destruction radius, so the material is in the form of gas.  However, if the truncation radius is close to the dust sublimation radius, this gas may entrain and uplift dust in the disk, causing a ``warp'' in the dusty disk structure.  Some visualizations of these processes can be found in new 3-D magnetohydrodynamic simulations \citep{romanova03,romanova04,romanova08,romanova11}.   We have added stellar hotspots and inner disk warps to our code to simulate these processes and allow users to test predictions of the magnetospheric model against data.   

The energy from accretion 
dissipated in the disk and at the stellar hotspots is \citep{pringle81,bjorkman97,hartmann09}
\begin{equation}
L_{\rm acc, disk} = {{3 G\, M_\star \,\dot M_{\rm disk}}\over{2\, R_{\rm dust}}} 
\left[1 - 2/3 \sqrt{{R_0}/{R_{\rm dust}}}\right],
\label{e_ldisk}
\end{equation}
%
%
\begin{equation}
L_{\rm acc, spot} =  G\, M_\star \,\dot M_{\rm disk} \left( {{1}\over{R_\star}} - {{1}\over{R_{\rm trunc}}} \right) \; ,
\label{e_lspot}
\end{equation}
where $\dot M_{\rm disk}$ is the disk accretion rate, $M_\star$ is the stellar mass, 
$R_\star$ is the stellar radius,  $R_{\rm trunc}$ is the inner edge of the disk from where the accreting material 
freefalls along magnetic field lines onto the stellar hotspots (accretion shocks), and $R_0$ is the radius for the zero-torque boundary condition, which we take to be equal to $R_{\rm trunc}$.
If $R_{\rm trunc}<R_{\rm dust}$, we are neglecting the accretion luminosity between these
radii, since we do not account for gas opacity inside the  dust sublimation radius, which is $R_{\rm dust}$.
If  $R_{\rm trunc}=R_{\rm dust}$, equation \ref{e_ldisk} simplifies to 
$L_{\rm acc, disk} = {{G\, M_\star \,\dot M_{\rm disk}}\over{2\, R_{\rm trunc}}}$.
Following \citet{calvet98} we set half of the hotspot luminosity to be emitted outwardly as X-rays from the shock, and half
goes into heating the stellar atmosphere at the hotspot.   

Note that the total luminosity is now
\begin{equation}
L_{\rm tot} = L_\star + L_{\rm acc, disk} + L_{\rm acc, spot} + L_{\rm ISRF}.
\end{equation}
The disk accretion luminosity is emitted via equation (4) of Paper I.   The spot temperature is calculated by equating the hotspot accretion luminosity with the area emitted, i.e., $4 f_{\rm spot} \pi \sigma T_{\rm spot}^4 R_\star^2 = 0.5 L_{\rm acc, spot}  + L_\star f_{\rm spot}$, giving
\begin{equation}
T_{\rm spot} = T_\star \left( 1+ {{L_{\rm acc, spot}}\over{2\, L_\star\, f_{\rm spot}}}\right)^{1/4} \;,
\end{equation}
where $T_\star$ and $L_\star$ are the stellar temperature and luminosity and $f_{\rm spot}$ is the 
fraction of the stellar surface covered by hotspots.  
Our default size for $f_{\rm spot}$ is 0.007, the median value estimated by \citet{calvet98} from models and observations of several T Tauri stars.
Based on the input value for $f_{\rm spot}$ and the number of spots (one or two), the code calculates the spot angular radius. 
For a single spot, this is 
$\theta_{\rm spot} = \cos^{-1}\left({1-2\, f_{\rm spot}}\right)$.
For two hotspots, which are placed diametrically opposite (that is, the second hotspot is at the negative of the latitude of the first, and its longitude is 180\arcdeg\ from the first), the angular radii are
$\theta_{\rm spot} = \cos^{-1}\left({1- f_{\rm spot}}\right).$
The stellar hotspot flux is emitted from a Planck function at the temperature $T_{\rm spot}$.
The X-ray flux is emitted evenly from
100-500 $\AA$.  
This is not the correct spectrum, but serves the purpose to heat up the disk when absorbed by it.
  
The disk structure can be modified to simulate warping of the inner dusty disk where the gas flows along magnetic field lines towards the dipole hotspots on the star \citep[see also][]{osullivan05,romanova08,romanova11}.  The dust is destroyed inside a radius
where the equilibrium temperature is above the estimated dust sublimation temperature (we assume 1600 K).  Thus in broadband
images that see only the dust and not the gas, the effect may be that
the dust is uplifted slightly with the gas, forming warps in the disk where the accretion
columns are.   
The normal disk structure is assumed to have a gaussian $z-$dependence, with a scale height that increases with radius via $h=h_0 (\varpi/R_\star)^\beta$ \citep[paper I, eqn. 3, see also][]{shakura73,lynden-bell74,pringle81,bjorkman97,hartmann09}.  
We adopt the following parametrized form for the variations in the disk scale height warping as a function of
cylindrical radius $\varpi$ and the longitude $\Omega$:
\begin{equation}
h_{warp}(\varpi , \Omega) = h_{fid} \left[H_{\rm warp}   \cos^w\left(\Omega/2 -a \right)
exp \left ( \frac {\varpi - R_{dust}}{L_{\rm warp}} \right)^2
 \right] ,\label{eqn_warp}
\end{equation}
where the input variables are the additional height of the warp $H_{\rm warp}$, the radial scale length of the warp $L_{\rm warp}$, and the exponent $w$ which determines the angular width of the warp; and $h_{fid}$ is the fiducial scale
height, $h_0 \left( {{\varpi}\over{R_\star}}\right)^\beta$.  The parameter $a$ is set to 0 to place the first warp at 0 longitude; $a$ is set to  $\pi/2$ if a second warp is desired at 180 longitude.  This function is added to the
fiducial scale height of the disk ($h_{fid}$).  The first warp extends in the $+z$ direction and the second warp extends in the $-z$ direction to line up with the stellar hotspots.    

\begin{figure}
\epsscale{1.0}
\includegraphics[angle=0,width=3.2in]{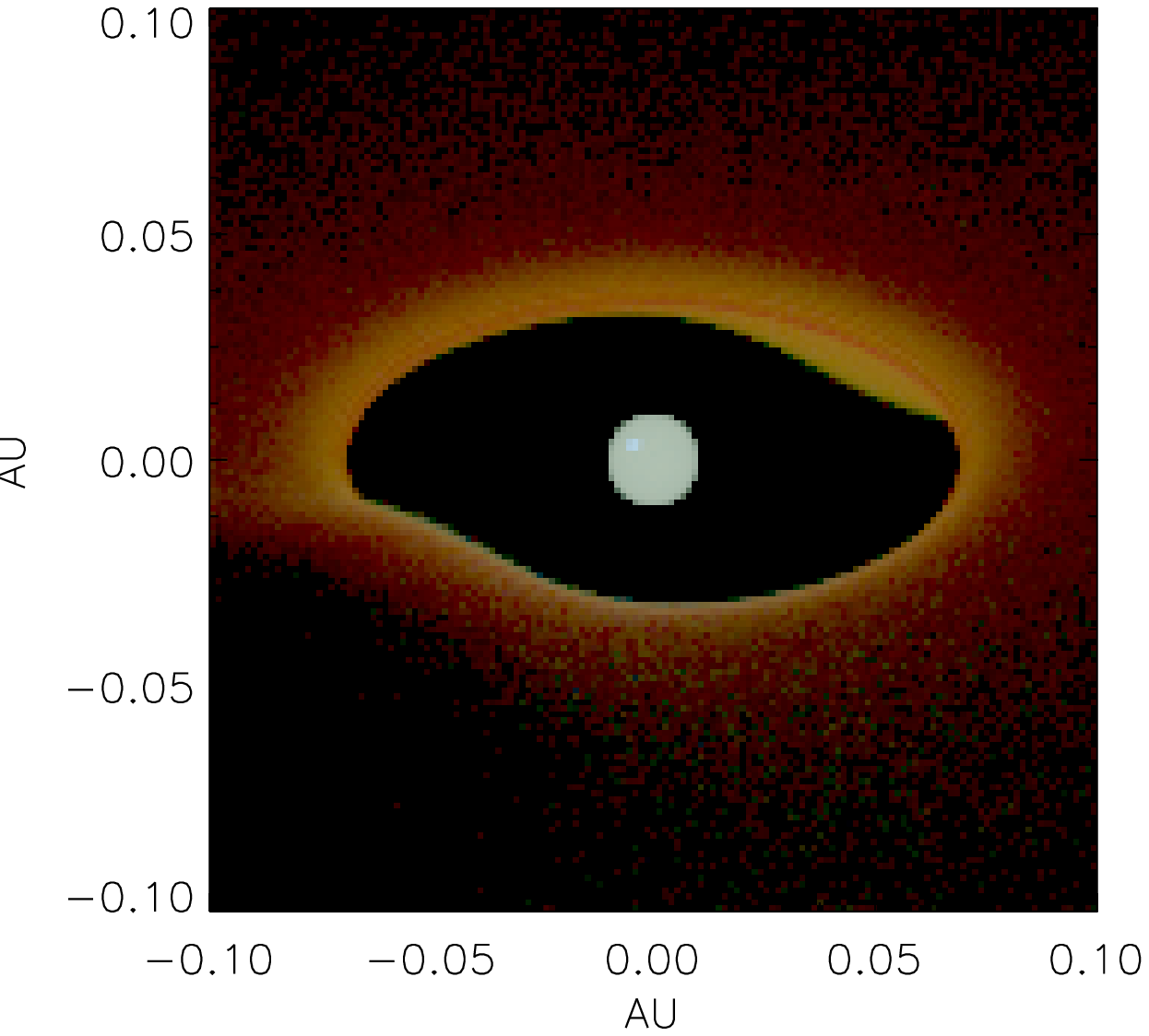}
\includegraphics[angle=0,width=3.2in]{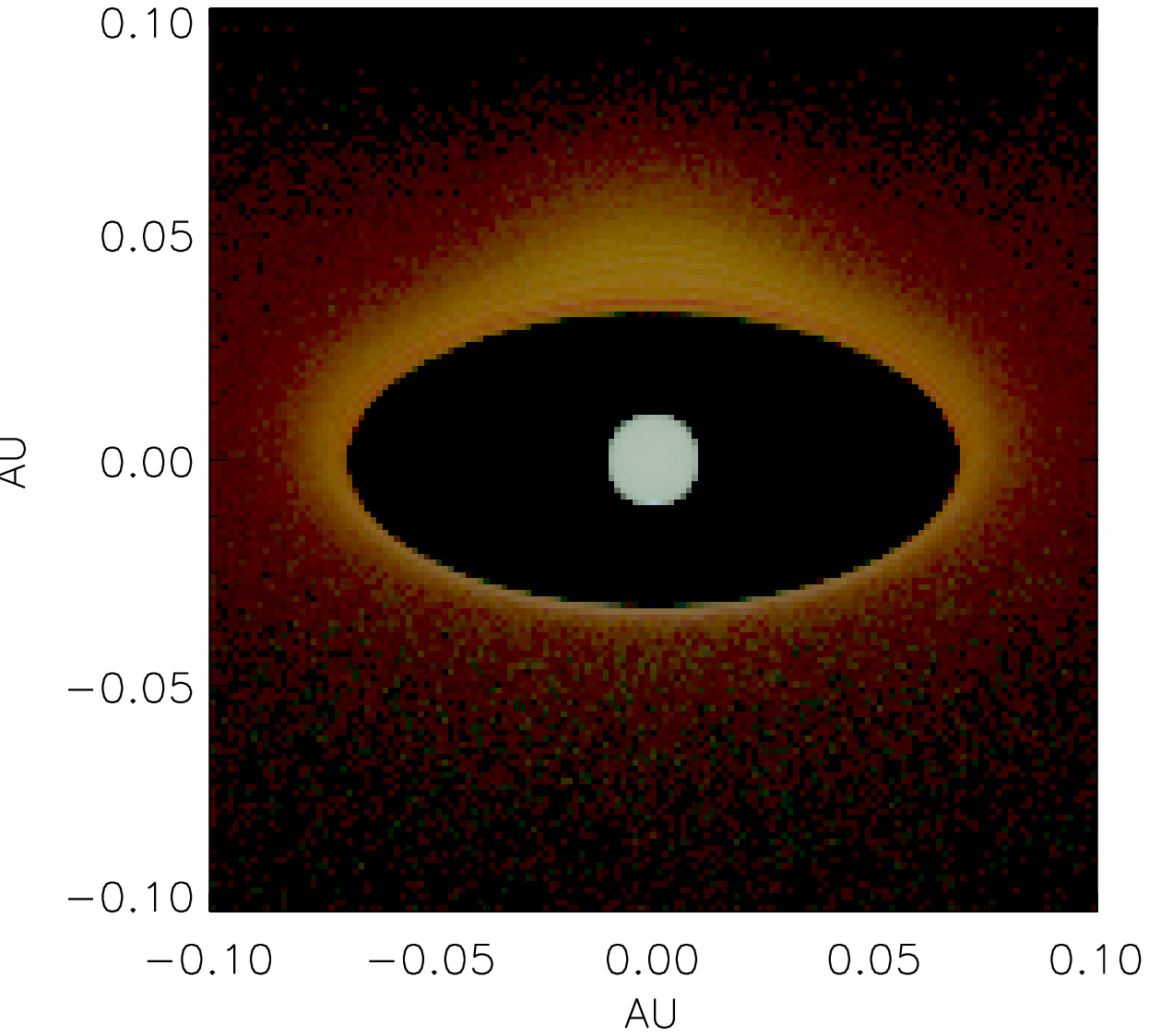}
\caption{
3-color images showing the geometry of the rotating hotspot model (\S\ref{s_hotspot}).  V-band (0.55 $\mu$m) is in blue, J-band (1.2 $\mu$m) is in green, and IRAC 4.5 $\mu$m is in red.  In the left planel, one hotspot is visible on the star at a latitude of 45$\arcdeg$; the other is behind the star, 180$\arcdeg$ away in longitude, and at a latitude of -45$\arcdeg$.  The inner disk is warped up and below at the same longitudes of the hotspots.  The outer disk flares as in the standard models.  The left panel shows an azimuthal angle of $\phi=40\arcdeg$ and the right panel shows $\phi=180\arcdeg$.
\label{f_hotspot_img}}
\end{figure}

\begin{figure}
\epsscale{1.0}
\includegraphics[angle=0,width=2.in]{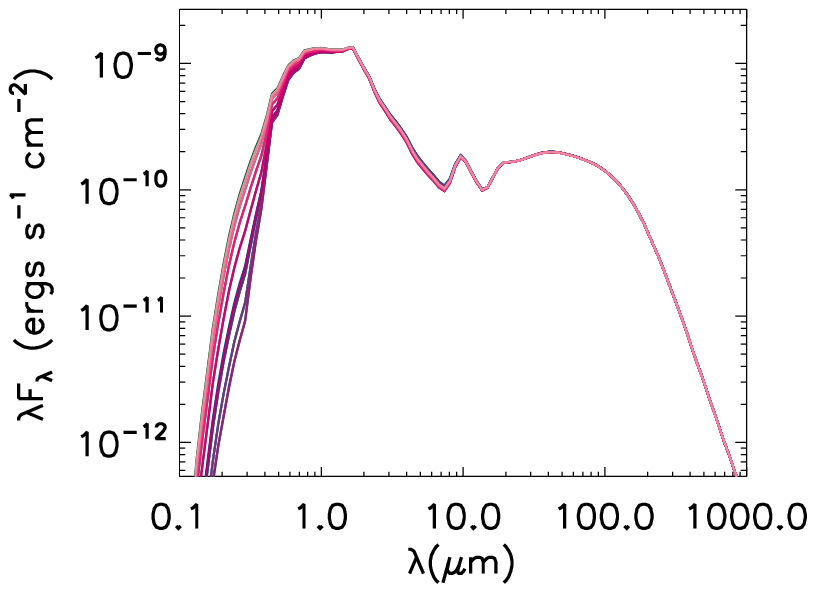}
\includegraphics[angle=0,width=2.in]{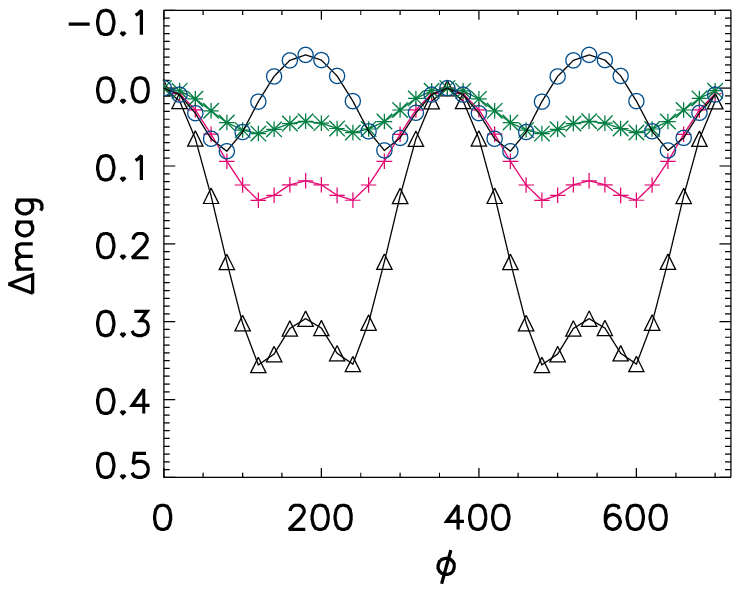}
\includegraphics[angle=0,width=2.in]{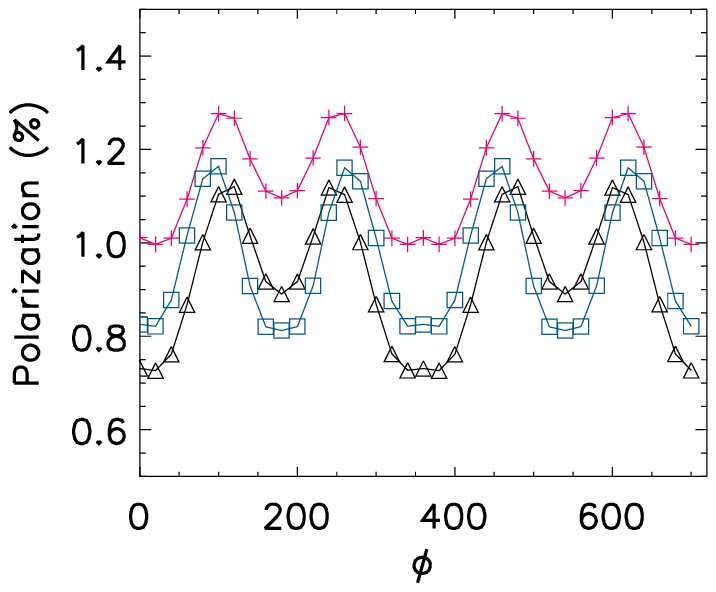}
\caption{
SEDs (left), light curves (middle) and polarization phase curves (right) for the rotating hotspot model.  The models are viewed at an inclination ($\theta$) of 60$\arcdeg$, and at 40 azimuth angles ($\phi$).   The variation with azimuth is the same as if the star and inner disk are rotating.  Two cycles of the rotation are plotted to show the periodicity.  The colors of the SEDs vary from pink to purple for the azimuthal angle.  The colors of the light curves correspond to different wavelengths:  black is V-band (0.55 $\mu$m), pink is I-band (0.8 $\mu$m), green is J (1.2 $\mu$m), and blue with circles is IRAC 4.5 $\mu$m.  For the polarization curves:  black is V-band, pink is I-band and light blue is K (2.2 $\mu$m).  
\label{f_hotspot_lc}}
\end{figure}

A large amount of data showing infrared variability has become available from the {\it Spitzer} YSOVAR project \citep{morales09,morales11}, and now Herschel \citep{billot12}.  We have produced simulations for these projects (Kesseli et al. in prep) and plan more in the future.
Other groups have also computed radiation transfer models producing variability from inner disk warps \citep{flaherty10}.
Figure \ref{f_hotspot_img} shows a model with two stellar hotspots at latitudes of $+45\arcdeg$ and $-45\arcdeg$ separated by 180$\arcdeg$ in longitude.  The temperature of the star is 4000 K and that of the hotspots is 7000 K (set by the disk accretion rate and size of the spots).  The emission contrast from the hotspot and star is greatest at wavelengths smaller than the peak of the stellar spectrum (on the Wien side of the Planck function) and is smaller for wavelengths longward of the peak.  Thus we should see the largest variations of the hotspot at visible wavelengths. The disk has warps at the same longitude as the hotspots above the equatorial plane for the upper hotspot, and below for the lower.  The disk warps should be noticeable at infrared wavelengths due to the variations in the emitting area.  Figure \ref{f_hotspot_lc} shows the SEDs, light curve, and polarization curves, as the model rotates--that is, as the azimuthal angle $\phi$ varies--for an inclination of 60\arcdeg.  

The light curve (middle panel of Figure \ref{f_hotspot_lc}) shows the largest variations at V-band ($0.55 \mu$m).  It is brightest when the upper hotspot is pointing directly at us.  As the phase angle goes past 90\arcdeg\, the upper hotspot goes out of view and the lower is too low in latitude to be seen at our viewing inclination.  The small brightness increase at 180\arcdeg\ is due to the increasing back-scattered light  from the upper warp in the disk coming into view (Figure \ref{f_hotspot_img}, right).  At 4.5 $\mu$m, the contrast between the stellar hotspot and the rest of the star is much smaller, so there is little stellar variation with phase.  All of the variation is from the warmed inner disk. In this case, the peak flux occurs at phase 180$\arcdeg$ where the viewed disk emitted area is largest (Figure \ref{f_hotspot_img}, right).  The I and J bands show intermediate effects from both the hotspot and the disk scattered and emitted areas.  

The polarization plot (right panel of Figure \ref{f_hotspot_lc}) shows the largest polarization at I band, though with less variation.  
The polarization peaks when the warps are off to the side, as this gives the largest asymmetry for scattering.  The polarization is due to a combination of dust properties:  scattering albedo, polarizing efficiency, and scattering phase function that all vary with wavelength.  At K-band the scattering phase function is more isotropic, allowing more of the asymmetric scattering from the warps to reach the observer near the 90\arcdeg\ phase angle.  Its peak is shifted to just shortward of 90\arcdeg\ and longward of 270\arcdeg\ where the scattering from the warp is more visible but also more asymmetric (than at 0 and 180\arcdeg).  The V-band polarization is lowest when the spot is in view because it is bright and unpolarized.  

\subsection{Gaps, Curved and Puffed Rims, Mis-Aligned Inner Disks, and Spiral Arms in Disks}

\label{s_gaps_etc}

Planet formation theories predict disk structures such as clumps, spiral arms, gaps and warps 
\citep{artymowicz94,quillen04,varniere06,jang-condell07,ireland08,jang-condell09,zhu11}.  
High spatial resolution observations
\citep{calvet02,fukagawa06,hughes09,thalmann10,hashimoto11,andrews11a,andrews11b}
as well as SED modeling 
\citep{calvet05,espaillat07,espaillat08}
have confirmed such structures.  We provide some simple analytic parameterizations of such structures
to allow users to model their data.  When good fits are attained, the parameterizations can be compared to
theoretical predictions to constrain physical parameters.
We include gaps in the disk and/or envelope.
The disk structure inside the gap can
be different from the rest of the disk; that is, with a different radial power law exponent,  flaring parameter, and fiducial
scaling.  

The inner disk and gap walls can be ``puffed-up.''  It is physically unclear what would provide this mechanism since
the hydrostatic equilibrium solution alone  increases the inner rim scale height by only a small amount (\S\ref{hseq}).  But some observations found improved fits with these structures \citep{eisner04}.  We use a similar parameterization as our warp (equation \ref{eqn_warp}), without the azimuthal dependence,
\begin{equation}
h_{rim}(\varpi) = h_{fid} \left[H_{\rm rim}
exp \left ( \frac {\varpi - R_{\rm in}}{L_{\rm rim}} \right)^2
\right].
\label{eqn_rim}
\end{equation}
This is an additive function to the scale height that peaks at the inner radius and falls off with radius with a Gaussian function.  The scale length $L_{\rm rim}$ and the additive scale height $H_{\rm rim}$ are input parameters.  The disk wall at the outer edge of the gap can be puffed-up as well.

Additionally, the inner rim and outer gap walls could be curved, due to dust sublimation or other effects \citep{kama09,isella05}.   We parameterize this similarly:
\begin{equation}
h_{curve}(\varpi) = h_{fid} \left[H_{\rm curve}
exp \left ( - \frac {\varpi - R_{\rm in}}{L_{\rm curve}} \right)^{\rm expcurve}
 \right] , \label{eqn_curve}
\end{equation}
where $H_{\rm curve}$, $L_{\rm curve}$, and $\rm expcurve$ are input.
This function is subtracted from the fiducial scale height of the disk.   These equations are applied in both the
$+z$ and $-z$ directions.  

\begin{figure}
\epsscale{1.0}
\includegraphics[angle=0,width=6.in]{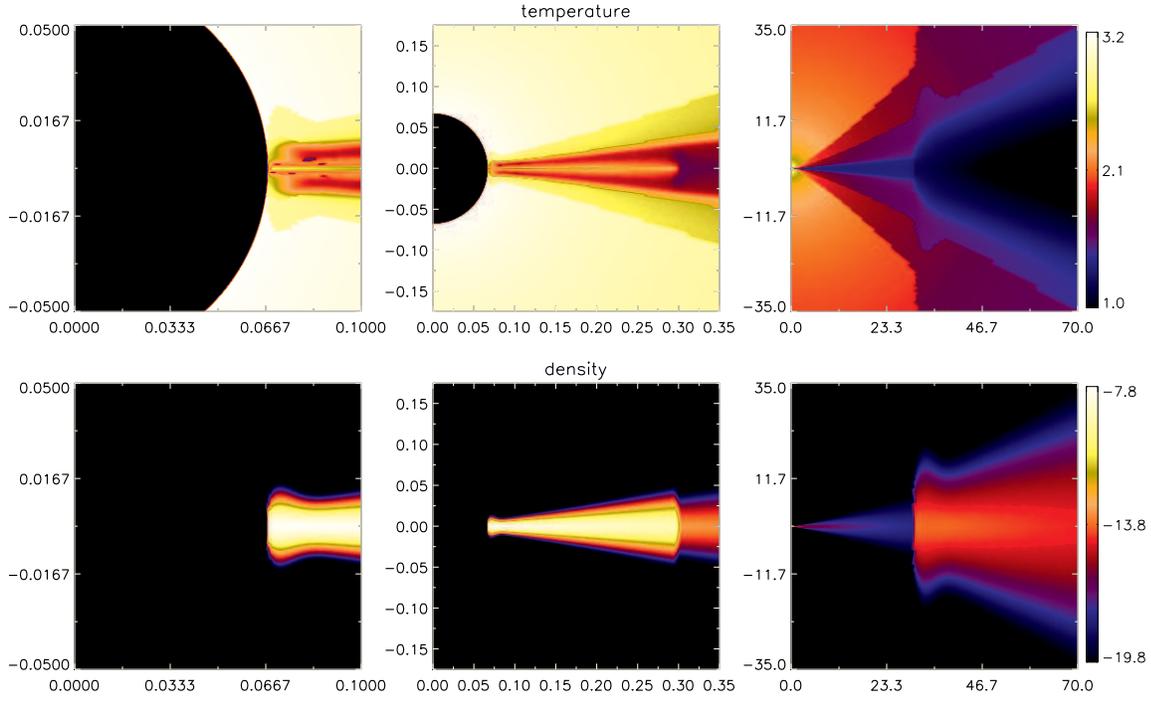}
\caption{
Azimuthal ($z$ vs $\varpi$) temperature (top) and density (bottom) slices for a disk with a gap and puffed up, curved inner rims.  From left to right are three different zooms to show the features at different radii.
\label{f_trho_gap}}
\end{figure}

\begin{figure}
\epsscale{1.0}
\includegraphics[angle=0,width=3.2in]{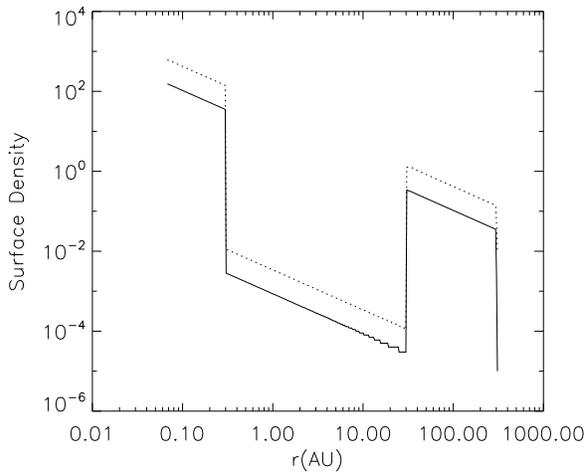}
\caption{
Surface density of the gapped disk.  The dotted line is for the ``small-grain'' disk and the dashed for the ``large-grain.''  The mass of the large-grain disk is 1/5 of the total.
\label{f_sigma_gap}}
\end{figure}

\begin{figure}
\epsscale{1.0}
\includegraphics[angle=0,width=3.2in]{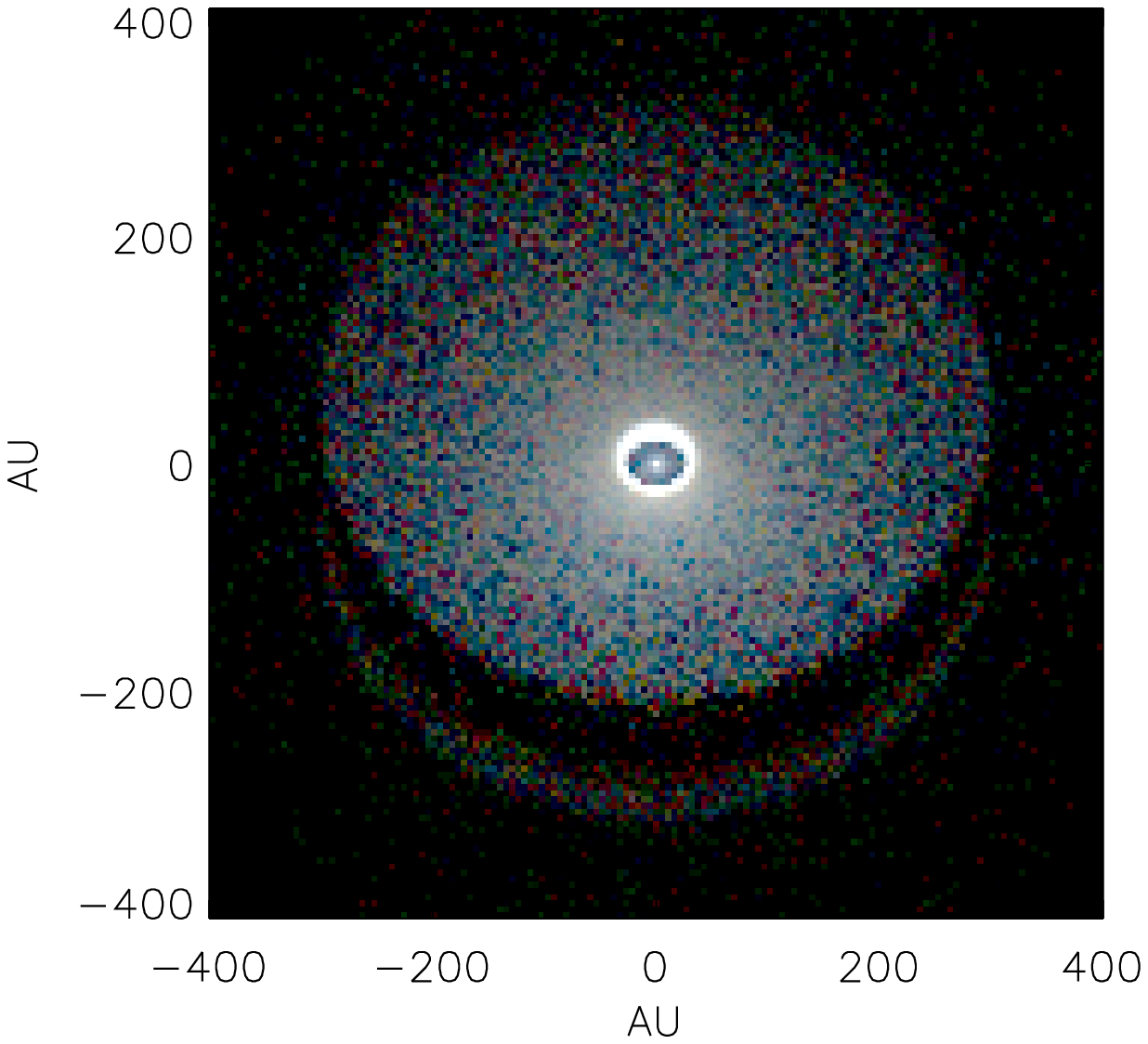}
\includegraphics[angle=0,width=3.2in]{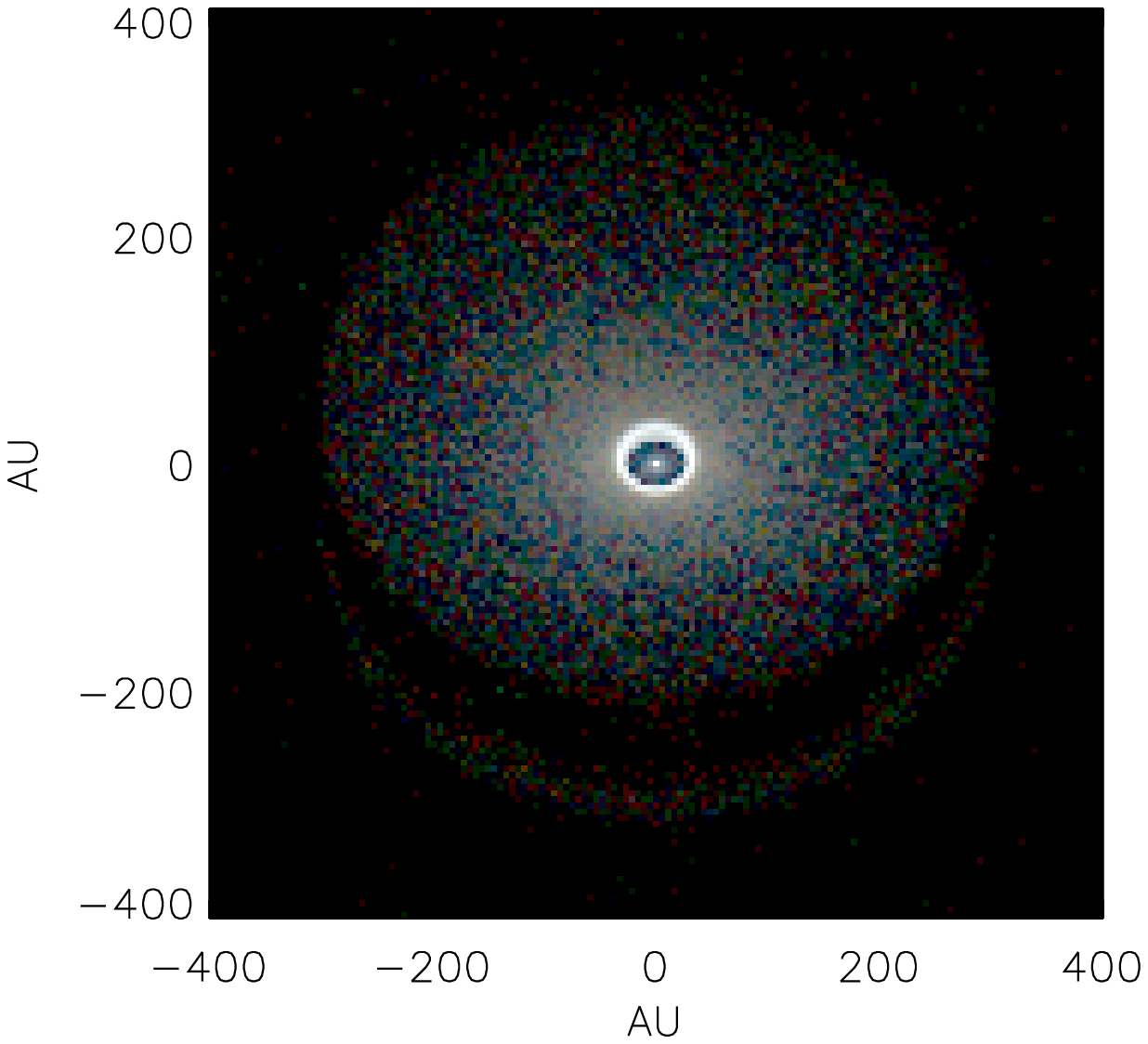}
\caption{
Left: Near-infrared 3-color (JHK) image of the gapped disk with puffed inner rims, viewed at 30\arcdeg\ inclination.
Right:  The same except polarized flux is displayed instead of intensity.  The images show similar features.  However, at left the central pixel has all the stellar flux and at right it has none.   Convolving with a stellar PSF would wash out the features in the intensity image but not as much in the polarized flux image.  This is the motivation for the SEEDS project which images disks in polarized flux.
\label{f_3col_gap}}
\end{figure}

\begin{figure}
\epsscale{1.0}
\includegraphics[angle=0,width=3.2in]{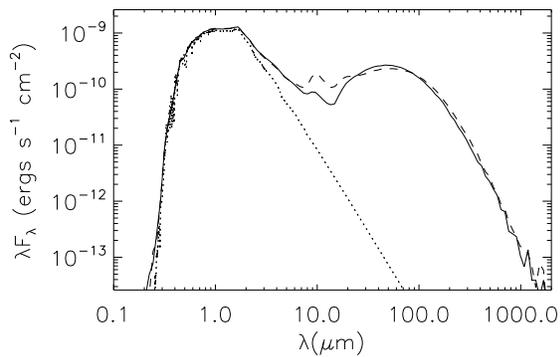}
\caption{
SED of the gapped disk (solid line) compared to the standard Class II model (dashed line) at a viewing angle of 32\arcdeg.
\label{f_sed_gap}}
\end{figure}

Figures \ref{f_trho_gap}-\ref{f_sed_gap} show results for a disk with a gap from 0.3 to 30 AU.  The density inside the gap is scaled to be 0.0001 times the density if there were no gap.  For illustration we puffed up both inner rims (the inner disk radius and at 30 AU) and curved their surfaces, using the formulae in equations \ref{eqn_rim} and \ref{eqn_curve}.  The geometry of the disk is shown in Figure \ref{f_trho_gap}, and is further illustrated in the surface density plot in Figure \ref{f_sigma_gap}.   Figure   \ref{f_3col_gap} shows a JHK image (left) and JHK polarized flux image (right) for the disk viewed at an inclination of 30\arcdeg.  The images show the brightening from the large wall at the outer edge of the gap at 30 AU.  Note that this is an idealized image where all the stellar flux is in the central pixel.  Convolving with a realistic point spread function for the stellar image would wash out these features.
However, an instrument designed to produce polarized flux images, as in the SEEDS project \citep{tamura09}, can mitigate this effect
since the central star is unpolarized.  Our code is well-suited to model the SEEDS data \citep[e.g.,][]{dong12a,dong12b} and these in turn have motivated the geometric structures now available in the code.  Figure \ref{f_sed_gap} shows the SED of this model compared to the standard Class II model without a gap.  It shows the characteristic decrease in emission at mid-IR wavelengths due to the gap \citep{espaillat07}.

\begin{figure}
\epsscale{1.0}
\includegraphics[angle=0,width=2.15in]{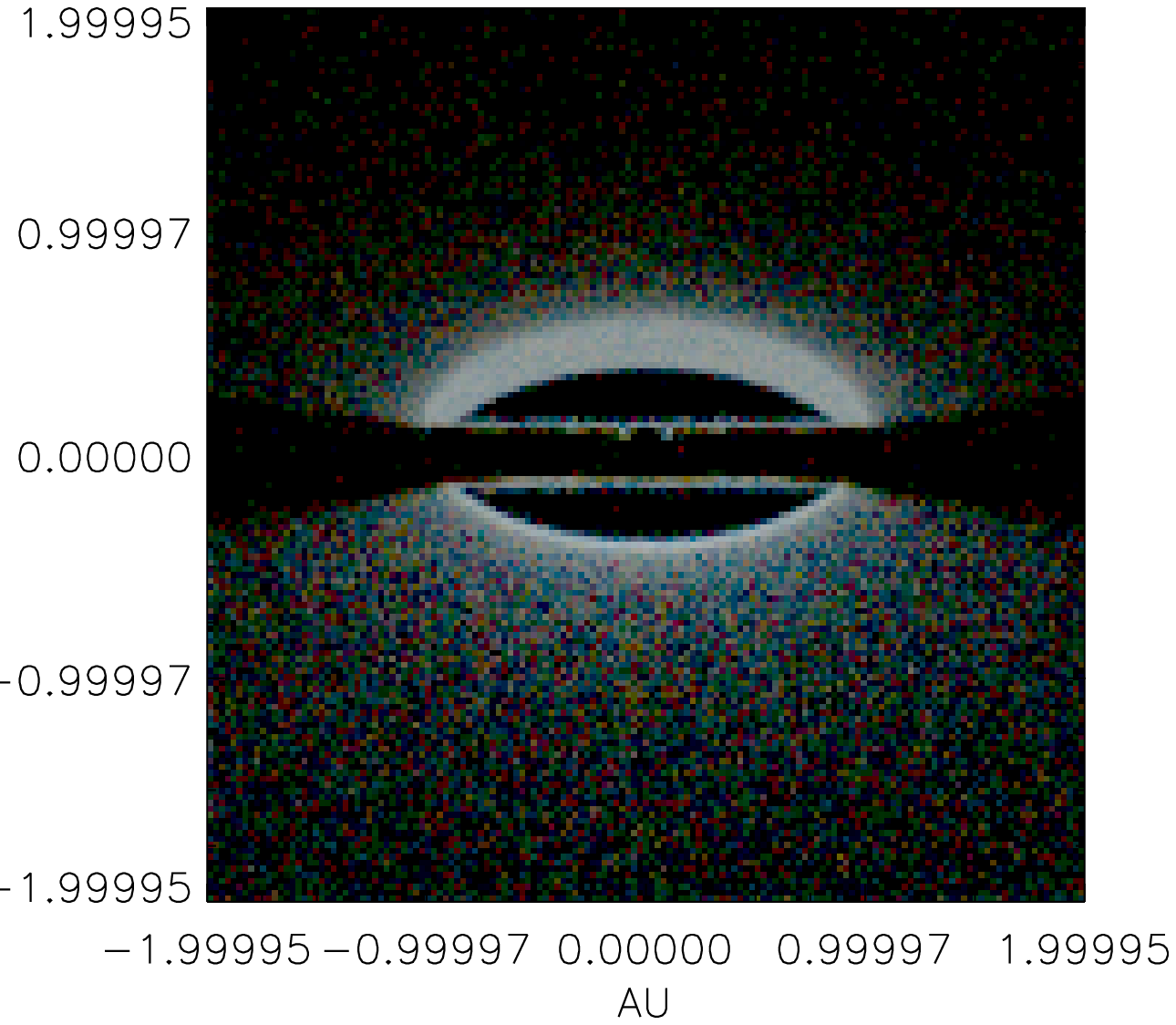}
\includegraphics[angle=0,width=2.15in]{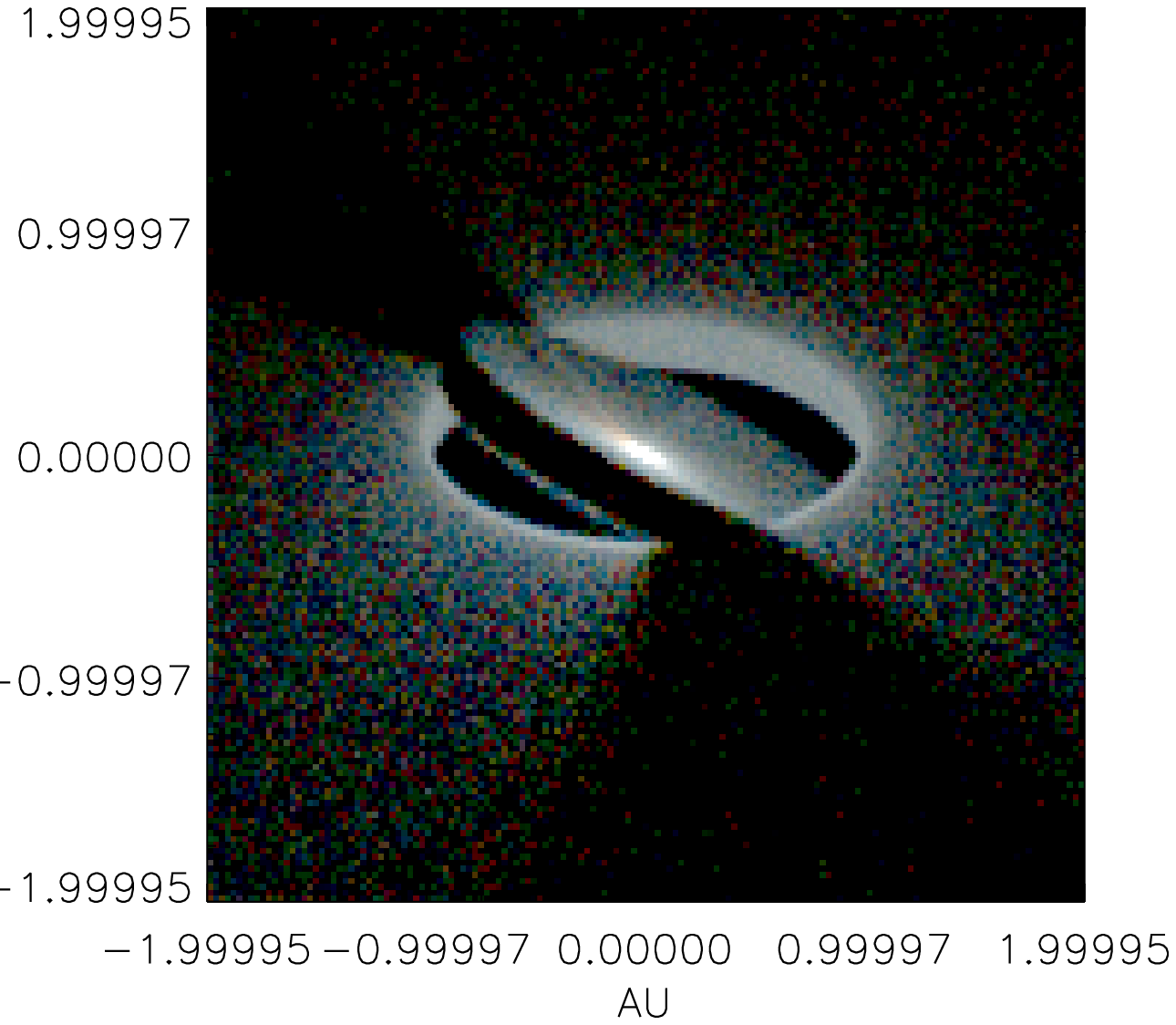}
\includegraphics[angle=0,width=2.15in]{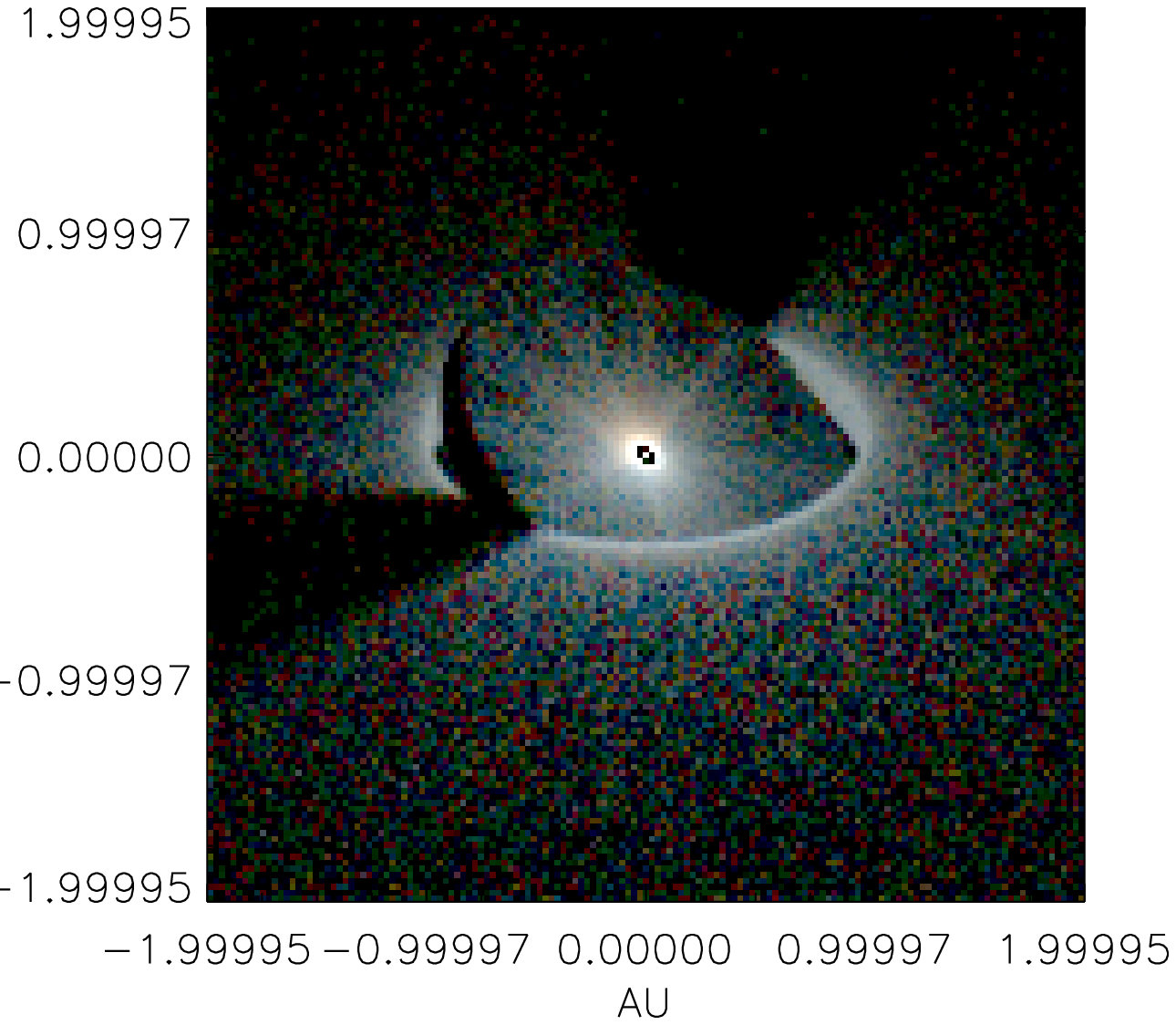}
\includegraphics[angle=0,width=2.15in]{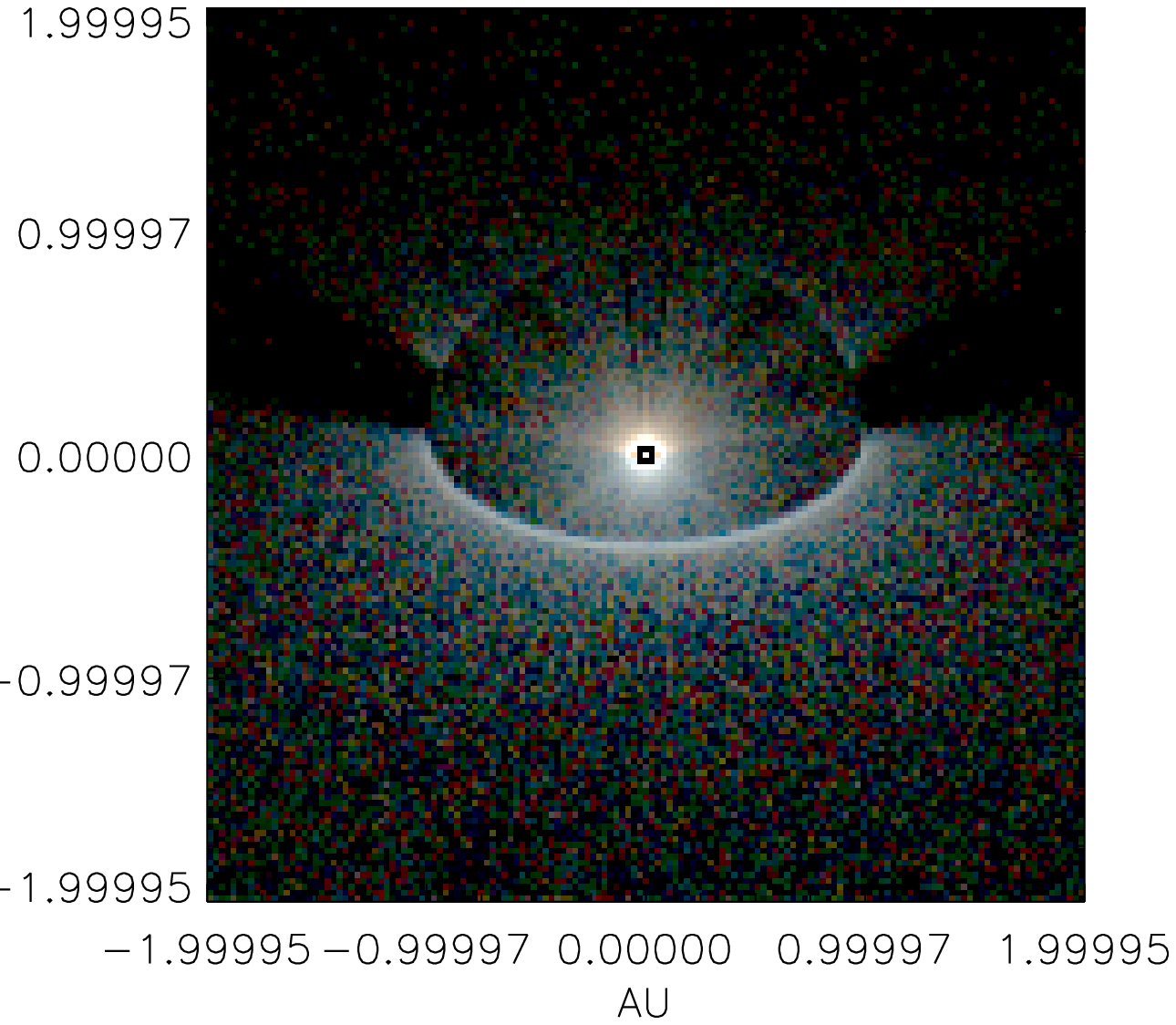}
\includegraphics[angle=0,width=2.15in]{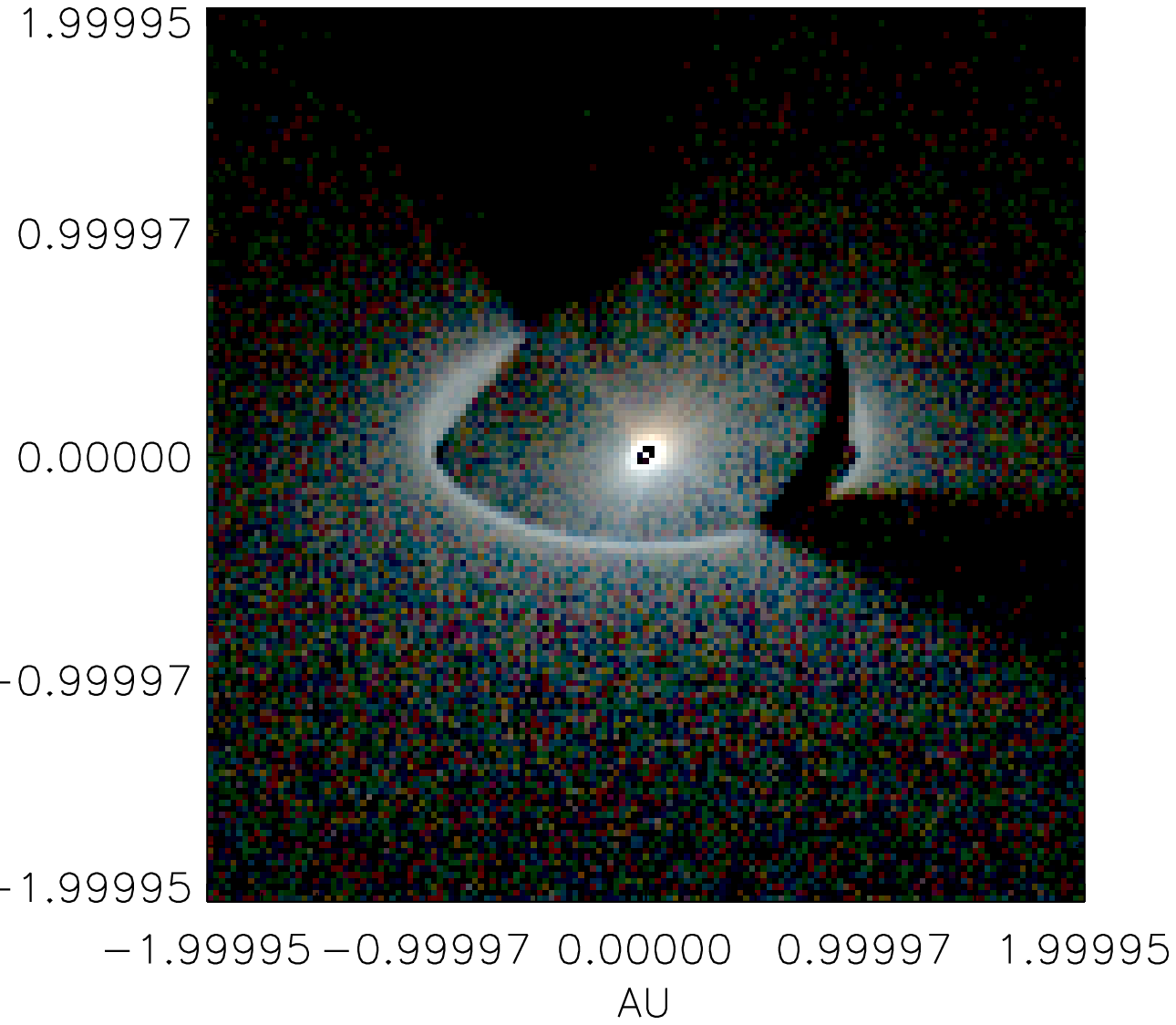}
\includegraphics[angle=0,width=2.15in]{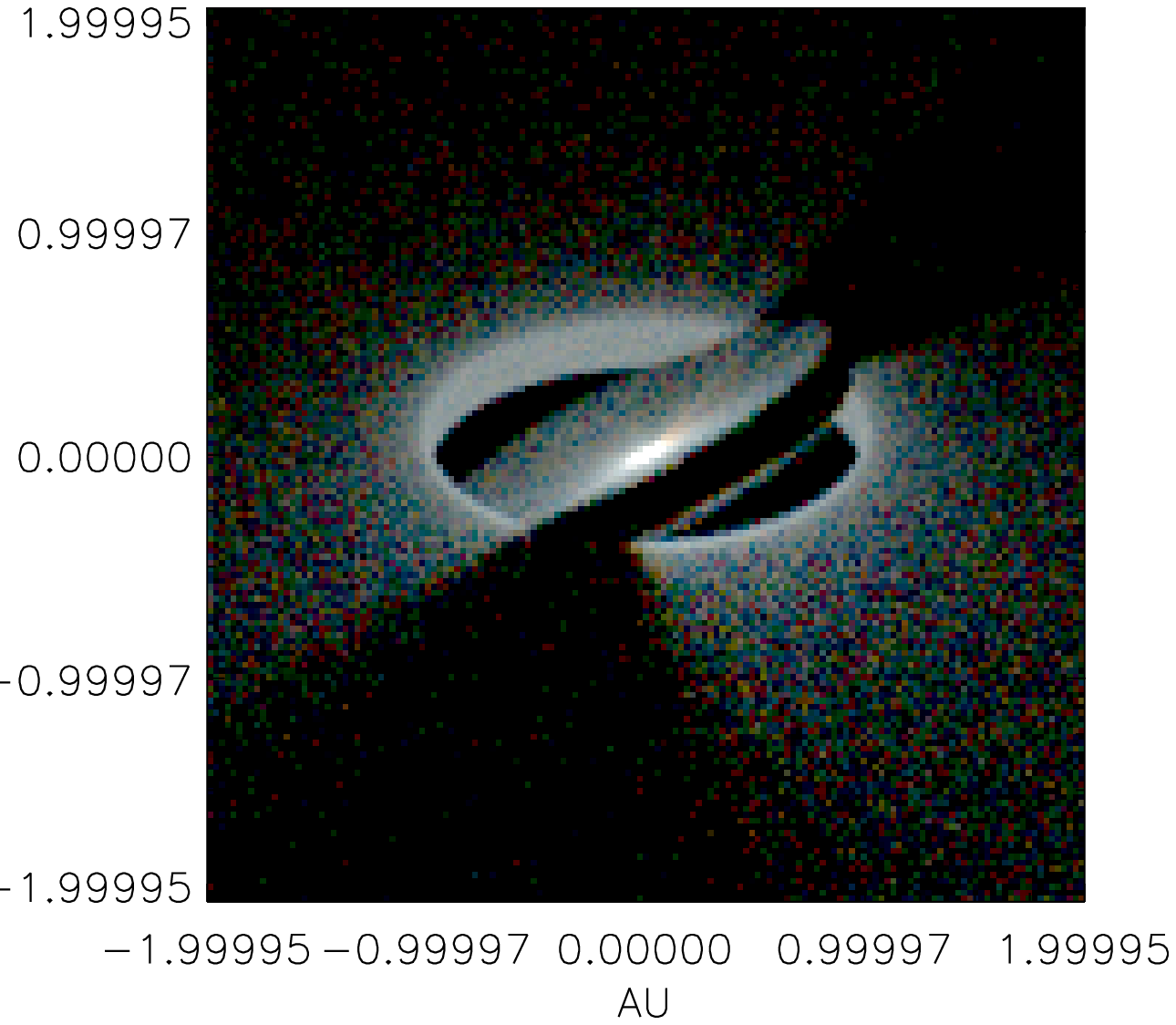}
\caption{
Near-infrared 3-color (JHK) images of the Class II model in which the inner 1 AU of the disk is misaligned at 30\arcdeg\ with respect
to the rest of the disk.  These are viewed at an inclination of 60\arcdeg\ and at six azimuthal angles:  0, 60, 120, 180, 240, and 300\arcdeg.
The  images are zoomed into a radius of 2 AU.  At this inclination, the central source is blocked from view at several azimuthal angles.  This is seen in the light curves of the rotating disk in Figure \ref{f_sed_misalign}.
\label{f_3col_misalign}}
\end{figure}

\begin{figure}
\epsscale{1.0}
\includegraphics[angle=0,width=2.in]{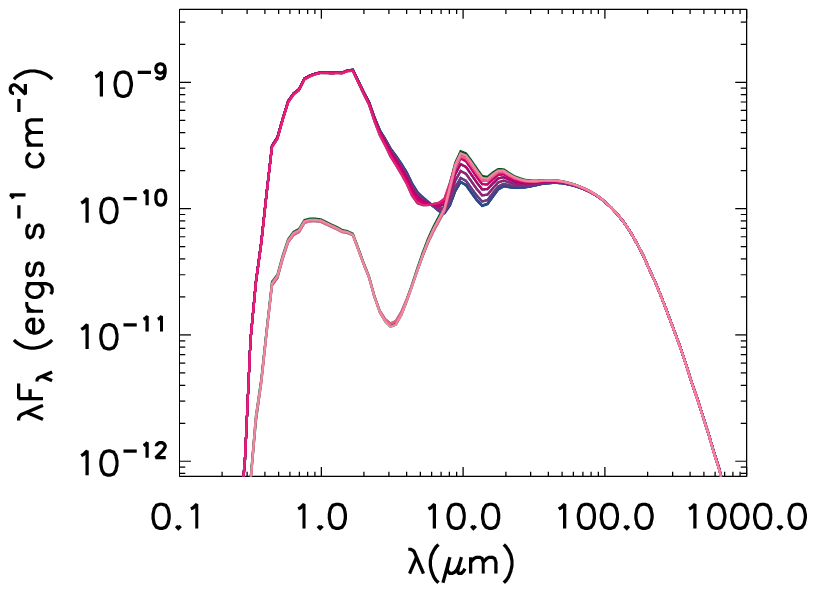}
\includegraphics[angle=0,width=2.in]{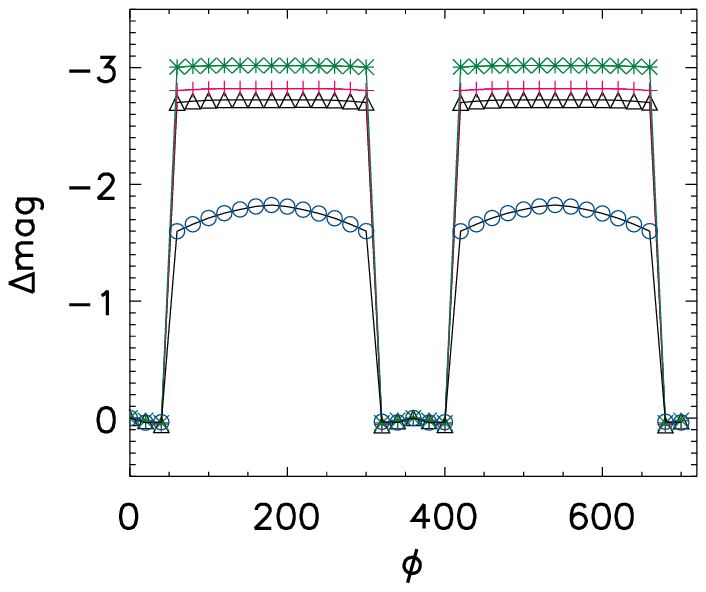}
\includegraphics[angle=0,width=2.in]{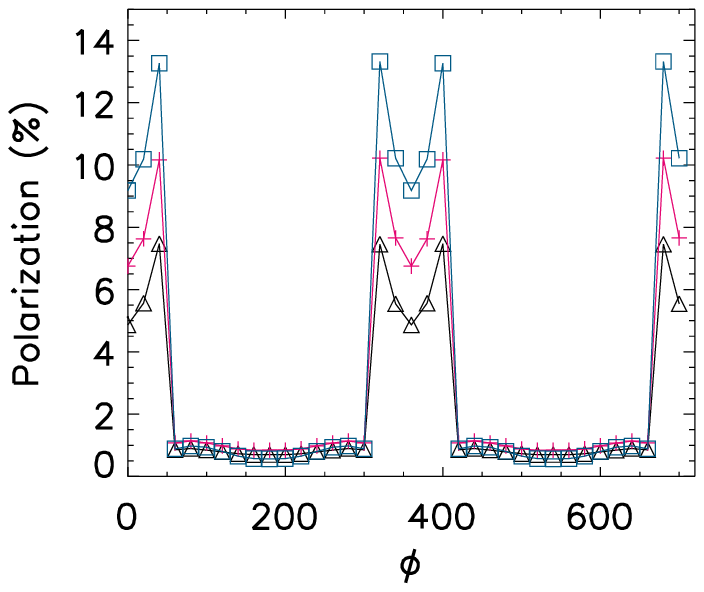}
\caption{Same as Figure \ref{f_hotspot_lc} for the misaligned inner disk.  The light curves and polarization jump as the disk rotates into directions that occult the inner star and disk regions.
\label{f_sed_misalign}}
\end{figure}

High spatial resolution observations of disks have hinted at misalignment of an inner portion of the disk with respect to the outer \citep{grady09,heap00,golimowski06,ahmic09}, so we allow for this option in our code.
A 3-D coordinate transform for the inner disk density is done during the initial
grid setup.
 We show an example model in Figures \ref{f_3col_misalign} and \ref{f_sed_misalign}.   Here the inner 1 AU of the disk is misaligned 30\arcdeg\ with respect to the outer.  As the disk rotates, the central star and disk region is blocked from view over several rotation (or azimuthal) angles, at the inclination shown (60\arcdeg).   This gives large dips in the light and polarization curves.

\begin{figure}
\epsscale{1.0}
\includegraphics[angle=0,width=3.2in]{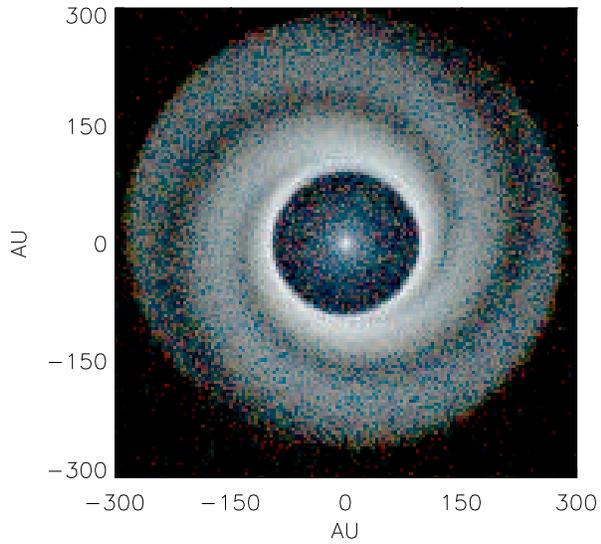}
\caption{
Near-infrared 3-color (JHK) image of the Class II model with a gap and spiral structure outside the gap, viewed at an inclination of 10\arcdeg.
\label{f_3col_spiral}}
\end{figure}

Various dynamical processes (accretion and planet formation) can lead to spiral structures, and recent observations
are suggestive of this or other very asymmetric structures \citep{fukagawa04,fukagawa06,hashimoto11,muto12}.  Following
\citet{schectman-rook12}, who were modeling galaxy images, we parameterize  as follows:
\begin{equation}
f_{\rm spiral}=\left[1-w+\prod_{n=2,n+2}^{N}\frac{n}{n-1}w\sin^{N}\left(\frac{\ln(\varpi)}{\tan(p)}
-\Omega+45^{\circ}\right)\right]
\label{eqn_spirality}
\end{equation}
This function is a multiplicative factor on the density (as opposed to the additive factor for the warps and rim variations),
and is applied beyond a user-specified radius, $R_{spiral}$.
$N$ determines the width of the arms,  $w$ is the fraction of mass in the arms, and $p$ is the pitch angle which determines
how ``wound-up'' the arms are.  
Figure \ref{f_3col_spiral} shows an image of a disk with a gap and spiral density enhancements outside the
gap radius.  The spiral arm parameters are $N=2$, $w=0.7$, and $p=10$.

To account for all the effects in \S\ref{s_hotspot} and \ref{s_gaps_etc}, the scale height is calculated as:

\begin{equation}
h(\varpi , \Omega) = h_0 \left( {{\varpi}\over{R_\star}}\right)^\beta \left(1+ h_{warp}+h_{rim}-h_{curve}    \right)
\end{equation}
and the density is
\begin{equation}
\rho(\varpi , z) = \rho_{fiducial} * f_{spiral}/(1+f_{warp}+f_{rim}-f_{curve}),
\label{eqn_rhoscale}
\end{equation}
where $f_{warp}=h_{warp}/h_{fid}$, etc.   The denominator in equation \ref{eqn_rhoscale}
scales the density to keep the surface density (integral along $z$) the same as if there are no modifications to scale height.



\subsection{Vertical Structure of the disk in hydrostatic equilibrium}\label{hseq}

\label{s_hseq}

Our default disk density structure is parameterized with a radial power law and vertical Gaussian structure \citep{shakura73}.  In hydrostatic equilibrium, if the temperature is assumed to be isothermal in the vertical direction and decreasing radially via a power law, the  vertical scale height increases with cylindrical radius as a power law.  That is, 
\begin{equation}
\rho=\rho_0 
\left( {R_\star \over \varpi} \right)^\alpha\
\exp\left\{{ -{1\over 2} \left[{z\over h(\varpi )}\right]^2  }\right\}\; ,
\label{diskdensity}
\end{equation}
where $\varpi$ is the cylindrical radius, and the
scale height is
$h=h_0\left ( {\varpi /{R_\star}} \right )^\beta$.
%
The surface density profile is the integral of the density over $z$, and therefore follows:
\begin{equation}
\Sigma = \Sigma_0 \varpi^{-p}, \label{eqn_sig}
\label{surfdens}
\end{equation}
 where $p = \alpha - \beta$, and $\Sigma_0$ is set by the mass of the disk:
\begin{equation}
M_{disk}=2 \pi \int_{R_{min}}^{R_{max}} \Sigma~ \varpi d\varpi. \label{eqn_mass}
\end{equation}
The exponent $p$ is usually chosen to be 1 following \citet{dalessio01} or 1.5 following the ``minimum solar nebula'' value \citep{weidenschilling77}.

We can solve for the hydrostatic equilibrium density structure by iterating on the temperature calculated from
the radiative equilibrium solution of the radiation transfer equation, and the density calculated from the equation
of hydrostatic equilibrium.  
Since we use a spherical polar grid, we solve the equation for hydrostatic equilibrium along the polar $\theta$ direction:
\begin{equation}
\frac{1}{r} \frac{dP}{d\theta}=-\rho \frac{\cos\theta}{\sin\theta} \frac{v_\phi^2}{r},
\end{equation}
where  $P= \rho c_s^2$ is the gas pressure, $c_s^2 = kT/(\mu m_H)$ is the isothermal sound speed, and $v_\phi$ is the
Keplerian velocity given by
\begin{equation}
v_\phi^2 = \frac{G M_\star \varpi^2}{r^3} = \sin^2\theta \frac{GM_\star}{r}.
\end{equation}
Note that this formulation is equivalent to the more familiar solution along $z$ ($dP/dz=-\rho g_z$, where $g_z$ is the vertical component of gravity ($=G M_\star z/\varpi^3)$).

Substituting for $\rho$ and reformulating into a difference equation, we calculate
\begin{equation}
\rho_i = \rho_{i-1} exp \left (  -\frac{GM_\star}{2r} \frac{1}{c_{s,i}^2} \left [ \cos^2 \theta_i -\cos^2 \theta_{i-1}  \right ]     \right ).
\label{densi}
\end{equation}
With density specified in the midplane at a given radius $r$, all the densities along the polar $\theta$ coordinate can be computed from equation \ref{densi}.  Next we have to normalize this density profile along $\theta$.  We do this
by requiring the integral along $\theta$ to equal that calculated in the initial grid setup using the input parameters for the power-law disk structure in equation \ref{diskdensity}.  This integral as a function of $r$ is nearly identical to the surface density (equation \ref{surfdens}), so in effect, we are normalizing the density so that the surface density remains constant after each iteration.

We have found that the temperature and density converge within about 8 iterations for typical T-Tauri disk parameters.  A noisy temperature solution can lead to noise spikes in the density solution so a higher number of photons should be processed during each iteration than in the power-law disk models.  We have found that 100 photons per grid cell is sufficient (that is, using the default 2-D setup, with 400 radial grid cells (NRG) and 197 theta cells (NTG), setting the iteration photons NPMIN equal to 10,000,000 will usually give stable hydrostatic equilibrium solutions for typical disk masses).  We plan to implement a temperature smoothing algorithm to speed up the process of convergence.  

\begin{figure}
\epsscale{1.0}
\includegraphics[angle=0,width=3.2in]{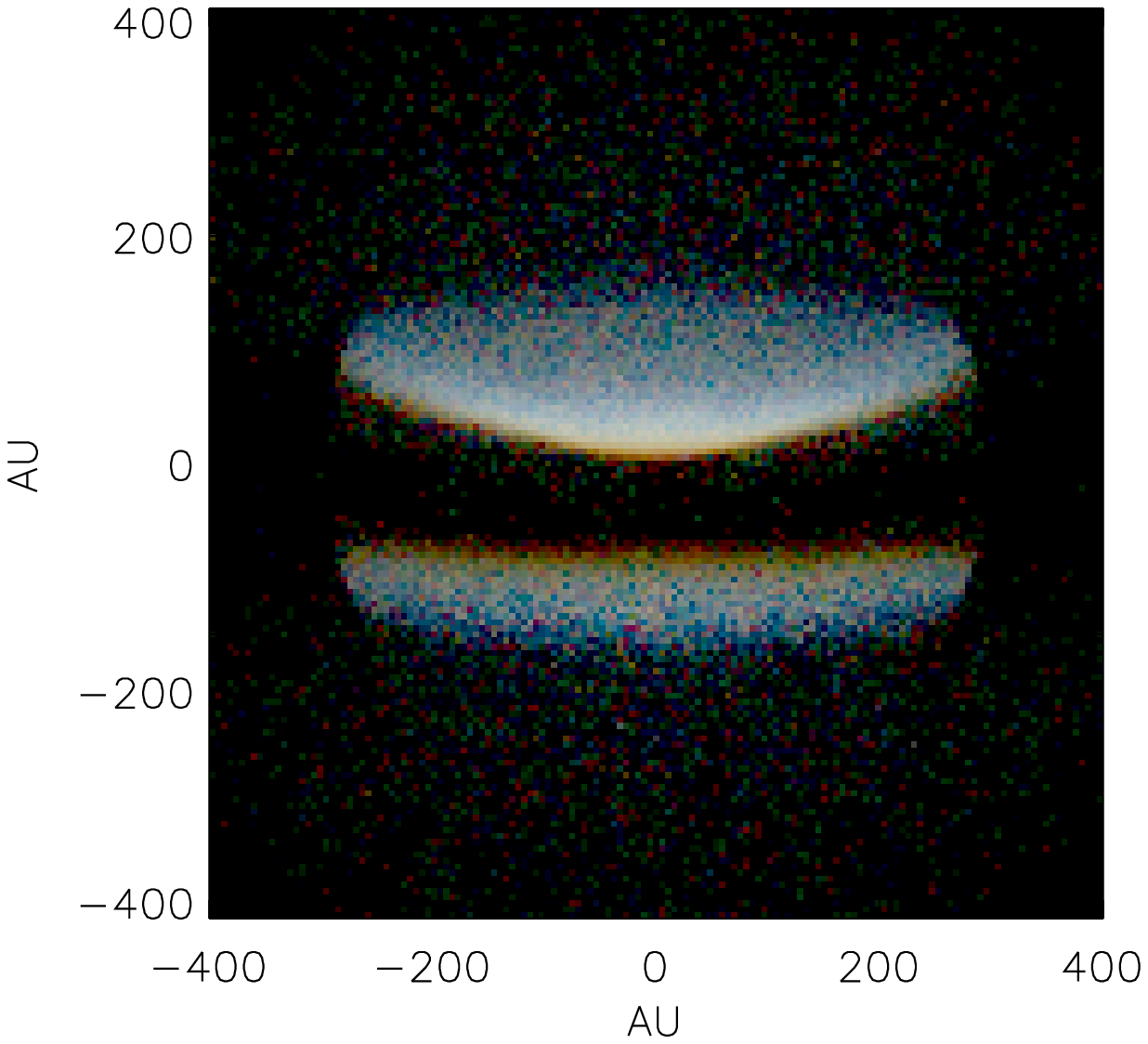}
\includegraphics[angle=0,width=3.2in]{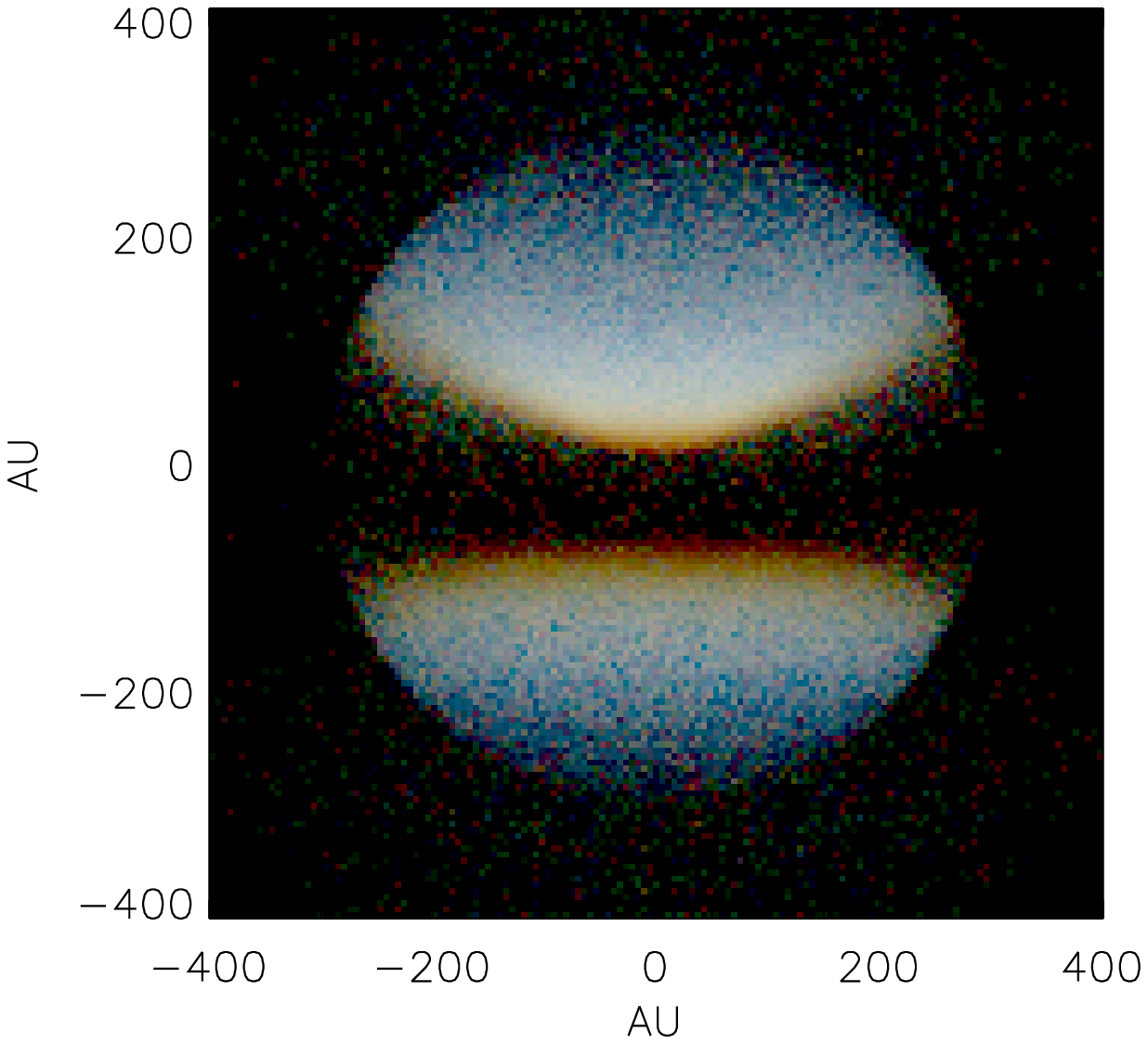}
\caption{
Left: Near-infrared 3-color (JHK) image of the standard Class II model (left) compared to the hydrostatic equilibrium (HSEQ) solution for the small-grain disk (right).   The HSEQ solution gives a more flared disk.
\label{f_3col_hseq}}
\end{figure}

Our hydrostatic equilibrium (HSEQ) solution results in density structures with significant
vertical extent, or flaring, as shown
in Figure \ref{f_3col_hseq}.  This figure compares a near-edge on image of our fiducial Class II model with the HSEQ solution.
Our models produce more flaring than those of \citet{dalessio97}.  Their models were not fully 2-D and did not
include radial transport of radiation, just irradiation from above.  
Our models include radiation from above as well as the horizontal transport inside the disk which leads to higher temperatures in the disk, and therefore higher vertical extent.

These models add more weight to the argument that in most protoplanetary disks, the dust is settled with respect
to the gas \citep{dullemond04,dubrulle95,schrapler04,furlan05}.  Since the gas dominates the mass of the disk (usually taken to be 100 times the dust mass), the HSEQ 
solution applies predominantly to the gas, and the dust is likely settled.   Our models can be used to estimate the amount
of dust stratification by first modeling images and/or SEDs with the parameterized disk:  varying the dust properties and scale heights in the two disk grids provided;  this will determine the actual extent of the dust flaring.  Then  calculating the HSEQ structure of both of the disks (large-grain and small-grain) shows what the dust extent would be without settling.

Figure \ref{f_trho_hseq} shows azimuthal temperature and density slices for the HSEQ disk and the standard Class II
model.  The density shows a slight puffing up of the inner wall of the disk, as shown by \citet{dullemond02}.  


\section{Conclusions}

We summarize here some of the new features in \textsc{Hochunk3d} and interesting results from initial investigations.

The  code includes new additional emission processes: PAH emission, and external illumination.  Now there are 3 sources of luminosity available:  stellar, disk accretion, and external illumination by the Interstellar Radiation Field (ISRF).
The code is parallelized, and includes two disk grids to enable simple dust stratification models.  Each grid
cell currently has 8 grain types for the 4 geometric structures: 2 disks, envelope, outflow; and for thermal vs PAH processes.  Sample grain model files are provided.  More can be created using a publicly available code \citep{bh83}.\footnote{available at \url{https://github.com/hyperion-rt/bhmie}}

The code includes several new 2-D and 3-D structures:  a simple power-law envelope can be substituted for the infall+rotation solution; fractal clumping in any or all of the disks, envelope and bipolar cavity; multiple cavity walls in the envelope; stellar accretion hotspots and inner disk warps; gaps, curved and puffed-up inner rims; misaligned inner disks, spiral arms;  and the hydrostatic equilibrium solution.

Some initial results of the code are as follows:

1)  High-mass main sequence stars in the vicinity of optically thin dust can show large IR excesses, with rising infrared
SEDs that resemble Class I sources, even more apparent if foreground dust extincts the optical flux from the central star.  This is because the stars heat up such a large volume due to their luminosity (\ref{s_pah}).  Thus these stars can ``masquerade'' as YSOs.  The presence of bright PAH features, no silicate absorption, and an optically bright source can be an indicator of this situation.   
These could be compared to
datasets, for example the SAGE-SPEC classifications of mid-IR sources in the Large Magellanic Cloud \citep{woods11, kemper10} to help determine evolutionary sequences for high-
and intermediate-mass YSOs.   PAH models can also determine in what regimes models that do not include PAHs
are appropriate.  For example the YSO SED grid\footnote{\url{http://www.astro.wisc.edu/protostars/}} \citep{robitaille06} does not include PAH emission.  Models with PAH emission can
verify what ``errors'' should be applied to the data when fit with models that do not include PAH emission. 

2) Distant externally illuminated clumps  also exhibit rising infrared SEDs that resemble YSOs.  However, they
tend to fall in a certain region of mid-IR color-magnitude space \citep{sewilo12}, so their colors in combination with radiation transfer models
can distinguish them.

3) In comparing our clumpy models to high resolution images \citep{padgett99}, we get the best agreement with small
deviations in the envelope and large deviations (more clumpiness) in the outflows.  Perhaps the outflows are more turbulent and the infall clumps are smeared out by the infall and rotation.  Comparing clumpy models to image observations from visible to radio (including velocities) could determine the amount of clumpiness in envelopes and outflows and provide a measure of turbulence in these regions.  Of course, in addition to clumpiness, there could be smooth non-axisymmetric structures, so clumpiness is not the only way to make our models more ``realistic.''



4) Our initial HSEQ models find the gas and dust to be more vertically extended than most observations indicate.
This is in agreement with a wealth of observations and theories suggesting that dust is settled
\citep{dalessio97,dalessio06,furlan05,dullemond04,mulders12}
Our models are even more vertically extended than those of \citet{dalessio97,dalessio06}.  We suggest this is due 
to radial transport of heat, which is not included in their models.
 
\acknowledgments

This work was supported 
by the NASA Astrophysical Theory Program (NNG05GH35G, BAW);
 by NASA's Spitzer Space Telescope Fellowship (TR), the Spitzer GLIMPSE projects (RSA 1368699 \& 1367334, BAW),
 and the Spitzer YSOVAR project (RSA 1368444, BAW).


\clearpage


\clearpage

\begin{figure}
\epsscale{1.0}
\includegraphics[angle=0,width=3.2in]{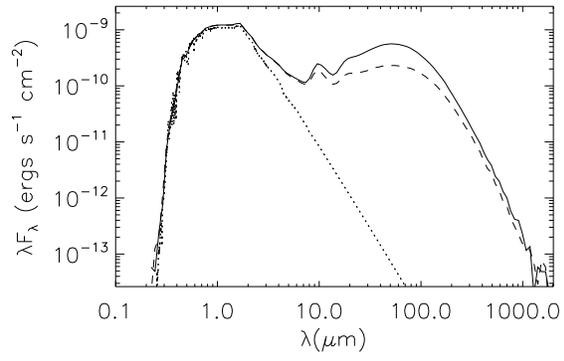}
\caption{
SED of the HSEQ disk (solid line) compared to the standard Class II model for a viewing angle of 32\arcdeg (dashed line).  Since the HSEQ disk is more flared, it has more mid-infrared emission.
\label{f_sed_hseq}}
\end{figure}

\begin{figure}
\epsscale{1.0}
\includegraphics[angle=0,width=6.in]{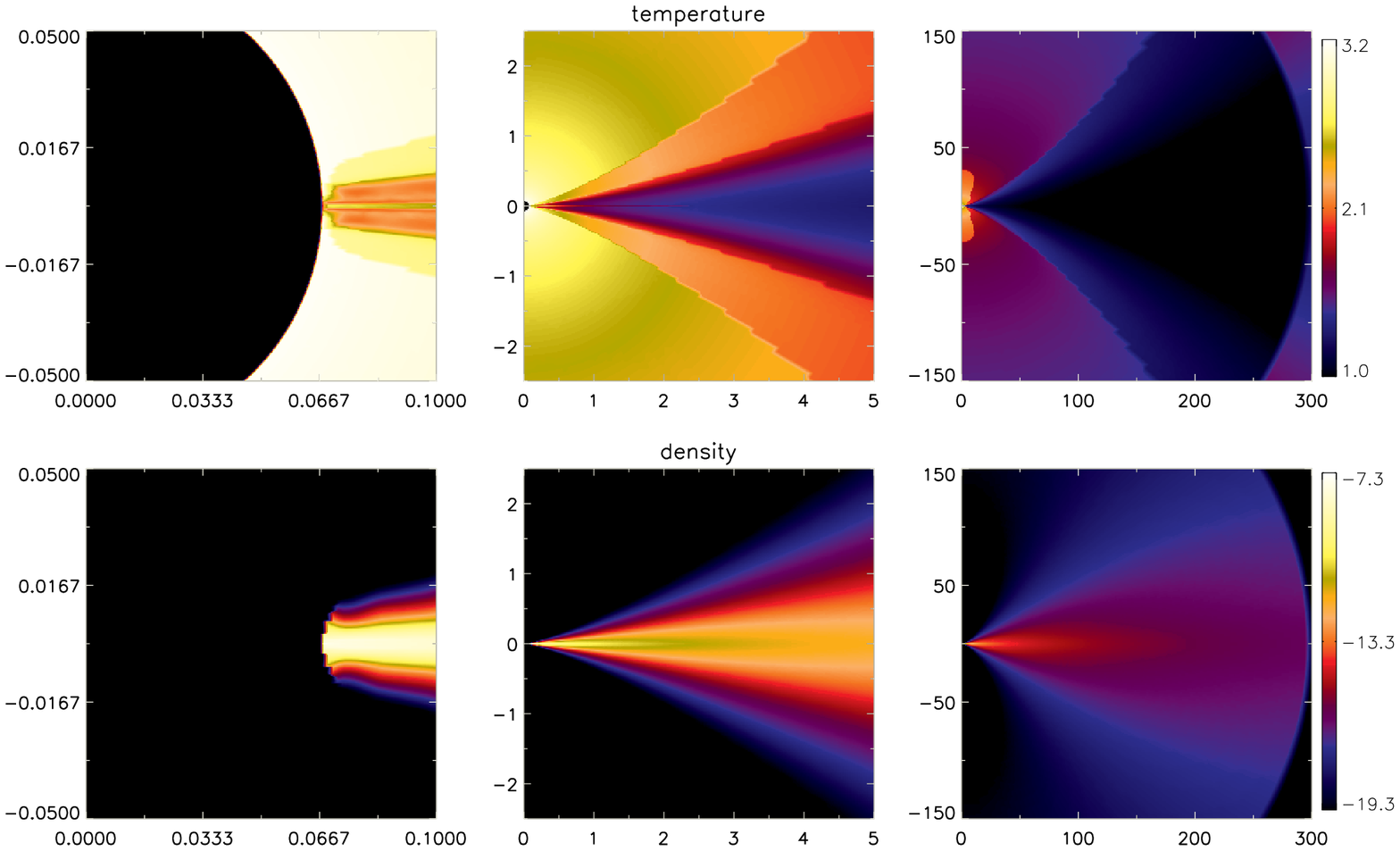}
\includegraphics[angle=0,width=6.in]{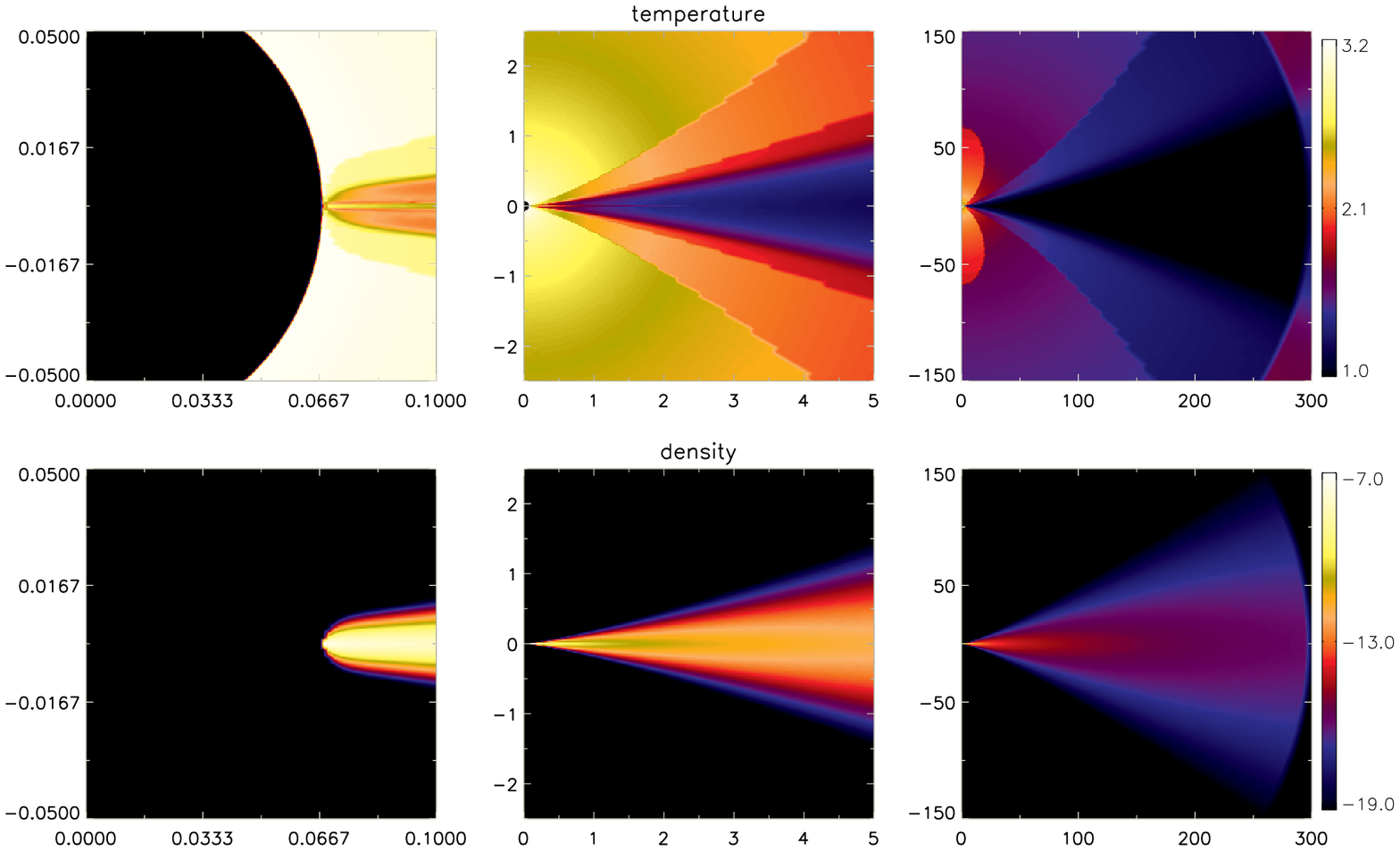}
\caption{
Azimuthal temperature and density slices for the HSEQ disk (top two panels) and the standard Class II model (bottom
two panels).    From left to right are three different zooms to show the features at different radii.
\label{f_trho_hseq}}
\end{figure}

\clearpage


\begin{deluxetable}{llll}
\tablecaption{Dust models and their locations.\label{t_dust}}
\tablehead{
\colhead{Dust array index} &  \colhead{Description} & \colhead{Region} & \colhead{File} 
}
\startdata
1 & large grain & disk 1 (flatter)\tablenotemark{a}  & www003, www006, www005\tablenotemark{b} \\
2 & small grain & disk 2 & kmh\tablenotemark{c} \\
3 & molecular cloud model & envelope & r400\_ice095\tablenotemark{d} \\
4 & small grain & outflow cavity & kmh \\
5 & PAH/VSG & disk 1 & draine\_opac\_new\tablenotemark{e} \\
6 & PAH/VSG & disk 2 & draine\_opac\_new \\
7 & PAH/VSG & envelope & draine\_opac\_new \\
8 & PAH/VSG & outflow cavity & draine\_opac\_new \\
\enddata
\tablenotetext{a}{Dust stratification can be simulated by giving disk 1 flatter structure than disk 2, either with a lower scale height,
smaller flaring, or both}
\tablenotetext{b}{These correspond to Models 1, 2, 3 respectively in \citet{wood02_hh30}}
\tablenotetext{c}{This is an average Galactic ISM grain model \citep{kmh94}}
\tablenotetext{d}{This is the ``envelope'' grain model in \citet{whitney03p2} (Table 3, \S2.3, Fig. 2), with a ratio of total-to-selective extinction, $r_V = 4.3$ (typical of star forming regions) and water ice mantles covering the outer 5\% in radius}
\tablenotetext{e}{This  model was computed by Bruce Draine \citep{draine07}, and is discussed in \citet{wood08}}
\end{deluxetable}

\end{document}